\documentclass[12pt,letterpaper]{article}
\usepackage[a4paper, total={7in, 10in}]{geometry}

\usepackage{graphicx}
\usepackage{helvet}
\usepackage{authblk}
\usepackage{hyperref}
\usepackage{amsmath} 
\usepackage{amssymb} 
\usepackage{orcidlink} 
\usepackage[super,comma,sort&compress]  
   {natbib}
\usepackage[right]{lineno} \linenumbers

\usepackage{lipsum}
\usepackage{subcaption}

\usepackage{caption}
\usepackage{booktabs}
\usepackage{longtable}
\usepackage{array}

\usepackage[normalem]{ulem}
\usepackage{cancel}
\usepackage[table]{xcolor}
\newif\ifshowrevisions
\showrevisionsfalse
\newcommand{\added}[1]{\ifshowrevisions{\color{blue!75!black}#1}\else#1\fi}
\newcommand{\deleted}[1]{\ifshowrevisions{\color{red!70!black}\ifmmode\bcancel{#1}\else\sout{#1}\fi}\else\fi}
\newcommand{\replaced}[2]{\deleted{#1}\added{#2}}

\makeatletter
\renewcommand{\maketitle}{\bgroup\setlength{\parindent}{0pt}
\begin{flushleft}
  \textbf{\@title}
  
  \@author
\end{flushleft}\egroup}
\makeatother

\title{Technology configurations for decarbonizing residential heat supply through district heating and implications for the electricity network}

\date{}

\author[1,6,*,\orcidlink{0009-0002-1905-8221}]{Christian Doh Dinga}
\author[2,\orcidlink{0000-0002-7624-5886}]{Francesco Lombardi}
\author[3]{Roald Arkesteijn}
\author[4]{Arjan van Voorden}
\author[5, \orcidlink{0000-0001-6159-041X}]{Sander van Rijn}
\author[2, \orcidlink{0000-0002-4014-9294}]{Laurens J. de Vries}
\author[1, \orcidlink{0000-0002-5169-7368}]{Milos Cvetkovic}

\affil[1]{Department of Electrical Sustainable Energy, Delft University of Technology, 2628 CD, Delft, the Netherlands}
\affil[2]{Department of Engineering Systems and Services, Delft University of Technology, 2600 GA, Delft, the Netherlands}
\affil[3]{Eneco, District Heating Network Operator, 3068 AV, Rotterdam, The Netherlands}
\affil[4]{Stedin, Electricity Distribution Network Operator, 3011 TA, Rotterdam, The Netherlands}
\affil[5]{Environment and Sustainability Section, Netherlands eScience Center, 1098 XH, Amsterdam, the Netherlands}
\affil[6]{Lead contact}

\affil[*]{Correspondence: c.dohdinga@tudelft.nl}

\begin{document}

\maketitle

\section*{SUMMARY}

District heating networks (DHNs) have significant potential to decarbonize residential heating and accelerate the energy transition. However, designing carbon-neutral DHNs requires balancing several objectives, including economic costs, social acceptance, long-term uncertainties, and grid-integration challenges arising from electrification. By combining modeling-to-generate-alternatives with power flow simulation techniques, we develop a decision-support method for designing carbon-neutral DHNs that are cost-effective, socially acceptable, and impose minimal impacts on the electricity grid. Applying our method to a Dutch case, we find substantial diversity in how carbon-neutral DHNs can be designed. The flexibility in technology choice, sizing, and location enables accommodating different real-world needs and achieving high electrification levels without increasing grid loading. For instance, intelligently located heat pumps and thermal storage can limit grid stress even when renewable baseload heat sources and green-fuel boilers are scarce. Using our method, planners can explore diverse carbon-neutral DHN designs and identify the design that best balances stakeholders' preferences.




\section*{INTRODUCTION}

Heating is the largest energy end-use globally, accounting for half of total energy consumption \cite{iea2014_heating_global_warming, iea2025_renewables}. Most of the heat generated is used in the residential sector\textemdash second only to industry\textemdash to meet space heating and domestic hot water needs in buildings \cite{iea2021_heating, Gilbert2023_joule}. In the European Union (EU), residential heating is responsible for 40\% of total energy consumption, with roughly two-thirds of the heat supply still based on fossil fuels \cite{Hofmann2025_energy, jrc_2023_heatpump_wave}. The reliance on fossil fuel for residential heating is even more pronounced in some individual EU Member States; for example, in the Netherlands, around 85\% of residential heating is supplied by natural gas \cite{Gross2019_ne, kaandorp2023_urbanheating}. Consequently, the decarbonization of residential heating has become a central pillar for accelerating the energy transition globally \cite{Henry2020_nature_energy, iea2023_netzero, staffell2023_global_space_heating_ne}, and particularly in the EU \cite{Thomaen2021_ep, 2025_jrc_energies, mathiesen2025_heat_roadmap_eu}. Besides climate goals, decarbonizing residential heating will also reduce the EU's reliance on imported fossil fuels and increase its energy security in an era of fading geopolitical stability \cite{iea2022_avoid_gas_dependence, Draghi_2025_part_a}.



There are several potential pathways for decarbonizing residential heat supply \cite{jrc2019_decarbonizing_heat, Narula2019_energy, ZEYEN2021120784_energy, Friebe2023_energy, berrill2022_use_ne}. However, two main pathways have received particular attention in policy debates and academia: decentralized heating using individual building-level heat pumps \cite{Waite2020_joule, Peacock2023_ae, Mascherbauer2025_ae}, and centralized heating through carbon-neutral \added{(net-zero)} district heating networks (DHNs) \cite{jrc_2022_efficient_dhc, FALLAHNEJAD2024_ae, jrc_2025_smart_thermal_networks}. While decentralized heating using individual heat pumps has significant potential to decarbonize residential heat supply \cite{iea2022_heatpumps}, large-scale deployment faces several practical limitations, such as substantial space requirements and the need for higher building insulation \cite{Liu2021_jcp, ROCAREINA_jrc_2024_eb}. This poses a significant challenge in the EU, where most of the residential building stock is old, poorly insulated, and difficult to retrofit at scale \cite{jrc_2023_heatpump_wave}. Centralized heating through carbon-neutral DHNs offers an alternative or complementary decarbonization pathway, as it can meet user preferences across buildings with different characteristics \cite{jrc_2022_efficient_dhc, jrc_2025_smart_thermal_networks}. Moreover, DHNs enable the integration of diverse heat sources and thermal energy storage, which increases system flexibility and resilience \cite{Malcher2025_rser}. The deployment of DHNs is particularly appealing in the EU since approximately 70\% of buildings are located in urban areas with high heat demand density, making district heating more cost-efficient than decentralized heat pumps \cite{Connolly2014_ep, hansen2018_competitiveness_dhn, MOLLER2019_heat_roadmap_energy}. As highlighted in the Heat Roadmap Europe reports, district heating will play a key role in the transition to a sustainable, resilient, and affordable energy system in Europe, with its market share in the heat supply mix set to expand from 13\% today to 55\% by 2050 \cite{MOLLER2019_heat_roadmap_energy, mathiesen2025_heat_roadmap_eu}.




Given its key role in the heat transition, many studies have investigated different strategies to design cost-effective DHNs that, unlike traditional DHNs relying on fossil-fuel combustion, use carbon-neutral heat sources. Among these strategies are power-to-heat (P2H) technologies such as large-scale centralized heat pumps and electric boilers that rely on the\textemdash increasingly carbon-neutral\textemdash electricity system for heat production, as well as geothermal and other low-temperature heat sources that require electricity to upgrade heat supply temperatures. Consequently, the decarbonization of DHNs will further accentuate the interactions between heat and electricity systems, making their operation and planning increasingly interdependent \cite{Leitner2019_energy, Palensky_2024_PEM}. This growing interdependence has increased the need for integrated heat–electricity system planning \replaced{\cite{XU2025_joule}}{\cite{XU2025_joule, Lund_2018_smart_energy_systems_vs_smart_grid}}. Prior research on the decarbonization of DHNs has ignored these interactions \cite{HUCKEBRINK_2022_energy, Zhang2022_ecm, PAGANI2020115177_ae, Malcher2024_energy, Popovski2019_energy, Liu2024_ne, Siddique2024_ee, Thommessen2025_ae}, or mostly focused on electricity generation capacity adequacy challenges, to assess the additional generation capacity required to support the decarbonization of DHNs \cite{Ashfaq2018_energy, Zhang2018_ae, jrc2019_decarbonizing_heat}. These studies therefore provide limited insight into electricity network bottlenecks, as they typically rely on highly simplified models that lack spatial resolution.

The electricity network is increasingly recognized as a critical enabler or bottleneck of the energy transition \cite{Few2024_ne, Navidi2023_joule}, and must be modeled with high spatial resolution to determine where and to what extent network upgrades are needed to support decarbonization. In an attempt to address this, recent studies used integrated optimization frameworks to jointly optimize the planning of DHNs and electricity networks \cite{ZEYEN2021120784_energy, Friebe2023_energy, Ehsan2022_energy, Aunedi2020_ae}. However, these studies have two major limitations. First, although spatially resolved, they typically rely on a simplified representation of the electricity network due to computational limitations in including detailed power flow equations in large-scale optimization models. As a result, such models are inadequate for identifying operational bottlenecks, such as transformer and line loadings, which are crucial for the management and planning of electricity network infrastructure \cite{ILYUSHIN_2024_re, RAJAKOVIC_2025_ecmx}. Second, \replaced{integrated optimization assumes common ownership, which is intrinsically misaligned with real-world settings since these networks are typically owned and operated by different stakeholders.}{integrated optimization assumes a central planner perspective \cite{DeCarolis_2017_AE, SHU_2024_iScience}, which is intrinsically misaligned with real-world settings since these networks are typically owned and operated by different stakeholders.} Hence, integrated optimization may result in impractical and misleading conclusions \cite{WIDL2022_segn}. In the literature, model coupling has been proposed as a technique to address these limitations of integrated optimization \cite{lund_2023_model_coupling_energy}. Model coupling links different computational models through interface variables, enabling high-resolution analysis with high-fidelity models while respecting system boundaries \cite{Palensky_2024_PEM, lund_2023_model_coupling_energy}. Despite its advantages, the application of model coupling to investigate the electricity network implications of decarbonizing DHNs is lacking in the literature. Therefore, a high-resolution assessment of electricity network operational bottlenecks arising from the decarbonization of DHNs\textemdash particularly when accounting for more realistic planning practices\textemdash remains largely unexplored.

Another gap persists in the literature: existing studies strive to identify the ‘‘least-cost’’ technology configuration\textemdash under a handful of narrative scenarios\textemdash required to decarbonize DHNs \cite{Ashfaq2018_energy, Popovski2019_energy, Aunedi2020_ae}. However, not looking beyond the ‘‘least-cost’’ solution can be short-sighted due to the inability of cost-optimization models to capture harder-to-model dimensions such as structural uncertainty, social acceptance, system resilience, and other political aspects \replaced{\cite{VOLL2015_optimum_not_enough, NEUMANN_2023_broad_range, sinha2024_nc}}{\cite{VOLL2015_optimum_not_enough, NEUMANN_2023_broad_range, sinha2024_nc, Lund_2017_simulation_vs_optimization}}. This is particularly relevant in the context of decarbonizing DHNs as it requires long-term investment decisions under deep uncertainty, and involves multiple stakeholders with potentially conflicting objectives. For instance, while residual heat from industries is a cost-effective, low-carbon heat source, its long-term availability is uncertain as industries decarbonize or may relocate \cite{LYGNERUD_2018_residual_heat_risks}. Public opposition may limit the deployment of geothermal heat sources \cite{COUSSE_2021_geothermal_resisitance}, while space limitations may restrict the build-out of pipeline infrastructure needed to connect remote heat sources to demand \cite{SNEUM_2025_pipeline_constraints}. These multi-dimensional uncertainties are difficult to parameterize in a model\added{; even sensitivity and parametric uncertainty analysis methods \cite{Lund_2018_beyond_sensitivity} are ill-suited to capture these structural uncertainties,} making the insistence on a single ‘‘least-cost’’ solution of limited practical use to real-world stakeholders \cite{lombardi_2023_what_is_redundant, lombardi_2025_mga_perspective}.

Moreover, studies that focus solely on the ‘‘least-cost’’ system design provide limited generalizable insights for accelerating the decarbonization of DHNs in other regions. This is because DHNs are inherently local systems, and technology choices are strongly shaped by site-specific or local conditions. For instance, the uneven distribution of geothermal resources across Europe \cite{mathiesen2025_heat_roadmap_eu} implies that geothermal heat might be technically infeasible in some regions, despite being a cost-effective heat source. The potential for exploiting residual heat also varies across regions, as its feasibility depends on proximity to demand centers and the temporal alignment between industrial activities and local heat demand. Regions with a highly congested electricity network may experience a significant slowdown in the uptake of P2H technologies, regardless of their economic attractiveness. All these factors play out at a more local scale and define what is technically and socially feasible before cost considerations. Therefore, the lack of analysis beyond purely economic assessment leaves planners with a limited understanding of how disparities in local conditions reshape DHN design, specifically regarding the trade-offs that arise when the deployment of specific technologies is constrained by local conditions.

This paper addresses the above gaps by conducting the first high-resolution assessment of a broad range of technology configurations for decarbonizing DHNs and their potential impacts on the electricity network. Specifically, we address the following questions: (i) What is the range of alternative, economically-comparable technology configurations for decarbonizing DHNs, and to what extent do these configurations differ in their electricity network impacts? (ii) What trade-offs are involved when the deployment of certain low-carbon heat supply technologies is constrained by local conditions, and what are the spillover effects on the electricity network?

In doing so, we develop a model coupling framework that links a capacity-expansion and operation-optimization model of a DHN to an AC power flow simulation model of an electricity network. Model coupling enables a high-resolution assessment of temporally and spatially resolved electricity network bottlenecks that are difficult to capture with monolithic integrated optimization frameworks. Also, model coupling more closely reflects real-world practices as it retains individual ownership of each network model. Our approach follows a sequential workflow (see section~\hyperref[methodology_framework]{Methodology framework} in experimental procedures) in which we first apply the Modeling to Generate Alternatives (MGA) technique to the DHN model to generate many near-optimal carbon-neutral DHN designs. MGA allows us to explicitly account for structural uncertainty by systematically exploring a broad option space of similarly costly but technologically diverse carbon-neutral DHN designs. By moving beyond prescribing a single “least-cost” design to uncovering an option space of alternatives, we reveal the flexibility in choice, location, and sizing of heat supply technologies, thermal storage, and heat pipeline infrastructure required to achieve carbon neutrality. This option space can help planners understand the trade-offs required to accommodate harder-to-model dimensions\textemdash such as social acceptance and long-term uncertainty in resource availability\textemdash and to balance potentially conflicting preferences across different stakeholders.

Subsequently, we quantify the electricity network impacts of each carbon-neutral DHN design by using its spatially resolved electricity profiles as inputs to the AC power flow simulation. We synthesize the outputs from both models into an integrated heat-electricity network planning decision space that maps DHN design characteristics to their corresponding electricity network loadings. Network operators can explore this decision space to examine cross-network trade-offs and support coordinated network expansion planning. In sum, using our method, energy system planners can identify economically comparable DHN designs that are \replaced{more robust to future risks}{less vulnerable to future risks} and impose minimal impacts on the electricity network.

To showcase the potential of our method, we apply it to a case study in the Dutch region of South Holland. We focus on one of its largest existing DHNs (see Figure~\hyperref[heat_network_fig]{S1}), which is projected to expand substantially in the future, reaching an annual heat demand of about 2700 GWh and a peak capacity of 1400 MW by 2050 (see Figure~\hyperref[heat_demand_fig]{S3}). For the electricity system, we model the current regional network (see Figure~\hyperref[electricity_network_fig]{S2}). In addition to the spatial and temporally resolved electricity profiles from the DHN model, we also include profiles from other consumers connected to the same network. This allows us to assess the overall grid impact induced by the decarbonization of the DHN in the context of broader regional electrification. As a starting point, we apply MGA to generate 515 near-optimal solutions that represent alternative DHN designs that are maximally different from the least-cost solution while total system cost remains within 10\% of the least-cost case. Additionally, we examine the effects of tightening and further relaxing this cost budget as a proxy for technology cost uncertainty, and also investigate scenarios with different weather years and demand projections (see section~\hyperref[sensitivity_analysis_results]{Sensitivity analysis results} in supplemental information). This yields a total of 3,605 alternative carbon-neutral DHN designs, each of which is subject to a full-year, hourly-resolution AC power flow simulation to quantify their corresponding electricity network impacts using transformer and line loadings.  Given the extent of our analysis, we cannot examine all results with the same degree of detail here. However, we release all our results freely on Zenodo \cite{doh_dinga_2026_data}, and also make our code open source \cite{doh_dinga_2026_code} for transparency and reproducibility.

 
Our contribution extends beyond the methodology as we also advance the broader discussion on planning future carbon-neutral DHNs. First, our systematic exploration of the near-optimal design space captures the interactions between technology options and reveals untold design features and associated trade-offs. For instance, we show the existence of diverse strategies to achieve higher DHN electrification levels without increasing electricity network loading\textemdash such as spatially distributing the deployment of P2H units but at the expense of more heat pipeline infrastructure build-out, or by prioritizing heat pumps over electric boilers, with additional thermal storage to limit heat pipeline expansion. Second, by considering many technologies and explicitly analyzing cases in which certain options are unavailable, we make our findings generalizable by capturing trade-offs that could appear in any region, showing how location-specific constraints shape the design space. Finally, our extensive discussion provides insights into the potential role of different technologies in future carbon-neutral DHNs, particularly when interactions with the electricity network and resource constraints are considered.

We approach the presentation of our results in four steps. We first discuss the model results for the least-cost solution as a basis of comparison. Next, we present a broader option space of near-optimal carbon-neutral DHN designs to reveal the trade-offs it would entail to accommodate certain design preferences. We then assess these alternative DHN designs within an integrated heat-electricity network planning decision space to quantify their electricity network impacts. Finally, we investigate how local technology availability or deployment constraints reshape the feasible design space and associated electricity network impact outcomes.


\section*{RESULTS}

\section*{Cost-minimization concentrates investment in technology configurations with high deployment risks and electricity network loading}\phantomsection\label{potentially_problematic_least_cost_design}

The model results for the conventional cost-minimization run are shown in Figure~\hyperref[least_cost_result_figs]{1}. The least-cost system design for decarbonizing the DHN suggests a high investment in electric boilers and green gas boilers, each reaching roughly 400~MW of installed capacity. This is complemented by 200~MW of industrial residual heat and 185~MW of geothermal capacity, while heat pumps and hydrogen boilers play only a marginal role (also see Figure~\hyperref[heat_dispatch_duration_curve]{S11}), with a combined capacity of about 25~MW. The system further relies on short-term thermal storage in steel water tanks, providing 320~MW of discharge thermal power and 1920 MWh of energy buffer, together with a moderate build-out of heat pipeline infrastructure of around 415~MW.

Although cost-optimal and technically feasible, several features of the least-cost solution appear potentially undesirable when viewed through a broader planning lens. First, the least-cost design relies heavily on green gas\textemdash which is biogas upgraded to the Dutch natural gas quality standards such that it can be directly injected into the current natural gas grid \cite{Bekkering2023_greengas_ecm}. Energy transition scenarios in Europe anticipate strong cross-sectoral competition for biomass \cite{wu_and_stefan_2023}, which is needed for green gas production. As biomass is both resource-limited and contested \cite{millinger_2025_biomass_ne}, uncertainty remains regarding the amount of green gas that will be allocated for residential heating \cite{Grecu2025_energy}. Therefore, while cost-effective, large green gas capacity build-out could result in underutilized capacity or even undermine system resilience if green gas availability becomes more constrained or expensive than projected. Second, more than half of the baseload capacity is provided by industrial residual heat from a nearby petrochemical industry in Rotterdam. However, as for many similar industrial residual heat projects globally \cite{LYGNERUD_2018_residual_heat_risks, oldershaw_2016_residual_heat_barriers, LYGNERUD_2022_residual_heat_risks}, the long-term availability of this heat source is also uncertain, given societal opposition to fossil-fuel industries and the potential decline of refinery or similar heavy-industry activities as economies decarbonize. Third, the least-cost solution entails a substantial deployment of geothermal, which aligns with the real-world ambition to develop four geothermal wells in The Hague (see Figure~\hyperref[geothermal_deployment_lc]{S10}). Still, concentrating geothermal infrastructure in a single urban area may face practical barriers, including social acceptance and permitting challenges \cite{COUSSE_2021_geothermal_resisitance, SPAMPATTI_2022_geothermal_resisitance}.

Finally, another potentially undesirable feature is the higher deployment of electric boilers instead of more efficient heat pumps. This is because electric boilers are about five times cheaper than heat pumps in investment cost (see Table~\hyperref[dhn_technology_params_table]{S1}), and therefore\textemdash despite the lower efficiency\textemdash are an economically attractive heat electrification option for DHN operators. However, a high reliance on less efficient electric boilers imposes severe stress on the electricity grid, as shown by the transformer and line loading duration curves in Figure~\hyperref[least_cost_result_figs]{1}. Moreover, to minimize heat network expansion costs, the cost-optimal deployment of electric boilers is near demand centers, further increasing loading in already congested transformers and lines while those with more available capacity remain underutilized (see Figures~\hyperref[lc_loading_duration_all_trafos]{S12} and~\hyperref[lc_loading_duration_all_lines]{S13} for the loading duration curves of all transformers and lines in the electricity network). From an integrated system perspective, the implication is a shift in costs from the DHN to the electricity grid since accommodating the resulting peak electricity flows would require substantial investment in grid reinforcement. These outcomes are an artifact of cost-minimization, which hinges solely on economic performance to identify the ``least-cost" system design. Therefore, solely relying on cost-minimization to design carbon-neutral DHNs may result in technology pathways that are more vulnerable to future risks and uncertainties, and that impose substantial stress on the electricity grid.

\newpage

\begin{figure}[h!]
    \centering

    \begin{subfigure}[t]{0.75\linewidth}
        \centering
        \caption{Technology configuration for the least-cost design}
        \includegraphics[width=\linewidth]{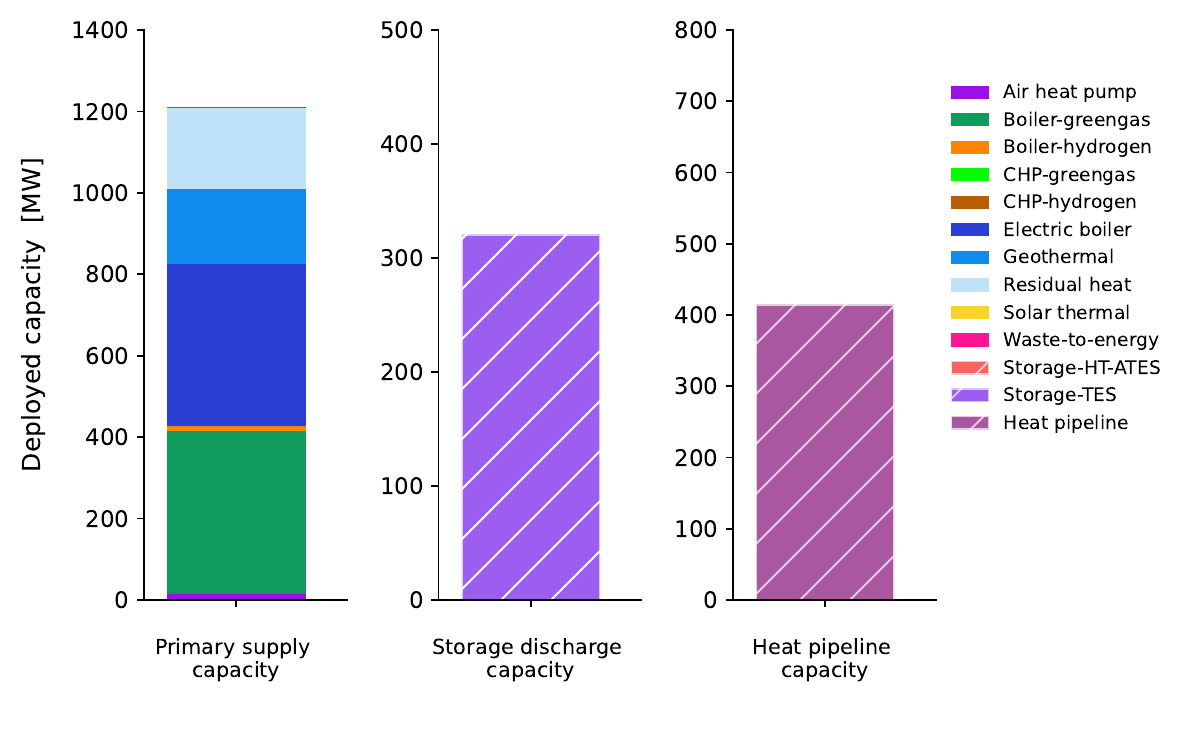}
        \label{fig:tech_mix}
    \end{subfigure}

    \vspace{-2em}

    \begin{subfigure}[t]{0.9\linewidth}
        \centering
        \caption{Transformer loading duration curve}
        \includegraphics[width=\linewidth]{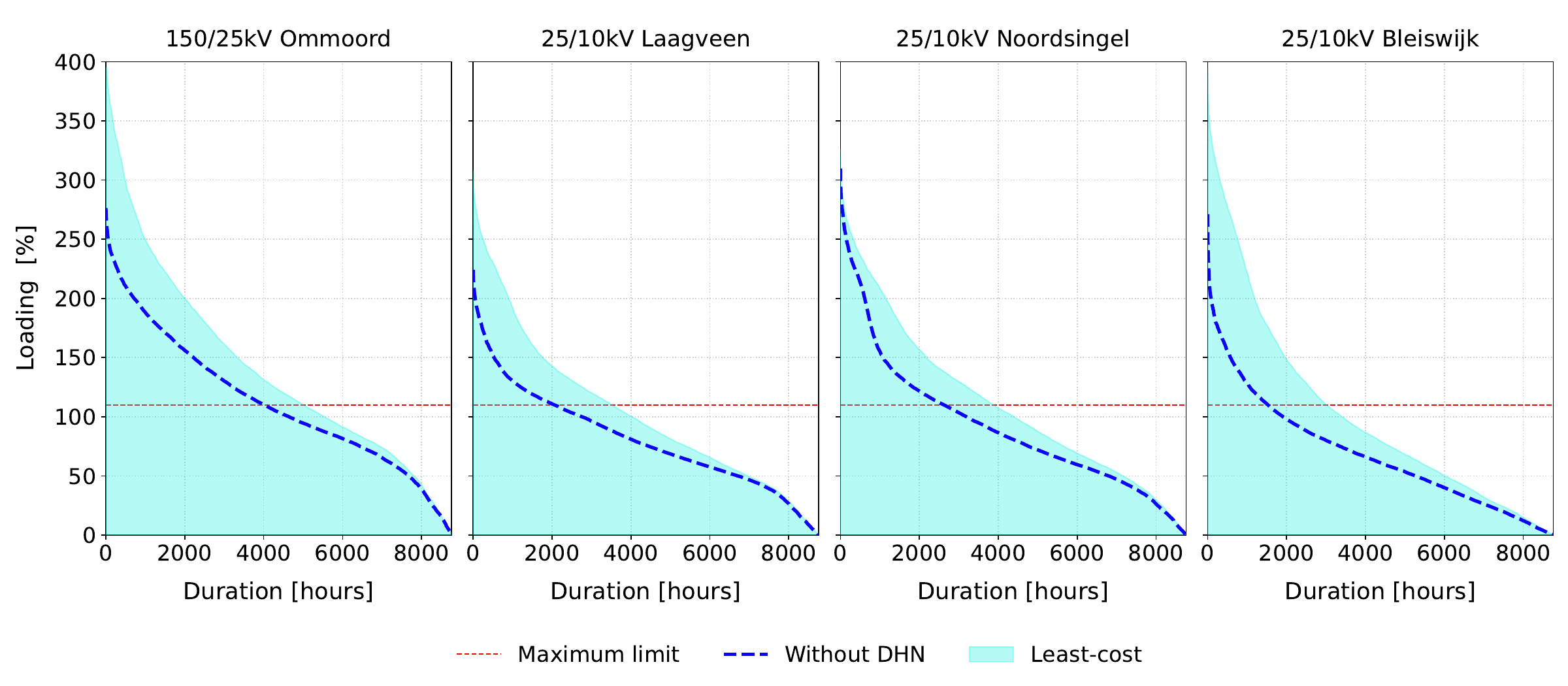}
        \label{fig:transformer_dc}
    \end{subfigure}

    \vspace{-1em}
    
    \begin{subfigure}[t]{0.9\linewidth}
        \centering
        \caption{Line loading duration curve}
        \includegraphics[width=\linewidth]{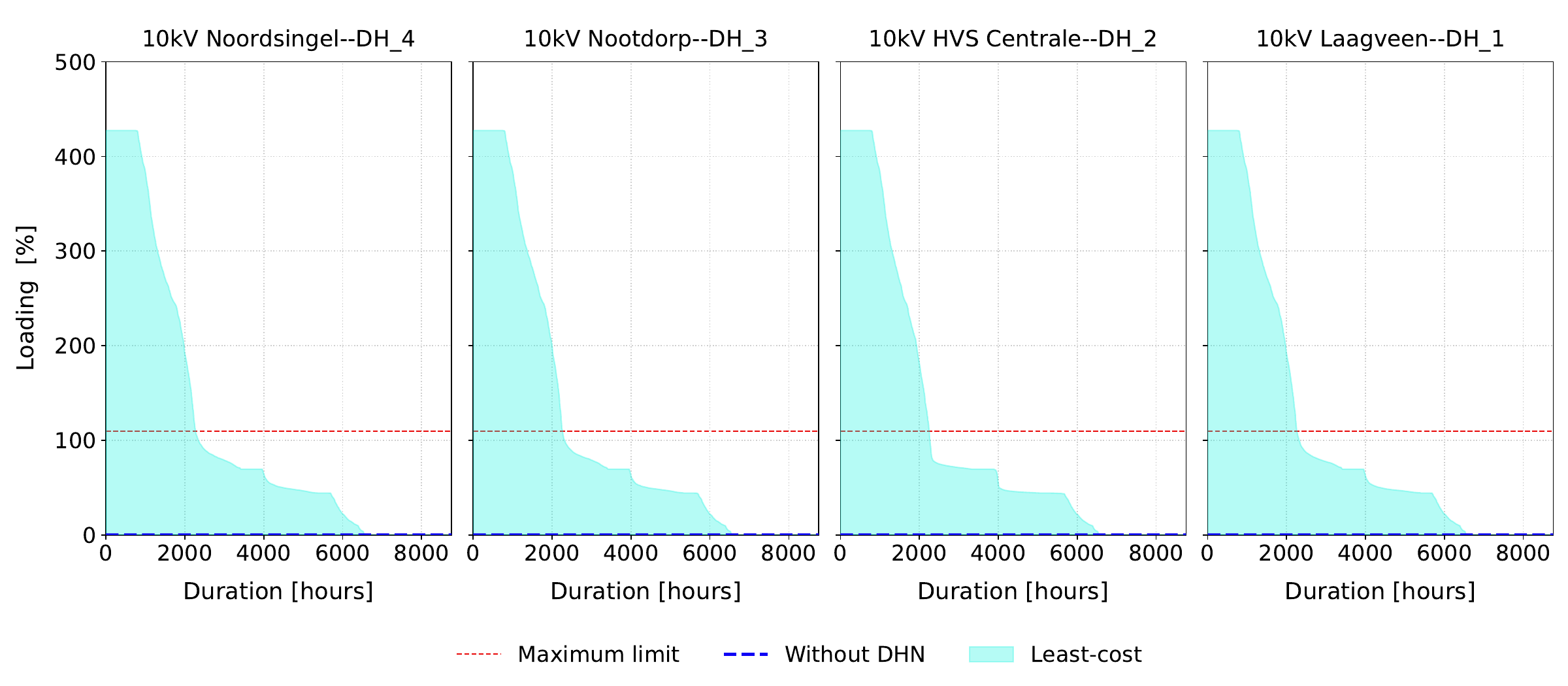}
        \label{fig:line_dc}
    \end{subfigure}

    \label{fig:lc_results}
\end{figure}

\subsection*{Figure 1. Least-cost model results}\phantomsection\label{least_cost_result_figs}
(A) Technology configuration for the least-cost DHN design showing cost-optimal deployed capacities of heat supply technologies, thermal energy storage, and heat pipeline infrastructure. \added{All technology capacities are reported in thermal terms (MW thermal).} CHP: combined heat and power; Storage-TES: short-term energy storage in water tanks; Storage-HT-ATES: seasonal energy storage using high-temperature aquifer thermal energy storage.\\
\replaced{(B) Transformer loading duration curve. (C) Line loading duration curve. Transformer and line loading duration curves}{(B-C) Transformer and line loading duration curves} for selected components in the electricity network (see Figures~\hyperref[lc_loading_duration_all_trafos]{S12} and~\hyperref[lc_loading_duration_all_lines]{S13} for the loading duration curves of all transformers and lines). The red dashed lines represent the maximum practical loading limit for transformers and lines, which is 110\% in the regional electricity network we used as case study. The blue dashed lines correspond to transformer and line loadings from the power flow simulation performed using only the electricity profiles of local consumers, without considering the DHN electricity profiles. The teal area plot shows transformer and line loadings after the DHN electricity profiles have been included, along with the electricity profiles from other local non-DHN consumers.




\section*{A broad range of diverse district heating system designs exists near the cost-optimum}\phantomsection\label{discovering_maneuvering_space}


        
        

Given that some features of the least-cost DHN design may be undesirable in practice and therefore warrant modification, we want to explore economically comparable design alternatives and investigate whether and to what extent these undesirable features can be mitigated, and the trade-offs that this would entail. To do so, we implement an upgraded version of the MGA SPORES algorithm (see section~\hyperref[generating_spores]{Generating SPORES} in experimental procedures) to generate 515 alternative carbon-neutral DHN system designs. These alternatives, which we call SPORES, represent feasible system designs, namely configurations of technology deployment, that are maximally different from the least-cost solution while total system cost remains within 10\% of the least-cost case.

Figure~\hyperref[diverse_spores_figs]{2(A)} illustrates the frequency distribution of potential capacity utilization across the near-optimal decision space. We see that, apart from pipelines, which are required to distribute heat throughout the network, all technologies can attain zero utilization of their potential capacity. This implies that apart from heat pipelines, all technologies are “real choices”; that is, they are not a strict “must-have” since they can be fully substituted by a combination of functionally equivalent alternatives, irrespective of the accepted cost relaxation (see Figures~\hyperref[05_unconstrained_maneuvering_space]{S23} and~\hyperref[15_unconstrained_maneuvering_space]{S24}). However, such substitutions entail significant trade-offs, revealing key interactions between technology choices. 

To illustrate these trade-offs, we highlight a few SPORES in Figure~\hyperref[diverse_spores_figs]{2(A)}, and also show their technology configurations in Figure~\hyperref[diverse_spores_figs]{2(B)}. We can see, for instance, that: (i) reducing reliance on green gas boilers (see \texttt{low-boiler greengas} SPORE) from 400~MW in the least-cost solution to 100~MW is possible by increasing the deployment of hydrogen (+240~MW) and electric (+100~MW) boilers. This additional peaking capacity reduces the need for storage (-25~MW). At the same time, geothermal capacity may increase slightly (+15~MW) to compensate for the decrease in heat pumps (-15~MW). (ii) Decreasing electric boiler capacity (see \texttt{low-electric boiler \& geothermal} SPORE) from 400~MW to 70~MW is possible by deploying higher heat pump (+160~MW) and hydrogen boiler (+~200 MW) capacity, and consequently, higher storage capacity (+130~MW) to exploit low price periods in electricity and hydrogen, which are more volatile than green gas (see Figures ~\hyperref[electricity_price_fig]{S4}, ~\hyperref[hydrogen_price_fig]{S5}, and~\hyperref[static_prices_fig]{S6}). This also increases system flexibility and reduces the need for baseload technologies such as geothermal (-45~MW). (iii) Reducing residual heat (see \texttt{low-residual heat} SPORE) from 200~MW to 90~MW, which provides cheap baseload capacity, can be compensated by higher deployment of waste-to-energy (+200~MW) and geothermal (+5~MW), coupled with higher heat pipeline expansion (+255~MW) to connect these remote heat sources to demand centers.

Interestingly, we find that technologies such as CHPs, seasonal storage (HT-ATES), and solar thermal play only a limited role and are therefore, not crucial to the system as their utilization remains relatively low across all SPORES, irrespective of the cost relaxation (see Figures~\hyperref[05_unconstrained_maneuvering_space]{S23} and~\hyperref[15_unconstrained_maneuvering_space]{S24}), weather year (see Figures~\hyperref[warm_weather_unconstrained_maneuvering_space]{S25} and ~\hyperref[cold_weather_unconstrained_maneuvering_space]{S26}), or heat demand scenario (see Figures~\hyperref[low_demand_unconstrained_maneuvering_space]{S27} and~\hyperref[high_demand_unconstrained_maneuvering_space]{S28}). For CHPs and HT-ATES, their limited utilization is primarily due to their high investment cost. For solar thermal, although its maximum potential capacity can be fully utilized even under small cost relaxations (see Figure~\hyperref[05_unconstrained_maneuvering_space]{S23}), its utilization frequency remains consistently low across most SPORES. This is because of the structural mismatch between its seasonal production and heat demand that requires substantial additional investment in storage and pipeline capacity (see Figures~\hyperref[low_deployment_frequency_techs]{S15} and~\hyperref[low_deploy_spores_technology_mix]{S16}), making it less competitive than other alternatives.

Assessing the effects of increasing the utilization of CHPs, HT-ATES, and solar thermal\textemdash by explicitly maximizing their deployment (see section~\hyperref[generating_spores]{Generating SPORES} in experimental procedures)\textemdash shows that under a higher cost budget, additional capacity in these technologies simply co-exists with the current least-cost deployment rates of other technologies (see Figures~\hyperref[low_deployment_frequency_techs]{S15} and~\hyperref[low_deploy_spores_technology_mix]{S16}). In other words, increasing the capacity of these technologies does not trigger significant substitutions but instead represents a parallel, higher-cost pathway. Further trade-offs associated with increasing other preferences can be explored using the full dataset of our results freely available on Zenodo \cite{doh_dinga_2026_data}. Overall, we find that a marginal increase in cost creates substantial flexibility in how a carbon-neutral DHN can be designed (also see Figures~\hyperref[05_unconstrained_maneuvering_space]{S23} and~\hyperref[15_unconstrained_maneuvering_space]{S24}). This flexibility in technology choice, sizing, and location provides maneuvering space to mitigate the potentially undesirable features of the least-cost solution, and to accommodate other unmodeled preferences or real-world needs, such as enhanced system resilience through greater technology and supply diversity.

\newpage

\begin{figure}[h!]
    \centering

    \begin{subfigure}[t]{0.99\linewidth}
        \centering
        \caption{Frequency distribution of potential capacity utilization in the near-optimal space}
        \includegraphics[width=\linewidth]{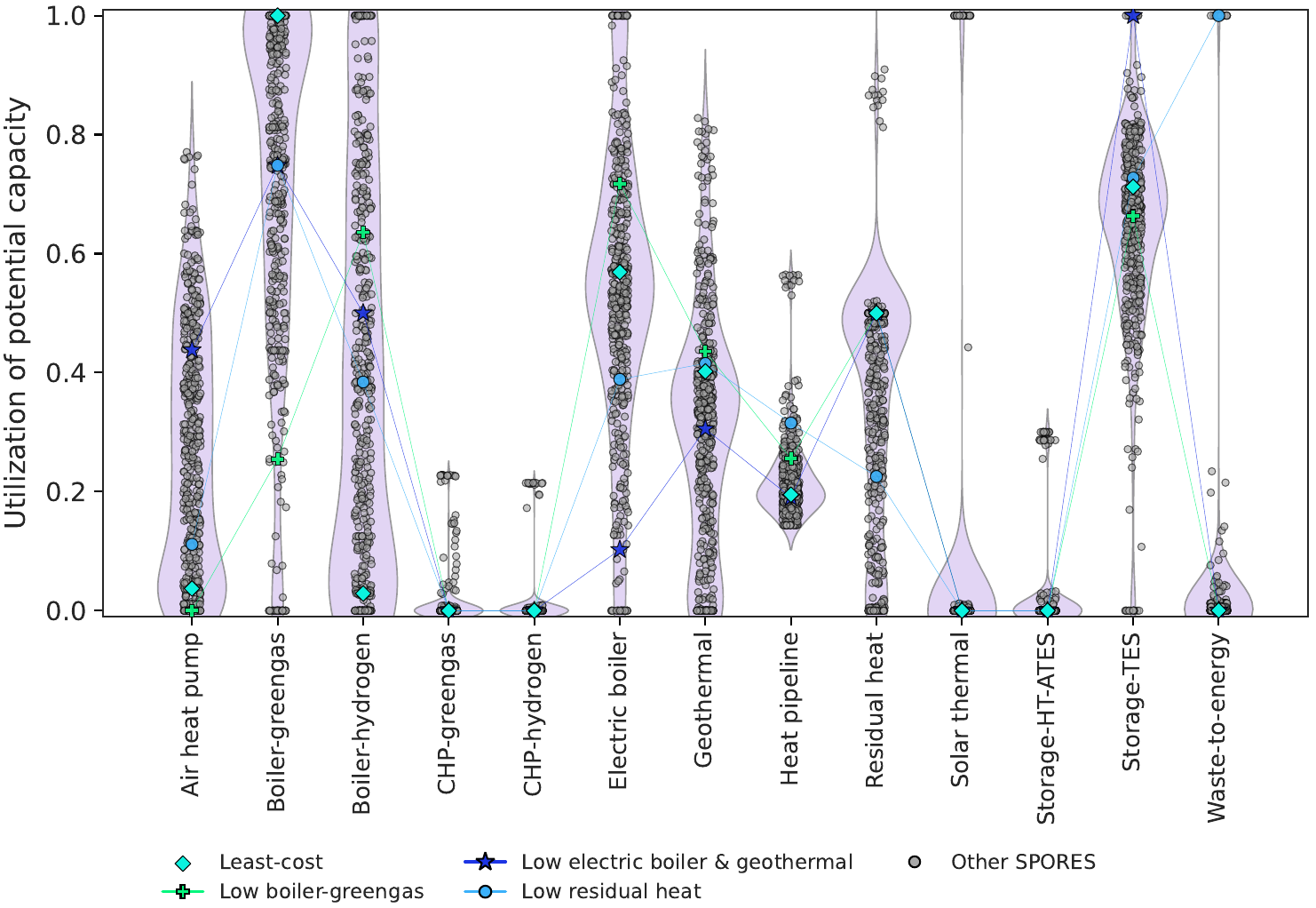}
        \label{fig:meneuvering_space_with_highlighted_spores}
    \end{subfigure}

    \vspace{1em}

    \begin{subfigure}[t]{0.97\linewidth}
        \centering
        \caption{Technological configuration of three highlighted alternative designs}
        \includegraphics[width=\linewidth]{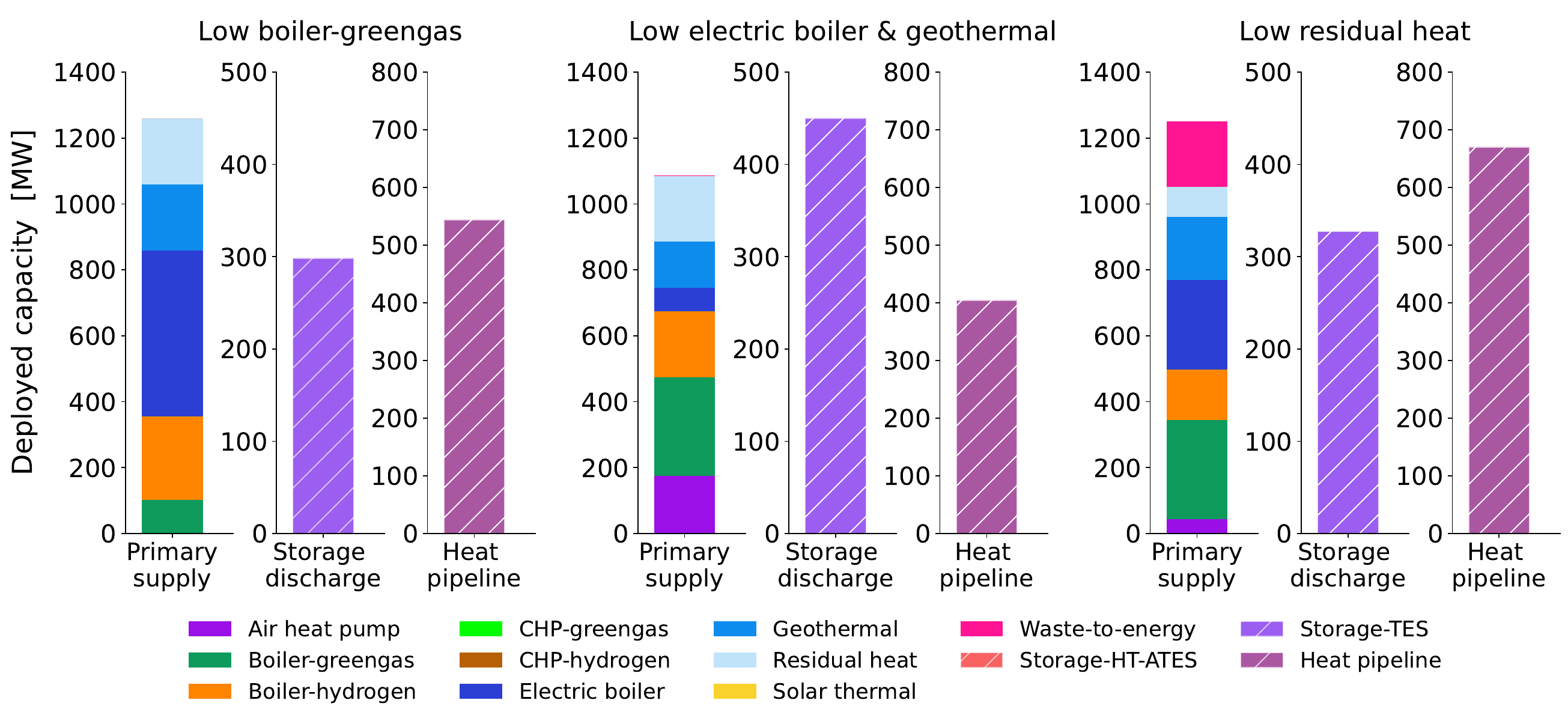}
        \label{fig:tech_mix_of_example_spores_in_maneuvering_space}
    \end{subfigure}

    \label{fig:diverse_spores}
\end{figure}

\subsection*{Figure 2. Near-optimal decision space with three highlighted alternative district heating network designs and their technology configuration}\phantomsection\label{diverse_spores_figs}
(A) Frequency distribution of potential capacity utilization across the whole set of alternative DHN designs (515 SPORES). Utilization is defined as the deployed technology capacity relative to the maximum technical potential capacity of the technology in the model. Three SPORES (low boiler-green gas, low electric boiler \& geothermal, and low residual heat) are highlighted to show the trade-offs it would entail to reduce the deployment of green gas boilers, electric boilers \& geothermal, and residual heat for these highlighted DHN designs (SPORES). \\
(B) Technological configuration of three highlighted alternative DHN (SPORES) showing the deployed capacity of primary heat supply technologies, thermal energy storage, and heat pipeline infrastructure.


\section*{Higher levels of electrification can be achieved without increasing electricity network loading}\phantomsection\label{integrated_decision_space}

We have shown in the previous section that maneuvering space exists to reduce or completely eliminate reliance on technologies associated with potentially undesirable features of the least-cost solution. So far, we assessed the trade-offs within this maneuvering space only from the DHN perspective. We now proceed to examine these trade-offs from an integrated heat–electricity system perspective to (i) determine the extent to which alternative DHN designs differ in their electricity network loading and (ii) identify key design features that impose impact loading on the electricity network. To achieve this, we subject each of SPORE to a full AC power flow simulation to quantify its impact on the electricity network, reflected by transformer and line loadings. We then synthesize the results from both the heat and electricity model runs for each SPORE into an integrated decision space to quantify cross-network trade-offs using the heat and electricity network assessment metrics defined in Table~\hyperref[high_level_network_metrics_table]{1}.

Examining the impact of each SPORE or near-optimal DHN design on the electricity network shows that, overall, SPORES exhibit significant variations in grid loadings (see Figures~\hyperref[range_trafo_loading_all_spores]{S19} and~\hyperref[range_line_loading_all_spores]{S20}). The integrated decision space, which maps the characteristics of SPORES to their corresponding electricity network loadings, is shown in Figure~\hyperref[integrated_decision_space_fig]{3}. We highlight the subset of generated SPORES with lower electricity network loading compared to the least-cost solution across all electricity network metrics. That is, these SPORES fall within all lower electricity network loading envelopes represented by the dashed rectangles. We further highlight selected SPORES in Figure~\hyperref[integrated_decision_space_fig]{3(A)}to illustrate the underlying cross-network trade-offs, and zero in on these SPORES in Figure~\hyperref[integrated_decision_space_fig]{3(B)} to compare their absolute electricity network impacts against the case without DHN electrification and the case with the least-cost DHN design. As expected, SPORES with low P2H capacity generally exhibit lower electricity network loading, as they rely more on non-electric heat supply technologies. For instance, the \texttt{low-electrification} SPORE has a P2H share of 25~\% of its primary supply capacity (-26~\% compared to the least-cost), resulting in higher deployment of gas-based capacity (+9~\%) and non-dispatchable heat sources (+20~\%). Since non-dispatchable heat sources are often located far from demand centers, they require additional (+355~MW) pipeline expansion (see Figures~\hyperref[electrification_techs_frequency]{S17} and~\hyperref[electrification_spores_tech_mix]{S18}).

A similar pattern is observed for the \texttt{lowest-electrification} SPORE; however, this design requires a more moderate expansion in pipeline capacity (+265~MW) due to its higher share of gas boilers (+70~\%), which can typically be sited closer to demand centers. Interestingly, the \texttt{lowest-electrification} SPORE does not fall within all lower electricity network loading envelopes, as it exhibits a higher number of transformer overload events. This is not driven by electrification of the DHN per se, but rather by insufficient local electric capacity to absorb power supply from local distributed PV generators. In this case, the absence of P2H capacity allows local renewable electricity to flow upstream, leading to reverse power flows and higher transformer overloads. Therefore, while excessive heat electrification can increase network loading, moderate levels of electrification may in fact alleviate overloading in networks with high shares of distributed PV generation (also see bus voltage profiles in Figure~\hyperref[lc_bus_voltage_histogram]{S14}).

Surprisingly, some SPORES that deploy P2H capacity above cost-optimal have even lower electricity network loading. For instance, the \texttt{high-electrification} SPORE deploys 20~\% more P2H capacity than the least-cost solution, and yet results in lower grid loading because P2H units are more spatially distributed across the network. This intelligent spatial deployment avoids concentrating P2H units in already congested locations, albeit at the expense of additional (+105~MW) pipeline expansion (see Figures~\hyperref[electrification_techs_frequency]{S17} and~\hyperref[electrification_spores_tech_mix]{S18}). However, where such large pipeline infrastructure build-out is not possible, for instance due to space constraints, higher levels of electrification can still be realized without further compromising grid loading by prioritizing\textemdash within P2H units\textemdash more efficient heat pumps combined with thermal storage over electric boilers, as shown by the \texttt{highest-electrification} SPORE in Figure~\hyperref[integrated_decision_space_fig]{3} (also see Figures~\hyperref[electrification_techs_frequency]{S17} and~\hyperref[electrification_spores_tech_mix]{S18}). In sum, we find that grid loading depends not only on the level of heat electrification but also on the interaction between technology choices and their spatial deployment patterns. Therefore, irrespective of the scenario (see Figures~\hyperref[05_integrated_decision_space]{S29} to~\hyperref[high_heat_demand_integrated_decision_space]{S34} in ~\hyperref[sensitivity_analysis_results]{Sensitivity analysis results}), higher levels of heat electrification can be achieved with even lower electricity network loadings.





\subsection*{Table 1. Definition of heat and electricity network assessment metrics and their range in the integrated decision space}\phantomsection\label{high_level_network_metrics_table}
\begin{tabular}{|p{3.2cm}|p{10cm}|p{2.8cm}|}
    \hline
    \textbf{Metric name} &
    \textbf{Metric description and interpretation} &
    \textbf{Metric range} \\ [1ex]
    \hline
    
    Power-to-heat capacity &
    Total share of heat pumps and electric boilers in the primary heat supply capacity. Reflects the degree of heat electrification and the dependence of the DHN on the electricity network (ELN). &
    0--74 \% \\ [1ex]
    \hline
    
    Gas-based capacity &
    Total share of heat supply technologies that use hydrogen or green gas as fuel (boilers and CHPs). Reflects DHN flexibility from fully dispatchable thermal capacity that can be rapidly ramped to meet peak demand. &
    0--62 \%  \\ [1ex]
    \hline
    
    Non-dispatchable capacity &
    Total share of baseload technologies (geothermal, residual heat, and solar thermal) in the primary heat supply capacity. Reflects inflexible heat sources since these technologies have limited ramping capability. &
    0--50 \% \\ [1ex]
    \hline
    
    Storage capacity &
    Total discharge capacity of heat storage technologies, including both short-term (TES) and seasonal (HT-ATES) storage. Reflects the temporal flexibility of the DHN and the decoupling of heat supply from demand. &
    0--456 MW \\ [1ex]
    \hline
    
    Pipeline capacity &
    Total capacity of heat distribution pipeline infrastructure. Reflects the spatial flexibility of the DHN to meet heat demand with supply units at different locations. &
    305--1200 MW \\ [1ex]
    \hline
    
    Levelized cost of heat &
    Average discounted long-run cost of heat supply from the DHN. Indicates the overall economic performance of the DHN. &
    70--79 €/MWh \\ [1ex]
    \hline
    
    Line overload events\textsuperscript{a} &
    Number of persistent overload events experienced by all lines in the ELN. Indicates the frequency and duration of line overloading across the entire ELN.&
    0--144 \\ [1ex]
    \hline
    
    Number of overloaded lines & 
    Proportion of lines in the ELN that experience persistent overloading. Indicates the spread or spatial extent of line overloading across the entire ELN. &
    0-65 \% \\ [1ex]
    \hline
    
    Line loading\textsuperscript{b} &
    The 75\textsuperscript{th} percentile of the distribution of the 90\textsuperscript{th} percentile loading values across all lines in the ELN. Indicates the typical severity of line loading across the entire ELN. &
    60--350 \% \\ [1ex]
    \hline
    
    Transformer overload events &
    Number of persistent overload events experienced by all transformers in the ELN. Indicates the frequency and duration of transformer overloading across the entire ELN. &
    104--149 \\ [1ex]
    \hline
    
    Number of overloaded transformers &
    Proportion of transformers in the ELN that experience persistent overloading. Indicates the spread or spatial extent of transformer overloading across the entire ELN. &
    60--95 \% \\ [1ex]
    \hline
    
    Transformer loading &
    The 75\textsuperscript{th} percentile of the distribution of the 90\textsuperscript{th} percentile loading values across all transformers in the ELN. Indicates the typical severity of transformer loading across the entire ELN. &
    160--285 \% \\ [1ex]
    \hline

\end{tabular}


\bigskip

Definition of heat and electricity network metrics used to quantify trade-offs in the integrated heat-electricity network decision space. The metric range represents the minimum and maximum computed values of each metric across all DHN designs in the near-optimal decision space, that is, across all 515 SPORES. These metrics are particularly relevant for heat and electricity network operators to help them understand how the different characteristics of alternative district heating network designs affect the electricity network.

\textsuperscript{a}Generally, it is not considered a severe problem if an electrical component (line or transformer) is slightly loaded above its rated capacity for a short duration of time. Therefore, we only track the number of ``persistent" overload events, which we compute as the number of counts an electrical component is loaded above its maximum limit (in this case, 110\%) for 7 consecutive simulation snapshots.

\textsuperscript{b}The electricity network loading is highly heterogeneous across components and time, as shown by the load duration curves in supplemental Figures~\hyperref[lc_loading_duration_all_trafos]{S12} and ~\hyperref[lc_loading_duration_all_lines]{S13}. To obtain a statistically representative measure of electricity network impacts, we first compute the $90^{th}$ percentile loading value over time for each individual line and transformer to capture high-loading conditions relevant for network planning. We then summarize these component-level values using the $75^{th}$ percentile across all components, since the differences between alternative DHN designs are most pronounced in this range, as shown by the box plots in Figure~\hyperref[integrated_decision_space_fig]{3(B)}.



\newpage

\begin{figure}[h!]
    \centering

    \begin{subfigure}[t]{0.9\linewidth}
        \centering
        \caption{Integrated heat-electricity network planning decision space}
        \includegraphics[width=\linewidth]{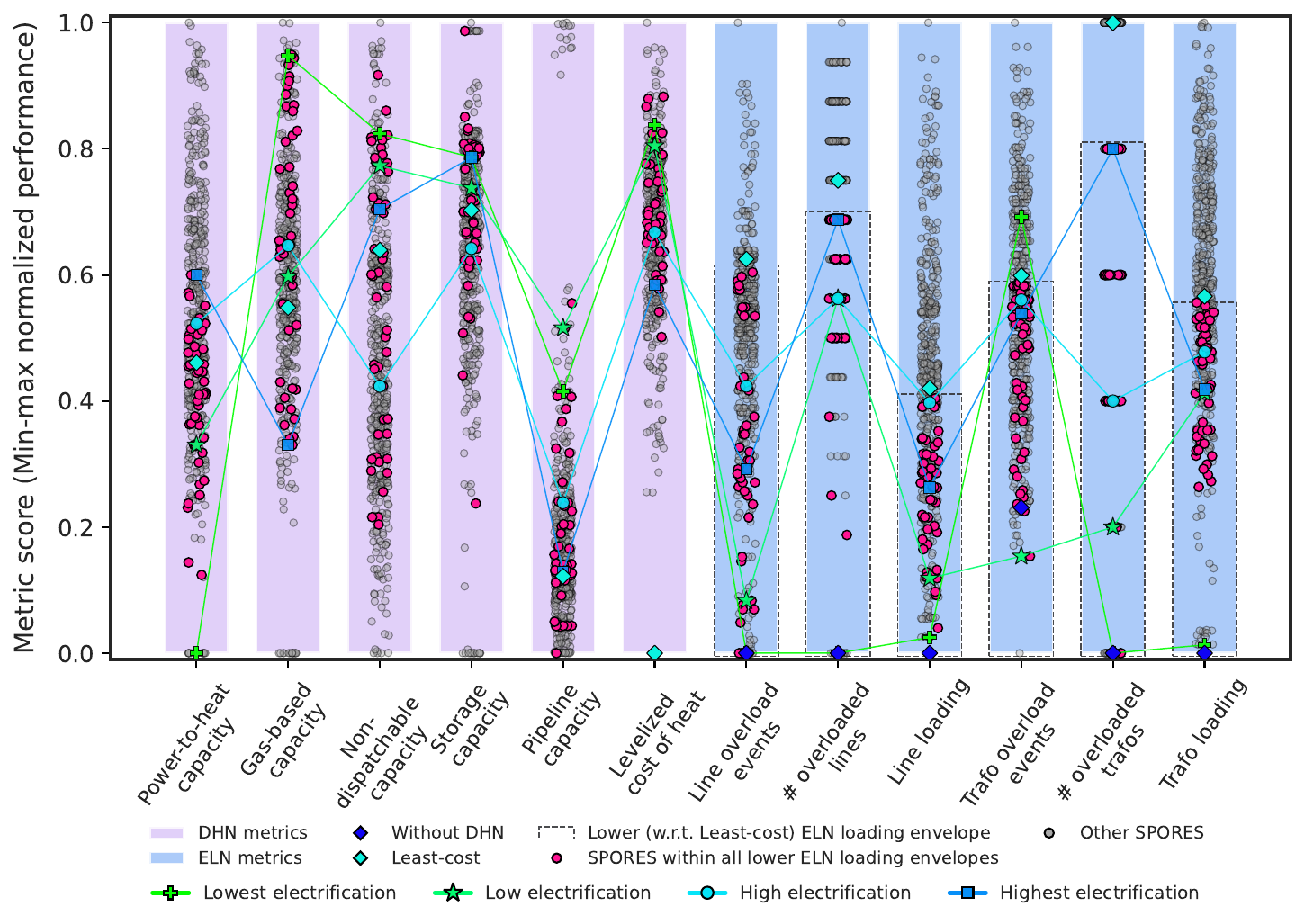}
        \label{fig:integrated_decision_space}
    \end{subfigure}


    \begin{subfigure}[t]{0.93\linewidth}
        \centering
        \caption{Electricity network impacts of the four highlighted SPORES}
        \includegraphics[width=\linewidth]{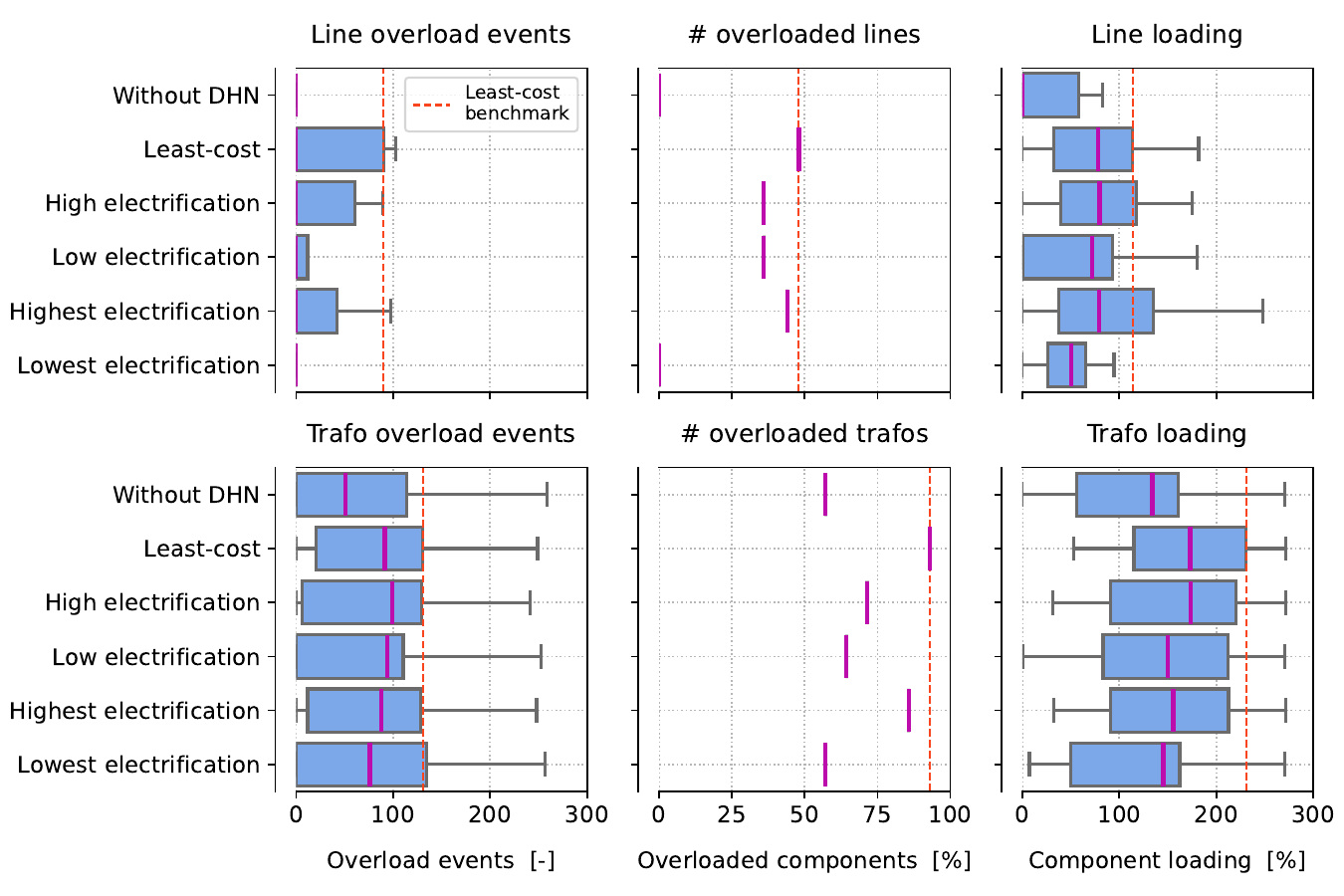}
        \label{fig:gird_impact_statistics}
    \end{subfigure}

    \label{fig:integrated_decision_space_and_gird_impact_statistics}
\end{figure}

\subsection*{Figure 3. Integrated heat–electricity network planning decision space with four highlighted district heating network designs and their grid impacts}\phantomsection\label{integrated_decision_space_fig}

(A) Integrated heat–electricity network planning decision space mapping different DHN characteristics to their corresponding electricity network (ELN) impacts. Light purple bars represent DHN metrics, and light blue bars represent ELN metrics. All SPORES within the lower grid loading envelope (dashed rectangle) correspond to designs with lower grid impacts for a given ELN metric compared to the least-cost DHN design. The deep pink dots denote SPORES that lie within the lower grid loading envelope across all ELN metrics\textemdash that is, the intersection of all dashed rectangles. Four SPORES are highlighted (lowest electrification, low electrification, high electrification, and highest electrification), spanning from minimal to very high heat electrification, to illustrate how different electrification levels affect the electricity network.\\
(B) Electricity network impacts of the four highlighted SPORES. The box plots show the absolute grid loading of the four highlighted SPORES against two benchmarks: the case without DHN electrification (first row in each box plot) and the least-cost DHN design (second row in each box plot).


\section*{Location-specific constraints transform real choices into practical must-haves}\phantomsection\label{local_technology_deployment_constraints}

The maneuvering space described in section~\hyperref[discovering_maneuvering_space]{A broad range of diverse district heating system designs exists near the cost-optimum} represents the purely techno-economic flexibility available to system planners to accommodate other practically desirable features or unmodeled preferences, assuming all the considered technology options are available. However, as previously mentioned, DHNs are inherently local, and therefore, the concrete technology configurations depend on local constraints. We now want to re-examine this option space to see how plausible local constraints may impact the trade-offs that we have revealed so far, thereby providing insights that may be generalizable to a broader range of DHNs. To do so, we consider the following set of plausible local constraints: (i) unfavorable subsurface conditions for the deployment of geothermal and HT-ATES; (ii) insufficient proximity to residual heat sources; (iii) restricted space availability for large solar thermal installations; (iv) limited availability or lack of existence of a reliable supply chain for less established technologies such as green gas and hydrogen boilers; (v) limited heat electrification in the case of a congested electricity network. We then filter the previously generated SPORES to retain only those that satisfy these constraints.

We see (Figure~\hyperref[local_technology_deployment_constraints_fig]{4}) that, unsurprisingly, the potential unavailability of technology options that already had a low deployment frequency in the unconstrained option space, such as solar thermal and  HT-ATES, does not dramatically affect the maneuvering space for designing the carbon-neutral DHN. Moreover, when explicitly considering the subset of SPORES that ensure limited grid impact, the changes are identical to those observed in the initial unconstrained option space (see Figure~\hyperref[unconstrained_maneuvering_space_with_grid_friendly_spores]{S21}), most notably, a reduced reliance on electric boilers. Other design decisions, such as whether to rely more on gas boilers, non-dispatchable (baseload) sources, or heat pumps, remain ``real choices". Contrarily, changes to the design space are more pronounced, instead, when the reliance on either green gas or hydrogen boilers is constrained by local social preferences or supply chain scarcity (also see Figure~\hyperref[constrained_integrated_decision_space]{S22}). When both such boilers are, for instance, limited to a capacity of no more than 20\% of peak demand, ensuring the economic competitiveness of the DHN requires limiting reliance on less cost-effective technologies, such as CHPs, solar thermal, HT-ATES, waste-to-energy, and\textemdash though to a lesser extent\textemdash also baseload sources, pipeline, and very large heat pump capacities. If mitigating grid impact is factored in, the design options to compensate for limited gas boiler capacity reduce further, with electric boilers\textemdash an obvious functionally equivalent substitute\textemdash almost ceasing to be a viable option. The consequence is a need to maximize the limited available capacity of cost-effective green gas boilers alongside as-high-as-possible (subject to budget constraints) deployment of efficient heat pumps and conventional TES far beyond least-cost levels (see Figures~\hyperref[05_slack_constrained_maneuvering_space_fig]{S35} to~\hyperref[high_demand_constrained_maneuvering_space_fig]{S40} in ~\hyperref[sensitivity_analysis_results]{Sensitivity analysis results}). However, constraining electric boilers does not nearly have the same effect; limiting their deployment mainly affects budget considerations as it marginally shrinks the overall maneuvering space, but substantially benefits grid-related considerations, irrespective of the scenario (see Figures~\hyperref[05_slack_constrained_maneuvering_space_fig]{S35} to~\hyperref[high_demand_constrained_maneuvering_space_fig]{S40}), consistent with the findings in section~\hyperref[integrated_decision_space]{Higher levels of electrification can be achieved without increasing electricity network loading}.

The most challenging limitation is dealing with the lack of baseload heat sources, such as geothermal and residual heat (also see Figures~\hyperref[05_slack_constrained_maneuvering_space_fig]{S35} to~\hyperref[high_demand_constrained_maneuvering_space_fig]{S40}). In line with the trade-offs we highlighted in prior sections, Figure~\hyperref[local_technology_deployment_constraints_fig]{4} shows that the complete absence of such sources still leaves several options for designing a cost-effective carbon-neutral DHN. However, factoring in the simultaneous need to mitigate grid impacts substantially reduces the maneuvering space for other decisions. If geothermal is not an option locally, then trading off such a heat source with moderate reliance on residual heat partly compensates for it. However, a simultaneous higher reliance on gas boilers is necessary to keep both costs and grid impact limited (also see Figures~\hyperref[05_slack_constrained_maneuvering_space_fig]{S35} to~\hyperref[high_demand_constrained_maneuvering_space_fig]{S40}). Meanwhile, higher heat pump capacities, alongside conventional thermal storage, must be prioritized, too, over economically convenient but less efficient electric boilers, when minimizing grid impact becomes a priority. What is more, a careful, intelligent spatial deployment of such P2H assets is critical, in line with our findings in section~\hyperref[integrated_decision_space]{Higher levels of electrification can be achieved without increasing electricity network loading}. When residual heat is the unavailable source, our results show a similar, albeit more pronounced, dynamic. This is because residual heat, in our case study (see section~\hyperref[technology_modeling]{Technology modeling} in experimental procedures), is already at high temperatures and therefore, unlike geothermal, requires no auxiliary electricity for temperature upgrades, making it more challenging to maintain low grid stress within the remaining technology options and budget (also see Figures~\hyperref[05_slack_constrained_maneuvering_space_fig]{S35} to~\hyperref[high_demand_constrained_maneuvering_space_fig]{S40}). This may differ for residual heat sources at lower temperatures. Our case study, therefore, plays a role in the exact magnitude of the remaining design maneuvering space in Figure~\hyperref[local_technology_deployment_constraints_fig]{4} when neither geothermal nor residual heat is an option. However, the high-level trade-offs we highlight remain valid and generalizable: a lack of baseload heat sources requires greater reliance on carbon-neutral gas boilers and, to the extent allowed by grid constraints, intelligently deployed heat pumps coupled with thermal storage.

It is worth noting that simultaneously removing multiple technology options is not always possible. For instance, the lack of baseload heat sources requires some reliance on gas boilers if mitigating grid stress is a concern. Moreover, even when using gas boilers, the simultaneous absence of local geothermal and residual heat does not enable the design of a DHN that meets the grid-integration requirements we have identified. This may, of course, be different for a DHN system whose local electricity grid has greater capacity and can accommodate more electrical assets than the one we used as a case study.

\newpage




\includegraphics[width=0.97\linewidth]{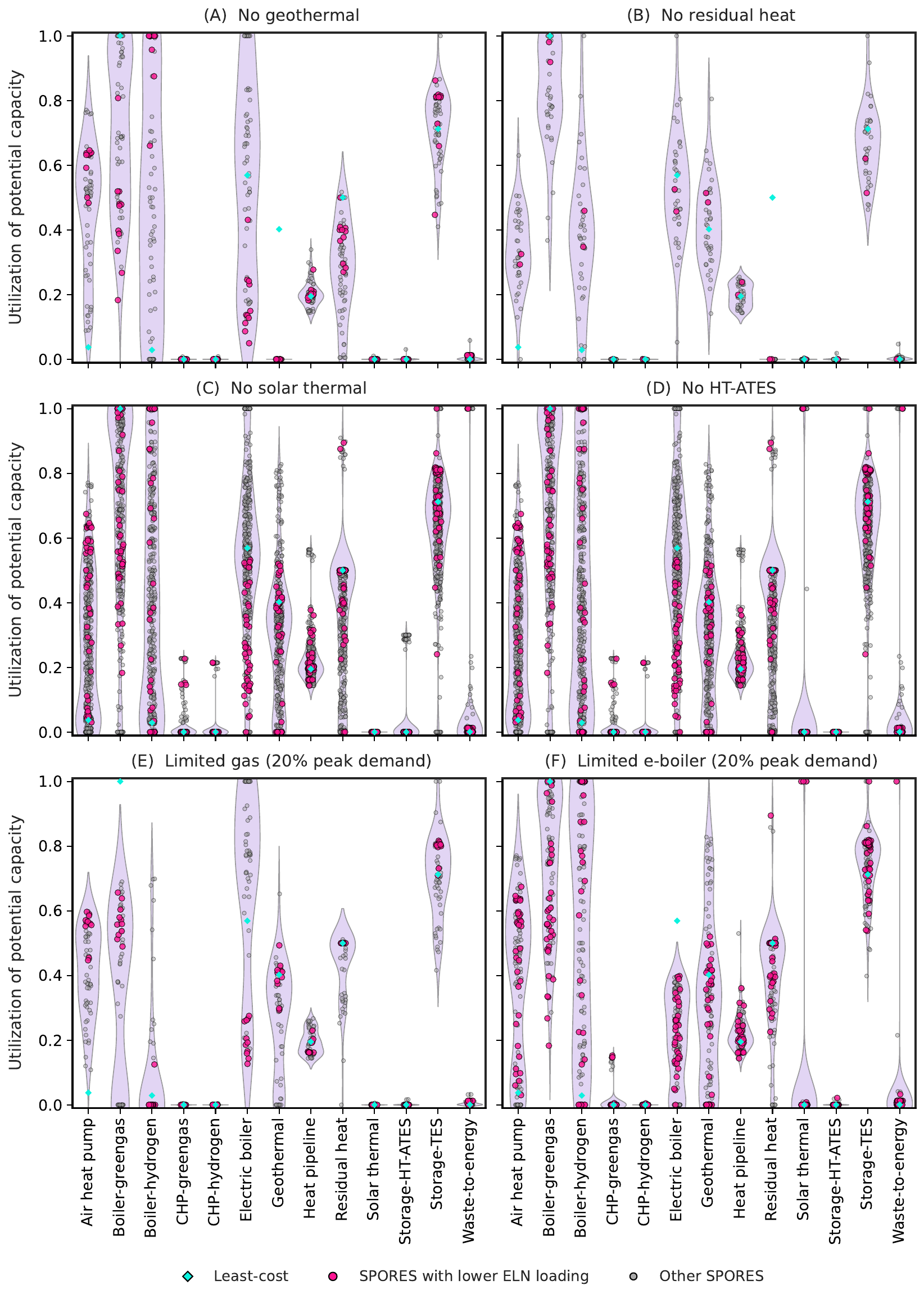} 
\subsection*{Figure 4. Trade-offs under local technology deployment constraints}\phantomsection\label{local_technology_deployment_constraints_fig}
(A-F) Alternative DHN designs under six technology deployment constraints. Gray dots represent feasible SPORES (DHN designs) under each specific constraint. \added{(A) No geothermal. (B) No residual heat. (C) No solar thermal. (D) No HT-ATES. (E) Limited gas (20\% of peak demand). (F) Limited electric boiler (20\% of peak demand).} The deep pink dots in each subplot indicate feasible SPORES that achieve lower grid impacts across all electricity network impact assessment metrics (see Table~\hyperref[high_level_network_metrics_table]{1}) compared to the least-cost DHN design. See supplemental Figure~\hyperref[unconstrained_maneuvering_space_with_grid_friendly_spores]{S21} for the corresponding trade-offs in the unconstrained “All options available” case. \\
(A-D) Represent cases with the complete absence of specific technologies such as geothermal heat, residual heat, solar thermal heat, and high-temperature aquifer thermal energy storage, whose deployment is constrained by local resource availability. \\
(E-F) Represent constraints to cap the deployed capacity of technologies, such as any type of gas boiler and electric boilers, that are likely to be always available in any location.

\section*{DISCUSSION}\phantomsection\label{discussion_section}


\added{This study concerns the design of carbon-neutral DHNs in the context of the 4th-generation district heating (4GDH) paradigm \cite{Lund_2018_4GDH_status}, which positions DHNs as a key component of climate-neutral smart energy systems \cite{Lund_2025_energy_balancing_climate_neutral} through the integration of low-temperature renewable heat sources and electrified heat production. While this expanded portfolio of low-carbon technologies increases the opportunities for decarbonizing DHNs, it also introduces a wider range of planning trade-offs related to resource availability, infrastructure requirements, social acceptance, and interactions with the electricity system.}

We have argued that designing a carbon-neutral DHN solely on the basis of minimum cost can lead to several potentially undesirable features, such as over-reliance on heat sources with uncertain availability, concentration of geothermal heat sources in a single location that could stir public opposition, and large electric boiler capacity build-out that could stress the electricity network. These features of cost-minimization are not only unique to our case. Broader energy system planning studies have likewise identified similar patterns in least-cost system designs, such as over-deployment of a technology in a single location \cite{lombardi_2020_SPORES}, or reliance on a large capacity of a few potentially problematic technologies \cite{greevenbroek_2025_ns} .

We concede that the type and severity of potentially undesirable features of a least-cost design may vary across DHNs in different regions. For instance, regions with abundant biomass resources may find it less problematic to deploy large capacities of biomass-based heating technologies. However, for many European countries, such as the Netherlands, which have limited domestic biomass availability \cite{millinger_2025_biomass_ne}, it remains highly uncertain where sufficient sustainable biomass would be sourced to produce biofuels or green gas at scale \cite{Bekkering2023_greengas_ecm}. Therefore, for such countries, designing a DHN that relies heavily on biomass-derived fuels represents a technology pathway that is more vulnerable to potential risks from supply uncertainty or import dependence, a concern that has already been raised by real-world stakeholders \cite{ii3050_Netbeheer}.



Irrespective of the region, designing a carbon-neutral DHNs requires long-term investment decisions under deep uncertainty, including uncertainties that are difficult to parameterize in a model, such as social acceptance \cite{lombardi_2025_mga_perspective}. Additionally, because electrification will play an important role in the decarbonization of DHNs \cite{lund2025_nrct}, DHN planners must ensure that system designs are not only robust to future uncertainties, but also impose minimal stress on the electricity network since reinforcing the grid is not only costly \replaced{\cite{roman_2023_utiliies_policy}}{\cite{roman_2023_utiliies_policy, Lund_2018_smart_energy_systems_vs_smart_grid}}, but also slow due to long lead times, permitting delays, material supply constraints, and workforce shortage \cite{ii3050_Netbeheer}.


This study, therefore, presents a decision-support method for designing carbon-neutral DHNs that are less vulnerable to future risks and have minimal impact on the grid while remaining near-optimal in cost. By applying MGA within a framework that couples a DHN capacity expansion and operation model with a power flow simulation model of the electricity network, we develop a method to systematically explore similarly costly but technologically diverse DHN designs, and quantify their impacts on the electricity network. Applying our method to a Dutch case study, we demonstrate that with a marginal cost increase above the optimum, there is substantial diversity in how a carbon-neutral DHN can be designed, and the diverse designs exhibit significant variations in grid impact. Within the near-optimal decision space, almost all technologies can be fully substituted by functionally equivalent alternatives, and thus are “real choices”. This flexibility in choice, sizing, and location of technologies\textemdash while system cost remains near optimal\textemdash provides planners with substantial maneuvering space to mitigate potential risks and balance conflicting preferences across different stakeholders in order to reach a consensus solution.

Naturally, accommodating certain preferences requires significant trade-offs. For instance, we find that decreasing the deployment of electric boilers\textemdash to minimize grid impact\textemdash and reducing the concentration of geothermal heat sources in The Hague\textemdash to mitigate potential risks from public opposition\textemdash requires higher deployment of heat pumps and more thermal storage. Similarly, reducing reliance on residual heat from a nearby petrochemical industry can be compensated for by higher deployment of waste-to-energy heat sources and more pipeline expansion. Although the exact magnitude of these trade-offs will vary across DHNs and regions, the structural dynamics in technology interaction and substitution we uncover here \replaced{are likely to be consistent in other DHNs}{appear broadly consistent across DHNs}. For instance, we find that designs with lower green gas boiler deployment require an almost proportional increase in peaking technologies, such as electric and/or hydrogen boilers, or more baseload heat sources, coupled with higher storage levels. Similar interaction patterns between heating technologies have been observed in a study examining existing DHN technology mixes across different European countries \cite{EU_commission_2018_cost_efficient_dh}. Our analysis, however, reveals these dynamics systematically across the full design option space, enabling planners to understand the underlying substitution patterns and quantify the trade-offs required to accommodate diverse preferences and constraints that match the needs of different stakeholders.



The trade-offs associated with reducing reliance on specific technologies become more pronounced when we wish to keep grid impact minimal. For instance, under a 10\% cost budget above the least-cost case, green gas boilers cannot be reduced below 20\% of their potential capacity, while the deployment of electric boilers must be lower than the capacity suggested by the least-cost design across all investigated scenarios if minimizing grid impact is a priority. Using our integrated decision space that maps DHN characteristics to their corresponding grid loading, it is possible to identify DHN designs with low grid impacts. While it is straightforward that such designs will generally have lower levels of electrification, we find that higher levels of electrification can, in fact, be achieved with even lower grid loading. The high-electrification and low-grid loading designs generally (i) have a more intelligent spatial distribution of P2H technologies across the network, thereby reducing concentration in already congested locations but at the expense of more pipeline expansion; (ii) prioritize more efficient heat pumps\textemdash combined with higher thermal storage deployment\textemdash over electric boilers. These findings refine prevailing narratives in the literature that associate high heat electrification rates with grid stress \cite{Waite2020_joule, Mascherbauer2025_ae}. Our results show that grid overloads depend not only on the degree of electrification but also on the interaction between technology choices and their spatial deployment patterns.

Another, perhaps counterintuitive finding enabled by our model coupling approach is the interaction between heat electrification and distributed PV generators. We find that some moderately electrified DHN designs reduce transformer overloads compared to low-electrification designs, since P2H units avoid reverse peak power flows upstream by absorbing surplus generation from local PVs. This suggests that heat electrification may alleviate grid congestion resulting from peak supply, and also increase the efficiency of PVs in energy systems with high shares of distributed PV generation. However, the temporal mismatch between heat demand for space heating and PV generation limits the scale of this solution. Perhaps, this could scale up if heat electrification is coupled with higher (seasonal) thermal storage, particularly for designs dominated by heat pumps, since heat pumps are most efficient when it is sunnier. A few studies claim that coordinating the operation of PV, heat pump, and thermal storage can accelerate decarbonization \cite{Leitner2019_energy, PenaBello2021_ecm}. However, more work is needed to determine whether relying on heat electrification and thermal storage to absorb excess generation from PVs is economically beneficial from a broader energy system perspective, compared to direct curtailment.

A particularly noteworthy discussion concerns hydrogen boilers, whose potential use for residential heating is highly debated in the literature \cite{Bekkering2023_greengas_ecm, THOMAS_2023_hydrogen_for_heating}. Most studies conclude that hydrogen is unlikely to play a role in future DHNs for several reasons, including costs, efficiency losses, and competition with other difficult-to-electrify sectors such as industry and transport \cite{ROSENOW_2022_h2_dream_joule, KURSES_2025_hydrogen_ae}. While our least-cost solution indeed aligns with this claim, the near-optimal option space, however, shows that hydrogen boilers emerge as a valuable alternative when reliance on biomass-derived fuels is reduced, and even more so when avoiding electricity network overloads becomes a priority. This implies that for DHNs in regions with limited biomass availability and a highly congested grid, hydrogen boilers could be a potential backup option for security of supply. Therefore, it may be premature to categorically exclude hydrogen boilers from future DHN planning. Instead, their potential role should be assessed in light of regional resource constraints, grid conditions, system resilience, and long-term infrastructure development strategies.

Another question concerns the future role of sustainable-fuel CHPs in carbon-neutral DHNs. We find that within the near-optimal option space, both the utilization frequency and deployed capacity of CHPs remain low across all designs. This is primarily because CHPs become more capital-intensive under frequent low electricity price periods in a broader decarbonized energy system dominated by wind and solar. Our finding aligns with a recent EU-wide study that suggests a declining role for CHPs in future DHNs \cite{mathiesen2025_heat_roadmap_eu}. However, some studies emphasize that CHPs may retain a strategic value for power system decarbonization, generation adequacy, and security of supply when peak demand periods coincide with low wind and solar availability \cite{jrc2019_decarbonizing_heat}. Our sensitivity analysis using a cold weather year with low solar and wind output\textemdash the so-called cold dark doldrums\textemdash shows only a marginal increase in CHP deployment across designs, as the few periods of high electricity prices are insufficient to offset investment costs. This suggests that if CHPs are to play a role in power system decarbonization, relying solely on energy-only market revenues may not provide sufficient incentive; instead, dedicated policy support or capacity remuneration mechanisms might be required to incentivize their deployment.

We also observe a similar pattern for seasonal storage (HT-ATES) and solar thermal across all designs. In our case, these technologies are only deployed when additional budget is made available, without any significant technology substitution, suggesting that their deployment mostly represents a parallel, higher-cost alternative design. Similar conclusions have been drawn in other studies, where HT-ATES is found to be too costly and to significantly increase the levelized cost of heat \cite{DANIILIDIS_2022_HT_ATES_LCOH}. For solar thermal, although it can reach its maximum potential capacity in some designs, we find that potential capacity utilization frequency remains low\textemdash even in relatively warmer weather year\textemdash primarily due to the additional investments required in storage and complementary supply technologies to compensate for the seasonal mismatch between solar thermal production and heat demand. A similar dynamic has also been highlighted in the literature for DHNs with significant solar thermal capacity, such as the Gram Fjernvarme DHN in Denmark, where achieving higher solar thermal penetration has required large-scale seasonal storage and integration with multiple additional heat supply technologies \cite{EU_commission_2018_cost_efficient_dh}. However, further work is needed to assess the economic viability of these technologies within a broader energy system context, particularly for HT-ATES, which some studies have shown to increase energy system efficiency and reduce overall cost \cite{Bloemendal_2021_energies}.

Nevertheless, as we have argued previously, DHN design decisions should not be based solely on economic cost. Deploying technologies that appear consistently at lower capacity across designs, such as CHPs, HT-ATES, and solar thermal, could be justified as a strategy to enhance system resilience through greater flexibility and technological diversity. Such technologies may provide alternative supply pathways, reduce dependence on a few dominant resources, and increase the system’s ability to cope with unexpected disruptions in fuel supply, infrastructure constraints, extreme weather events, or unplanned maintenance. Indeed, the economic value of resilience is difficult to quantify, as its benefits typically materialize only after a shock occurs \cite{wri_2021_dividends_resilience}. Therefore, it is important to provide planners with the full option space for them to deliberate on what is most practically viable for society since the contribution of certain technologies may extend beyond short-term cost performance.

The inherently local nature of DHNs implies that some of the technologies we investigate here might not be an option in certain locations. For instance, certain regions may lack geothermal heat sources due to unfavorable geological conditions. To provide more insights for DHNs in other regions, we reexamined the previously described trade-offs under a set of plausible technology availability constraints. We find that a challenging local constraint is the limited availability of fuels, such as hydrogen or biomass, for fuel boilers. When this is the case, prioritizing heat pumps and thermal storage over other technologies is vital to ensure both cost effectiveness and limited grid impact.  Even more challenging is the potential local unavailability of geothermal or residual heat. While this local barrier does not prevent designing a carbon-neutral DHN per se, if minimizing grid impact is a priority, then at least one carbon-neutral baseload heat source (geothermal or residual heat) and a design combining a spatially-optimized large-scale deployment of heat pumps with thermal storage and some form of gas-based peaking technology become indispensable.





\deleted{As with all modeling studies, our work is not without limitations. First, we do not explicitly quantify the extent of grid reinforcement required to accommodate DHN electrification. Grid reinforcement needs depend not only on DHN-related loads but also on other local electricity demands and generation, which are treated as static in our analysis. In practice, some of these static loads will have some flexibility to respond to high network tariffs during potential grid congestion periods. Therefore, quantifying grid upgrades without accounting for demand-response and grid tariffs would be misleading. Accordingly, our results should be interpreted as indicating the isolated potential impacts of DHN electrification on the existing grid, rather than as estimates of actual reinforcement requirements.

Second, computational limitations shape our modeling choices. The DHN planning model is formulated as a linear program (LP) to avoid the computational complexity associated with non-convex formulations. As a result, hydraulic flow dynamics are not explicitly represented, and we modeled heat flows using a simplified transport model. Moreover, investment decisions allow for fractional technology capacities rather than discrete unit sizes due to the LP formulation. Even with this simplification, the large number of MGA runs required to thoroughly explore the near-optimal space was only feasible using high-performance computing resources \cite{surf_snellius}. For similar computational reasons, location-specific technology constraints were imposed ex post by filtering the previously computed option space, rather than by re-running the MGA for each new constraint. Endogenously including each constraint and recomputing the near-optimal space would shift the global optimal cost, which is used for setting the cost budget. Our sensitivity analysis on the cost budget can be interpreted as partially capturing the implications of such shifts. A potential solution to improve computational efficiency would be to narrow the search space through stakeholder engagement, for example, via more targeted human-in-the-loop MGA approaches \cite{lombardi_2025_human_in_the_loop}. All these limitations provide a foundation for future research to focus on.}


In conclusion, carbon-neutral DHNs are expected to play a key role in decarbonizing residential heating and accelerating the energy transition. Yet, designing carbon-neutral DHNs requires balancing several potentially conflicting objectives, including economic costs, social acceptance, robustness to future risks, and grid integration challenges arising from electrification. By combining modeling-to-generate-alternatives with power flow simulation techniques for the first time, this study offers broadly applicable insights as well as a decision-support method for designing carbon-neutral DHNs that are cost-effective, socially acceptable, \replaced{robust to future risks}{less vulnerable to future risks}, and impose minimal impacts on the electricity network.

\subsection*{Limitations of the study}\phantomsection\label{limitations_section}

\added{
As with all modeling studies, our work is not without limitations. First, we do not explicitly quantify the extent of grid reinforcement required to accommodate DHN electrification. Grid reinforcement needs depend not only on DHN-related loads but also on other local electricity demands and generation, which are treated as static in our analysis. In practice, some of these static loads will have some flexibility to respond to high network tariffs during potential grid congestion periods. Therefore, quantifying grid upgrades without accounting for demand-response and grid tariffs would be misleading. Accordingly, our results should be interpreted as indicating the isolated potential impacts of DHN electrification on the existing grid, rather than as estimates of actual reinforcement requirements.

Second, a limitation of this work is that the sequential coupling between the DHN optimization and power-flow simulation does not allow the DHN to respond to grid constraints. Because the DHN model is optimized independently of the electricity network, it cannot exploit its flexibility (e.g., from thermal storage operation) to actively respond to grid loading. As a result, our reported grid impacts represent an upper bound. In practice, stronger coordination between DHN and electricity network operation could alleviate some of these impacts. Future work could quantify the benefits of coordinated operation by performing a single least-cost integrated optimization, representing a textbook coordinated solution, and comparing the resulting lower-bound grid impacts with those reported here.

Third, electricity prices are exogenous to the DHN optimization model. As a result, the model does not capture price-feedback effects, for example that a system dominated by electric boilers would, in equilibrium, require more upstream generation capacity and could therefore increase electricity prices, potentially favoring more efficient heat pump-based configurations. Energy efficiency is therefore partially undervalued in the cost-optimization step, which may shift technology choices toward less efficient electrified options. Nevertheless, within our framework, the MGA cost-slack budget and sensitivity scenarios (e.g., warm and cold weather years with different electricity prices) partly mitigate this effect by revealing a range of higher heat-pump-share designs under different conditions, rather than prescribing a single least-cost design. A natural extension is an endogenous-price formulation that couples the DHN capacity-expansion problem with a broader national energy system model to internalize these feedbacks.

Finally, computational limitations shape our modeling choices. The DHN planning model is formulated as a linear program (LP) to avoid the computational complexity associated with non-convex formulations. As a result, hydraulic flow dynamics are not explicitly represented, and we modeled heat flows using a simplified transport model. Moreover, investment decisions allow for fractional technology capacities rather than discrete unit sizes due to the LP formulation. Even with this simplification, the large number of MGA runs required to thoroughly explore the near-optimal space was only feasible using high-performance computing resources \cite{surf_snellius}. For similar computational reasons, location-specific technology constraints were imposed ex post by filtering the previously computed option space, rather than by re-running the MGA for each new constraint. Endogenously including each constraint and recomputing the near-optimal space would shift the global optimal cost, which is used for setting the cost budget. Our sensitivity analysis on the cost budget can be interpreted as partially capturing the implications of such shifts. A potential solution to improve computational efficiency would be to narrow the search space through stakeholder engagement, for example, via more targeted human-in-the-loop MGA approaches \cite{lombardi_2025_human_in_the_loop}. All these limitations provide a foundation for future research to focus on.
}

\section*{RESOURCE AVAILABILITY}\phantomsection\label{resource_availability}

\subsection*{Lead contact}\phantomsection\label{lead_contact}
Requests for further information, resources, and materials should be directed to and will be fulfilled by the lead contact, Christian Doh Dinga (c.dohdinga@tudelft.nl).

\subsection*{Materials availability}\phantomsection\label{materials_availability}
This study did not generate new unique materials.

\subsection*{Data and code availability}\phantomsection\label{data_code_availability}
\begin{itemize}
    \item \textbf{Data:} All datasets generated in this study are deposited on Zenodo and are publicly available: the default scenario (\url{https://zenodo.org/records/18943509}), the 5\% and 15\% cost-slack scenarios (\url{https://zenodo.org/records/18981309}), the low and high heat-demand scenarios (\url{https://zenodo.org/records/18981324}), and the cold and warm weather scenarios (\url{https://zenodo.org/records/18981332}). These deposits are also listed in the key resources table.
    \item \textbf{Code:} All source code to reproduce the experiments and visualizations in this study is available on GitLab (\url{https://gitlab.tudelft.nl/demoses/demoses-network-interactions}) and archived on Zenodo (\url{https://zenodo.org/records/18943449}). These references also appear in the key resources table.
    \item \textbf{Other items:} Any additional information required to reanalyze the data reported in this paper is available from the lead contact upon request.
\end{itemize}

\section*{ACKNOWLEDGMENTS}



This publication is part of the DEMOSES project, financed by the Dutch Research Council (NWO) under grant ID ESI.2019.004. This work used the Dutch national e-infrastructure with the support of the SURF Cooperative, using grant no. EINF-11996.

\section*{AUTHOR CONTRIBUTIONS}


Conceptualization, C.D.D. and M.C.; methodology, C.D.D. and F.L.; investigation, C.D.D. and F.L.; writing-–original draft, C.D.D. and F.L.; writing-–review \& editing, C.D.D., F.L., R.A., A.V.V., S.V.R., L.J.D.V. and M.C.; funding acquisition, L.J.D.V. and M.C.; resources, R.A., A.V.V., L.J.D.V. and M.C.; supervision, L.J.D.V. and M.C.

\section*{DECLARATION OF INTERESTS}



The authors declare no competing interests.


\section*{STAR\texorpdfstring{$\bigstar$}{*}METHODS}
\smallskip
\noindent
$\bullet$~\hyperref[key_resource_table]{KEY RESOURCES TABLE}\\[2pt]
$\bullet$~\hyperref[resource_availability]{RESOURCE AVAILABILITY}\\[1pt]
\hspace*{1.2em}$\circ$~\hyperref[lead_contact]{Lead contact}\\
\hspace*{1.2em}$\circ$~\hyperref[materials_availability]{Materials availability}\\
\hspace*{1.2em}$\circ$~\hyperref[data_code_availability]{Data and code availability}\\[2pt]
$\bullet$~\hyperref[method_details]{METHOD DETAILS}\\[1pt]
\hspace*{1.2em}$\circ$~\hyperref[methodology_framework]{Methodology framework}\\
\hspace*{1.2em}$\circ$~\hyperref[finding_least_cost]{Finding the least-cost district heating network design}\\
\hspace*{1.2em}$\circ$~\hyperref[finding_near_optimal_alternatives]{Finding near-optimal district heating network designs}\\
\hspace*{1.2em}$\circ$~\hyperref[network_impacts]{Quantifying electricity network impacts}\\
\hspace*{1.2em}$\circ$~\hyperref[model_setup]{Model setup} \\
$\bullet$~\hyperref[quantification_and_statistical_analysis]{QUANTIFICATION AND STATISTICAL ANALYSIS}\\[1pt]
\smallskip

\section*{SUPPLEMENTAL INFORMATION INDEX}


\begin{description}
    \item Figures S1-S41 and their captions in a PDF.
    \item Table S1. Technology cost and technical parameters used in the district heating network model.
\end{description}

\bibliography{references}
\bigskip

\section*{STAR\texorpdfstring{$\bigstar$}{*}METHODS}

\subsection*{KEY RESOURCES TABLE}\phantomsection\label{key_resource_table}

\definecolor{krtbanner}{gray}{0.90}

\newcolumntype{L}[1]{>{\raggedright\arraybackslash}p{#1}}

\begingroup
\setlength{\tabcolsep}{6pt}
\renewcommand{\arraystretch}{1.15}
\footnotesize

\begin{longtable}{L{0.34\textwidth} L{0.22\textwidth} L{0.36\textwidth}}
\toprule
\textbf{REAGENT or RESOURCE} & \textbf{SOURCE} & \textbf{IDENTIFIER} \\
\midrule
\endfirsthead

\toprule
\textbf{REAGENT or RESOURCE} & \textbf{SOURCE} & \textbf{IDENTIFIER} \\
\midrule
\endhead

\rowcolor{krtbanner}
\multicolumn{3}{l}{\textbf{Deposited data}} \\
Default scenario (10\% cost slack) & This paper & \url{https://zenodo.org/records/18943509} \\
5\% and 15\% cost slack scenarios & This paper & \url{https://zenodo.org/records/18981309} \\
Low and high demand scenarios & This paper & \url{https://zenodo.org/records/18981324} \\
Cold and warm weather scenarios & This paper & \url{https://zenodo.org/records/18981332} \\
Energy prices and weather data & ETM \cite{quintel_etm} & \url{https://energytransitionmodel.com} \\
DHN demand and topology data & Eneco & Available upon request \\
ELN demand and topology data & Stedin & Available upon request \\

\rowcolor{krtbanner}
\multicolumn{3}{l}{\textbf{Software and algorithms}} \\
PyPSA & Brown et al.\cite{PyPSA__2018_JORS} & \url{https://github.com/PyPSA/PyPSA} \\
PandaPower & Thurner et al.\cite{pandapower_2018_TPWRS} & \url{https://www.pandapower.org} \\
Reproducibility code for this study & This paper & \url{https://gitlab.tudelft.nl/demoses/demoses-network-interactions} \\
Reproducibility code archive         & This paper & \url{https://zenodo.org/records/18943449} \\

\bottomrule
\end{longtable}
\endgroup


\section*{METHOD DETAILS}\phantomsection\label{method_details}


\subsection*{Methodology framework}\phantomsection\label{methodology_framework}

A visualization of the methodology framework developed in this study is shown in Figure~\hyperref[methodology_framework]{5}. We use model coupling to link the DHN capacity expansion and operation optimization model to the AC power flow simulation model of the electricity network. The two models are coupled together using the temporally and spatially resolved electricity profiles as interface variables.

\vspace{1em}

\includegraphics[width=0.9\linewidth]{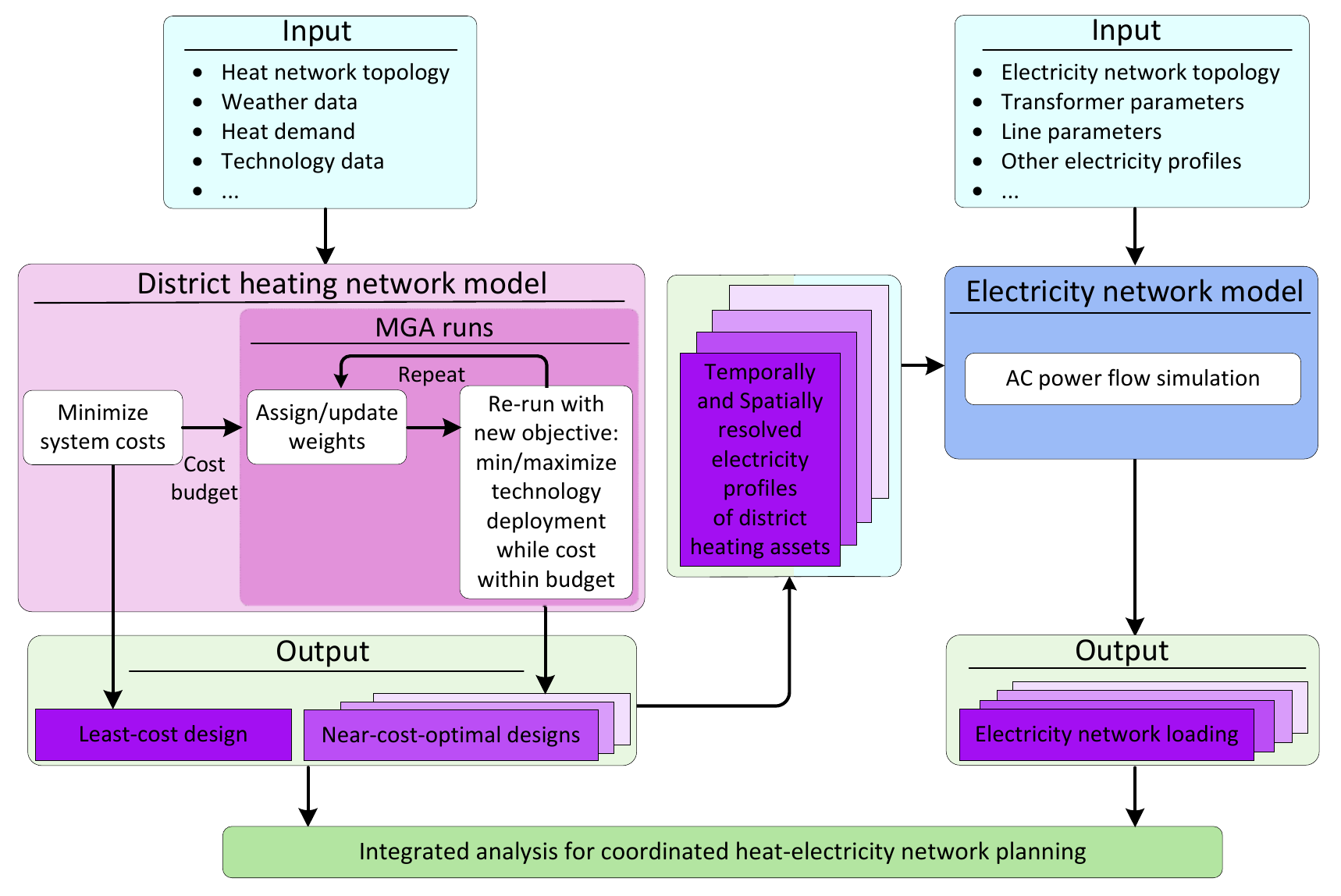}
\subsubsection*{Figure 5. Methodology framework}\phantomsection\label{methodology_framework_figure}

The model execution follows a sequential approach. First, we solve the DHN optimization problem to identify the least-cost system design. This yields the optimal capacities of all heat supply technologies, storage, and pipelines, along with their dispatch for each snapshot over the full simulation horizon. From the dispatch results, we extract the temporally and spatially resolved electricity consumption profiles for all P2H technologies (electric boilers and heat pumps) and other technologies, such as geothermal heat sources and HT-ATES, that have auxiliary electricity consumption due to booster heat pumps that upgrade heat from these sources. We also extract temporally and spatially resolved electricity-generation profiles for CHP technologies. These endogenously determined electricity profiles, along with the exogenous electricity profiles from other local loads/generation connected to the same electricity network, are subsequently used as inputs to an AC power flow model. The results of the second model run are then used to quantify the impacts of the least-cost DHN design on the grid using transformer and line loadings.

After analyzing the least-cost design and computing the least-cost of the system, we use it as input and apply a Modeling-to-Generate-Alternatives (MGA) algorithm (see section~\hyperref[finding_near_optimal_alternatives]{Finding near-optimal district heating network designs}) to generate a large set of near-optimal alternative DHN designs. For each alternative design, we repeat the full workflow: (i) capacity expansion and dispatch optimization, (ii) extraction of electricity profiles, (iii) AC power flow simulation. We use the open-source orchestrator tool Snakemake \cite{Snakemake_2021} to automate the entire workflow and ensure reproducibility. See supplemental  Figure~\hyperref[snakemake_dag_workflow_fig]{S41} for the resulting directed acyclic graph (DAG) showing all the jobs executed and their dependencies for the entire workflow per scenario. The final step of the workflow consists of synthesizing the outputs from both models into an integrated analysis to support coordinated network planning decisions.

\subsection*{Finding the least-cost district heating network design}\phantomsection\label{finding_least_cost}

The capacity expansion and operation optimization model of the DHN shown in supplemental Figure~\hyperref[heat_network_fig]{S1} is implemented in the open-source tool PyPSA \cite{PyPSA__2018_JORS}. An energy system infrastructure in PyPSA is represented using fundamental components such as buses $I$, generators $G$, links $F$, storage units $H$, and stores $E$. Based on this component representation, PyPSA automatically formulates a linear optimization problem that minimizes total system costs subject to technical and physical constraints.

The objective function of the DHN optimization problem is to minimize the total annual DHN system costs, including both investment and operational expenditures for generation, conversion, storage, and distribution infrastructure, and maximize revenues from electricity generated by CHPs. To express investment costs on an annual basis, we use the annuity factor $(1 - (1 + \tau)^{-n}) / \tau$, which converts the upfront overnight investment cost of an asset into annual payments, considering its lifetime $n$ and cost of capital $\tau$. The objective function therefore includes the annualized capital costs $c_{*}$ for investments at node $i$ in generator capacity $G_{i,r} \in \mathbb{R}^{+}$ of technology $r$, storage energy capacity $E_{i,s} \in \mathbb{R}^{+}$ of technology $s$, and energy conversion and transport capacities $F_{k} \in \mathbb{R}^{+}$ (links). In addition, it includes variable operating costs $o_{*}$ associated with generator dispatch $g_{i,r,t} \in \mathbb{R}^{+}$ and link dispatch $f_{k,t} \in \mathbb{R}^{+}$. Accordingly, the objective function can be written as shown in Equation~\ref{eq:lc_objective_function}:


\begin{equation}
\label{eq:lc_objective_function}
\left\{
\begin{split}
\underset{G, E, F, g, f}{\min} \Bigg[
\sum_{i,r} c_{i,r} \cdot G_{i,r}
+ \sum_{i,s} c_{i,s} \cdot E_{i,s}
+ \sum_{k} c_{k} \cdot F_{k} \\
+ \sum_{t} W_{t} \cdot \left(
\sum_{i,r} o_{i,r} \cdot g_{i,r,t}
+ \sum_{k} o_{k} \cdot f_{k,t}
\right) \\
- \left(\sum_{t,k} \lambda^{elec}_{t} \cdot \eta^{elec} \cdot \, f_{k=\text{chp}, t}\right)
\Bigg]
\end{split}
\right.
\end{equation}

Thereby, the representative time snapshots t are weighted by the time span $W_{t}$ such that their total
duration adds up to one year; $\sum_{t \in T} W_{t} = 365 \cdot  24\mathrm{h} = 8760\mathrm{h}$. The last term of the objective function represents revenues from electricity generated by CHPs, which is a product of electricity prices $\lambda^{elec}_{t}$, the electrical efficiency $\eta^{elec}$, and dispatch $f_{k={chp}, t}$ of CHPs.

In addition to the cost-minimizing objective function, as extensively described in prior work \cite{neumann_2023_h2_joule}, the model imposes a set of linear constraints including energy balance at each node and timestep, storage level consistency over time, limits on the capacities of generation, storage, conversion, and distribution infrastructure arising from technical potentials, and time-dependent availability constraints for variable heat sources.

\subsubsection*{Technology modeling}\phantomsection\label{technology_modeling}

In this section, we provide a high-level description of how different district heating technologies are modeled using PyPSA components. We present mathematical formulations only for technologies that require additional custom constraints beyond those already established by PyPSA. An exhaustive description of PyPSA’s mathematical formulation is provided in \citeauthor{neumann_2023_h2_joule} \cite{neumann_2023_h2_joule}, and our code is openly available online \cite{doh_dinga_2026_code}.

\textit{Air-source heat pumps}

\noindent Air-source heat pumps (ASHPs) are modeled as unidirectional links that convert electricity into heat with an efficiency given by the coefficient of performance (COP). The COP depends on the ambient temperature and on the temperature difference between source and sink, defined as $\Delta T = T_{\mathrm{sink}} - T_{\mathrm{source}}$, and is therefore time-varying. The COP for ASHPs is calculated as:

\begin{equation}
\mathrm{COP}(\Delta T) = 6.81 + 0.121 \Delta T + 0.000630 \Delta T^{2}
\end{equation}

The numerical coefficients correspond to the default values in the PyPSA-Eur model \cite{neumann_2023_h2_joule}, derived from regression analysis. The resulting time-varying COP in relation to ambient temperature for the default weather year is shown in supplemental Figures~\hyperref[temperature_fig]{S8} and~\hyperref[heatpump_cop_fig]{S9}.

\textit{Boilers and waste-to-energy}

\noindent Electric boilers, gas (green gas and hydrogen) boilers, and waste-to-energy are modeled using PyPSA's fundamental link component, and are configured as unidirectional energy converters that convert electricity, gas, and waste material, respectively, into heat with fixed conversion efficiencies.

\textit{Combined heat and power}

\noindent Combined heat and power (CHP) technologies are modeled as unidirectional multi-output links that convert gas (green gas or hydrogen) into heat and electricity. Similar to \citeauthor{neumann_2023_h2_joule} \cite{neumann_2023_h2_joule}, we assume back-pressure operation with a fixed ratio between electricity and heat output.

\textit{Geothermal heat}

\noindent A geothermal heat system is modeled as a compound component consisting of a generator and two unidirectional links. The generator represents a geothermal well producing low-temperature heat $q^{\mathrm{init}}_{t}$, one link represents a booster heat pump consuming electricity and producing heat $q^{\mathrm{booster}}_{t}$, and the final link supplies the upgraded heat $q^{\mathrm{final}}_{t}$ to the heat network. The seasonal performance factor (SPF) of the geothermal system, defined as the electricity consumption per unit of final heat supplied to the network, couples these variables. Therefore, we impose the following additional constraints:

\begin{equation}
\label{eq:geothermal_eq1}
    q^{final}_t = q^{init}_t + q^{booster}_t
\end{equation}

\begin{equation}
\label{eq:geothermal_eq2}
    q^{final}_t = SPF \cdot \frac{q^{booster}_t}{COP(t)}
\end{equation}

Combining Equation~\ref{eq:geothermal_eq1} and Equation~\ref{eq:geothermal_eq2} yields:

\begin{equation}
    q^{booster}_t = \frac{COP(t)}{SPF -COP(t)} \cdot q^{init}_t
\end{equation}

which guarantees a fixed proportional contribution of geothermal well heat and booster heat to the final heat supplied by the geothermal heat source to the heat network.

\textit{Solar thermal and residual heat}

\noindent Solar thermal and residual heat are modeled as generator components. For solar thermal, dispatch is constrained by the optimized capacity and a time-varying capacity factor derived from solar availability. High-temperature industrial residual heat is represented as a generator with a time-varying availability profile dependent on industrial activity, such that heat production is limited by both optimized capacity and temporal availability.

\textit{Seasonal thermal energy storage}

\noindent Seasonal storage in the form of high-temperature aquifer thermal energy storage (HT-ATES) is modeled as a compound component consisting of a store and four links. The store represents the thermal energy reservoir with energy level $e_{t}$, which has a standing loss or thermal decay defined as $1- \mathrm{exp}(-1/24t)$. The links correspond to charging $q^{charge}_{t}$, initial low-temperature discharge $q^{discharge}_{t}$, booster heat pump output $q^{booster}_{t}$, and final upgraded heat supplied to the network $q^{final}_{t}$. The relationship between $e_{t}$, $q^{charge}_{t}$, and $q^{discharge}_{t}$ is automatically established by PyPSA through intertemporal storage energy level balance constraints. Similar to geothermal systems, HT-ATES has a SPF that couples $q^{discharge}_{t}$, $q^{booster}_{t}$, and $q^{final}_{t}$. Accordingly, we introduce the following additional constraints:

\begin{equation}
\label{eq:ht_ates_eq1}
    q^{final}_t = q^{discharge}_t + q^{booster}_t
\end{equation}

\begin{equation}
\label{eq:ht_ates_eq2}
    q^{final}_t = SPF \cdot \frac{q^{booster}_t}{COP(t)}
\end{equation}

Combining Equation~\ref{eq:ht_ates_eq1} and Equation~\ref{eq:ht_ates_eq2} yields:

\begin{equation}
    q^{booster}_t = \frac{COP(t)}{SPF -COP(t)} \cdot q^{discharge}_t
\end{equation}

which guarantees a fixed proportional contribution of reservoir discharge and booster heat to the final heat supplied by the HT-ATES system to the network.

\textit{Short-term thermal energy storage}

\noindent Short-term thermal energy storage in small water tanks is modeled using PyPSA’s fundamental storage unit component with charging and discharging efficiencies and standing losses. The power and energy capacities in PyPSA are linked through a maximum discharge duration $h^{\max}$ parameter.

\textit{Heat pipelines}

\noindent Heat pipelines are modeled using PyPSA's fundamental link component, but are configured as bidirectional to allow flows in both directions between connected nodes. Heat distribution in pipelines is subject to efficiency losses proportional to pipeline length, and the flow through each pipe is limited by its optimized capacity.

\subsection*{Finding near-optimal district heating network designs}\phantomsection\label{finding_near_optimal_alternatives}
We use Modeling to Generate Alternatives (MGA) techniques \cite{lombardi_2025_mga_perspective} to compute near-optimal DHN designs that differ in choice, sizing, and location of technologies in the network. For notational brevity, let $c^{T}\mathrm{x}$ denote the linear objective function in Equation~\ref{eq:lc_objective_function} and $A\mathrm{x} \leq b$ the set of constraints that define the feasible region of the optimization problem in a space of continuous variables such that the minimized system cost $C^{*}$ can be represented by:

\begin{equation}
   C^{*} = \underset{\mathrm{x}} \min \{c^{T}\mathrm{x} | A\mathrm{x} \leq b \}
\end{equation}

To compute near-optimal alternatives, the least-cost solution $C^{*}$ is first identified as a starting point. Subsequently, the original cost-based objective expression is turned into a problem constraint such that the cost increase is limited by a slack $\epsilon$ as shown in Equation~\ref{eq:mga_slack_constraint}.

\begin{equation}
\label{eq:mga_slack_constraint}
   c^{T}\mathrm{x} \leq (1 + \epsilon) \cdot C^{*}
\end{equation}

In other words, the feasible space is cut to solutions that are at most $\epsilon$ more expensive than the least-cost solution $C^{*}$. With this additional constraint, we can reformulate the objective function to focus on iteratively identifying different feasible system designs, all of which are within a reasonable economic-viability bandwidth $\epsilon$. As discussed in prior work \cite{lombardi_2025_mga_perspective,lombardi_2023_what_is_redundant}, MGA objectives for such exploration of alternative near-optimal solutions can be set up in various ways, each with its own relative merits. In this work, we use an ad-hoc version of the MGA SPORES algorithm \cite{lombardi_2020_SPORES,lombardi_2023_what_is_redundant}, which is particularly suited to exploring system designs that differ not only in overall technology mix but also in the spatial deployment of technology, a key variable for the grid impact that we set out to assess in this study.

\subsubsection*{Generating SPORES}\phantomsection\label{generating_spores}
The key idea underlying the generation of Spatially explicit Practically Optimal Results (SPORES) is to use a multi-objective MGA search strategy that combines spatial and technological diversification with feature intensification, enabling heavy parallel computing and tailoring to problem needs. While this has been partly the case since the method's first appearance \cite{lombardi_2020_SPORES}, SPORES has evolved and has been used in various formulations in the literature \cite{lombardi_2023_what_is_redundant}, which may lead to confusion. Here, we provide a more generalizable and up-to-date illustration of SPORES, and we outline how we use it in this work.

The mathematical formulation of a generic problem for generating SPORES can be defined as follows.

\begin{equation}
\label{eq:mga_spores_formulation}
\left\{
\begin{split}
\min \mathrm{Z} = 
\mathrm{a} \cdot \left( \sum_{i,r} \omega_{i,r} \cdot Y^{cap}_{i,r} \right)
\pm  \mathrm{b} \cdot \left(  \sum_{i} \hat Y^{cap}_{i,r} \right) \\
\quad \quad \mathrm{s.t.} \quad \quad A\mathrm{x} \leq b \\
c^{T}\mathrm{x} \leq (1 + \epsilon) \cdot C^{*}
\end{split}
\right.
\end{equation}

The first term in Equation~\ref{eq:mga_spores_formulation} includes all technologies subject to diversification, where $Y^{cap}_{i,r}$ indicates the capacity decision variable at location $i$ for technology type $r$. Diversification is achieved by updating the dynamic weight $\omega_{i,r}$. This is similar to conventional `Hop-Skip-Jump' methods but is spatially explicit and offers several customization options for the weight formulation, with different relative merits \cite{lombardi_2023_what_is_redundant}, as described below. The second term includes only the subset of technologies under intensification, with $ \hat Y^{cap}_{i,r}$ the capacity decision variable under intensification (maximized or minimized). Also known as the `variable min-max' method when used standalone \cite{Lau_2024_IOP}, the intensification component of the objective ensures that all technological boundaries of the solution space are systematically mapped, while the diversification component ensures that in-between options are also sampled. The coefficients $a$ and $b$ represent static weights that configure the problem as a linear multi-objective optimization problem. When $b$ has a positive sign, the technology under intensification is minimized; when $b$ has a negative sign, the technology is maximized; and finally, if $b$ is null, the formulation collapses into a diversification-only run. 

In fact, the above formulation is designed to leverage heavy parallel computing. Equation~\ref{eq:mga_spores_formulation} is applied across many parallel runs, each intensifying a different key technology (e.g., hydrogen boilers) in a different direction (maximization or minimization), and generating a small sample of diverse designs around the intensified technology. In each parallel run, the first near-optimal solution is always generated with an exclusive focus on intensification (i.e., the $a$ coefficient is null for the first near-optimal solution). Subsequently, the diversification component is added to generate further near-optimal solutions that lie in the same region of the design space but are technologically or spatially distinct from such first near-optimal solution. In addition, a parallel run is carried out exclusively with a diversification component and no intensified feature ($b$ is null), in this case targeting the cost-optimal solution as the design to diverge from. In this study, we implement 32 parallel runs. 31 runs represent intensification directions, each generating 15 distinct solutions, while the final run is purely a diversification search, generating 50 distinct solutions. This yields a total of 515 near-optimal alternative system designs.

A distinct feature of the SPORES algorithm is its high degree of customization in updating the dynamic weight $\omega_{i,r}$, which drives the diversification component of the objective. As presented in prior work \cite{lombardi_2023_what_is_redundant}, SPORES supports up to four distinct weighting methods, each best suited to particular diversification goals. In this study, we adopt an \textit{evolving average} strategy to update $\omega_{i,r}$. The \textit{evolving average} strategy assigns a weight to each location–technology pair based on the distance from the average capacity deployed $\overline{Y}^{cap}_{i,r}$ for that pair across all previously found solutions, which is kept up to date\textemdash in other words, this average evolves. The \textit{evolving average} weight update strategy at iteration $n$ is defined mathematically as in Equation~\ref{eq:mga_evo_average}, and is particularly suited for deviating strongly from the reference solution\textemdash namely, the first near-optimal solution found in each parallel run, adopting a pure intensification objective. This ensures that, in each parallel run with a given intensification goal, we find solutions that progressively diverge from that goal, creating a smoother system design gradient instead of, for instance, solutions more tightly clustered around the intensified feature.

\begin{align}
\label{eq:mga_evo_average}
\left\{
\begin{aligned}
\omega^{n}_{i,r} \quad  \quad 
&= \left| \frac{\overline{Y}^{cap, n-1}_{i,r} - Y^{cap, n}_{i,r}}{\overline{Y}^{cap, n-1}_{i,r}} \right|^{-1} \\
\overline{Y}^{cap, n-1}_{i,r} 
&= \frac{\sum_{k=1}^{n-1} Y^{cap, k}_{i,r}}{n-1}
\end{aligned}
\right.
\end{align}

\subsection*{Quantifying electricity network impacts}\phantomsection\label{network_impacts}
As earlier described in section~\hyperref[methodology_framework]{Methodology framework}, to quantify the electricity network impacts of each DHN design, we use the resulting electricity profiles of the DHN assets as inputs to a power flow simulation. The full AC power flow model of the electricity network shown in supplemental Figure~\hyperref[electricity_network_fig]{S2} is implemented in the open-source tool pandapower \cite{pandapower_2018_TPWRS}.

Pandapower represents the network mathematically using the complex nodal admittance matrix $Y \in \mathbb{C}^{N \times N}$, whose elements are defined as $Y_{ij} = G_{ij} + j B_{ij}$, where $G_{ij}$ and $B_{ij}$ denote the conductance and susceptance between buses $i$ and $j$, respectively. Given nodal loads and generations for each timestep, pandapower establishes active $P_i$ and reactive $Q_i$ power balance equations at every bus $i$ in the network, defined as in Equation~\ref{eq:pf_active_power} and Equation~\ref{eq:pf_reactive_power}, respectively.

\begin{equation}
\label{eq:pf_active_power}
P_i =
V_i \sum_{j=1}^{N} V_j
\left(
G_{ij} \cos(\theta_i - \theta_j)
+
B_{ij} \sin(\theta_i - \theta_j)
\right)
\end{equation}

\begin{equation}
\label{eq:pf_reactive_power}
Q_i =
V_i \sum_{j=1}^{N} V_j
\left(
G_{ij} \sin(\theta_i - \theta_j)
-
B_{ij} \cos(\theta_i - \theta_j)
\right)
\end{equation}

where $P_i$ and $Q_i$ denote the  net active and reactive power injection at bus $i$, $V_i$ is the voltage magnitude, and $\theta_i$ is the voltage angle. The nodal injection is defined as the difference between generation and demand for active $P_i = P^{\text{gen}}_i - P^{\text{load}}_i$ and reactive $Q_i = Q^{\text{gen}}_i - Q^{\text{load}}_i$ power. The electricity consumption and generation profiles obtained from the DHN optimization enter the model as time-dependent active nodal power injections $P^{\text{load}}_i(t)$ and $P^{\text{gen}}_i(t)$, and pandapower computes their corresponding reactive power as $Q_i(t) = P_i(t) \cdot \tan(\phi)$. The phase angle $\phi$ is derived from the power factor $PF$ as $\arccos(PF)$.

To ensure a unique solution of this nonlinear system of equations, one bus is defined as the reference bus. At the reference bus, the voltage magnitude $V_{\text{reference}}$ and voltage angle $\theta_{\text{reference}}$ are fixed, while active and reactive power injections required to satisfy the system of equations are computed endogenously. For all remaining 
PQ buses (P and Q are given), pandapower solves for the unknown voltage magnitudes $V_i$ and voltage angles $\theta_i$ such that Equation~\ref{eq:pf_active_power} and Equation~\ref{eq:pf_reactive_power} are satisfied at each bus. The solution of the AC power flow provides nodal voltages, transformer loadings, and line loadings at each time step, which we used to quantify the impacts of each DHN design on the electricity network.

\subsection*{Model setup}\phantomsection\label{model_setup}

We simulate the year 2050, as the Netherlands aims to achieve carbon neutrality by this year. Most of our input data are obtained from the Energy Transition Model \cite{quintel_etm}, which provides rich data on the Dutch energy system at a national scale. This includes projected energy carrier prices for electricity, hydrogen, and green gas until 2050, as well as weather data such as solar thermal capacity factor and ambient temperature for different weather years. All time-varying data, such as energy prices (see Figures~\hyperref[electricity_price_fig]{S4} and~\hyperref[hydrogen_price_fig]{S5}),  solar thermal capacity factor (see Figure~\hyperref[solar_thermal_fig]{S7}), and ambient temperature (see Figure~\hyperref[temperature_fig]{S8}) are available at an hourly temporal resolution.

More detailed data on the current DHN topology and future expansion plans (see Figure~\hyperref[heat_network_fig]{S1}), projected heat demand for the studied DHN (see Figure~\hyperref[heat_demand_fig]{S3}), and techno-economic parameters of heat technologies\textemdash such as technology efficiencies, and technical potentials\textemdash and projected prices for other relevant energy carriers\textemdash such as waste materials and residual heat (see supplemental Figure~\hyperref[static_prices_fig]{S6})\textemdash are provided by the DHN operator. The investment costs of heat technologies (see supplemental Table~\hyperref[dhn_technology_params_table]{S1}) are sourced from the Netherlands Environmental Assessment Agency report \cite{pbl_sdepp_2025}, and we assume that operational costs are incurred solely from fuel and electricity consumption. For the DHN optimization model, we assume a greenfield overnight investment setting in which no existing assets are considered, as the current fleet of heat technologies is expected to be retired well before 2050. Annualized capital costs are calculated using a discount rate of 7\%. The optimization determines both investment and operational decisions at hourly resolution for a full year.

For the MGA runs, we use a 10\% cost slack relative to the least-cost solution for the default scenario. We intensify (maximize and minimize) each of the technologies defined in section~\hyperref[technology_modeling]{Technology modeling}, as well as selected subsets of technologies. These include electrification technologies (electric boilers and heat pumps), molecule-based technologies (gas boilers and CHPs), dispatchable technologies (e.g., boilers), and baseload or non-dispatchable technologies (e.g., geothermal and residual heat). Under each of these main search directions, a batch of SPORES is generated by adding a diversification term in the new objective function and updating its weights through iterations (see the \texttt{run\_heat\_model\_for\_each\_spores} loop in Figure~\hyperref[snakemake_dag_workflow_fig]{S41}) as described in section~\hyperref[generating_spores]{Generating SPORES}. In total, these runs result in 515 SPORES, or alternative DHN designs for the default scenario with a 10\% cost slack relative to the least-cost solution.

Given the large number of input parameters, we complement the MGA analysis with additional sensitivity analyses. Specifically, we vary the cost slack parameter to 5\% and 15\%. We also consider two alternative weather years, in addition to the default weather year 2015: a warm weather year (2004), characterized by several periods of excess solar generation, and a cold weather year (1987), with low solar and wind output\textemdash the so-called cold dark doldrums. Furthermore, to reflect uncertainty in future heat demand projections, we consider low (-20\%) and high (+20\%) heat demand scenarios. For each of these scenarios, we generate 515 SPORES, which, together with the default scenario, result in a total of 3,605 SPORES, which are all subject to a full AC hourly-resolution power flow simulation for the entire year (8760 snapshots).

For the electricity system, we model the current regional network topology (see supplemental Figure~\hyperref[electricity_network_fig]{S2}), and do not explicitly consider grid reinforcement. This is because grid reinforcement needs depend not only on DHN-related electricity profiles but also on other local electricity load and generation profiles\textemdash which are treated as static here but may in practice have some flexibility to respond to high network tariffs during potential grid congestion periods. Detailed data on network parameters for buses, transformers, and lines are provided by the electricity network operator. We combine this existing network infrastructure with projected 2050 exogenous electricity profiles for all loads and generators connected to it. The endogenously calculated electricity consumption and generation from the DHN are added separately based on the DHN optimization results. The power flow simulations are performed at hourly resolution. For solving the AC power flow equations, we assume a power factor of 0.95 and focus on active power to assess transformer and line loadings, which we used to quantify grid impacts.

\section*{QUANTIFICATION AND STATISTICAL ANALYSIS}\phantomsection\label{quantification_and_statistical_analysis}

This study is fully computational and does not involve statistical hypothesis testing, confidence intervals, or p-values. Each near-optimal DHN design (SPORE) is the deterministic output of a single linear optimization run, and each electricity-network impact metric is the deterministic output of a single hourly-resolution AC power-flow simulation over a full year (8{,}760 snapshots).

For each scenario, we generate $n = 515$ SPORES at a given cost slack (e.g., 10\% for the default scenario) relative to the least-cost solution. We additionally vary the cost-slack budget (5\% and 15\%), the weather year (cold: 1987; warm: 2004), and the heat-demand projection ($-20\%$ and $+20\%$), yielding a total of 3{,}605 alternative DHN designs across all scenarios. In all strip-plots and scatter-plots, each marker represents one near-optimal DHN design, and the visual ``spread'' along any axis represents the variation across the relevant set of designs ($n = 515$ per scenario) rather than statistical error or measurement uncertainty.

Summary metrics shown in the figures (e.g., the 75th-percentile in-distribution line/transformer loading and the spread of in-distribution overload events) are computed across the empirical distribution over the 8{,}760 hourly snapshots of the AC power-flow simulation for each design; the construction of the lower-grid-loading envelope highlighted in Figure~\hyperref[integrated_decision_space_fig]{3} is described in the \hyperref[network_impacts]{Quantifying electricity network impacts} subsection.

All computations were performed in Python using PyPSA for the DHN capacity-expansion and dispatch optimization \cite{PyPSA__2018_JORS} with Gurobi as the LP solver, PandaPower for AC power-flow simulation \cite{pandapower_2018_TPWRS}, and Snakemake for workflow orchestration \cite{Snakemake_2021}.

\newpage

\clearpage
\section*{SUPPLEMENTAL INFORMATION}
\setcounter{figure}{0}
\renewcommand{\thefigure}{S\arabic{figure}}

\vspace{3em}





\vspace{12em}
\noindent
\begin{minipage}{\linewidth}
    \centering
    \includegraphics[width=0.95\linewidth]{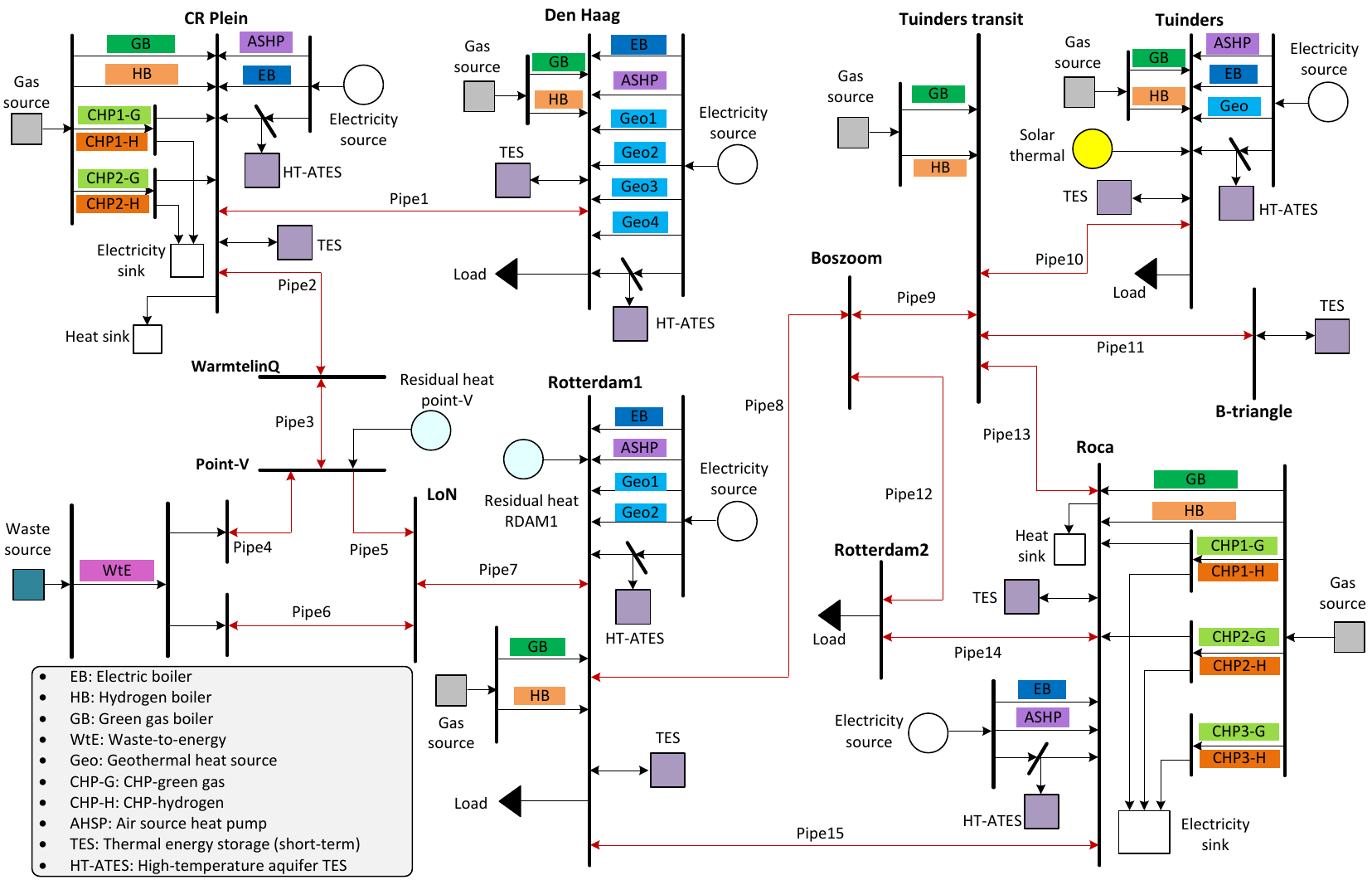}
    
    \subsubsection*{Figure S1. District heating network}
    \phantomsection
    \label{heat_network_fig}
\end{minipage}

\vspace{3em}

\noindent
\begin{minipage}{\linewidth}
    \centering
    \includegraphics[width=0.97\linewidth]{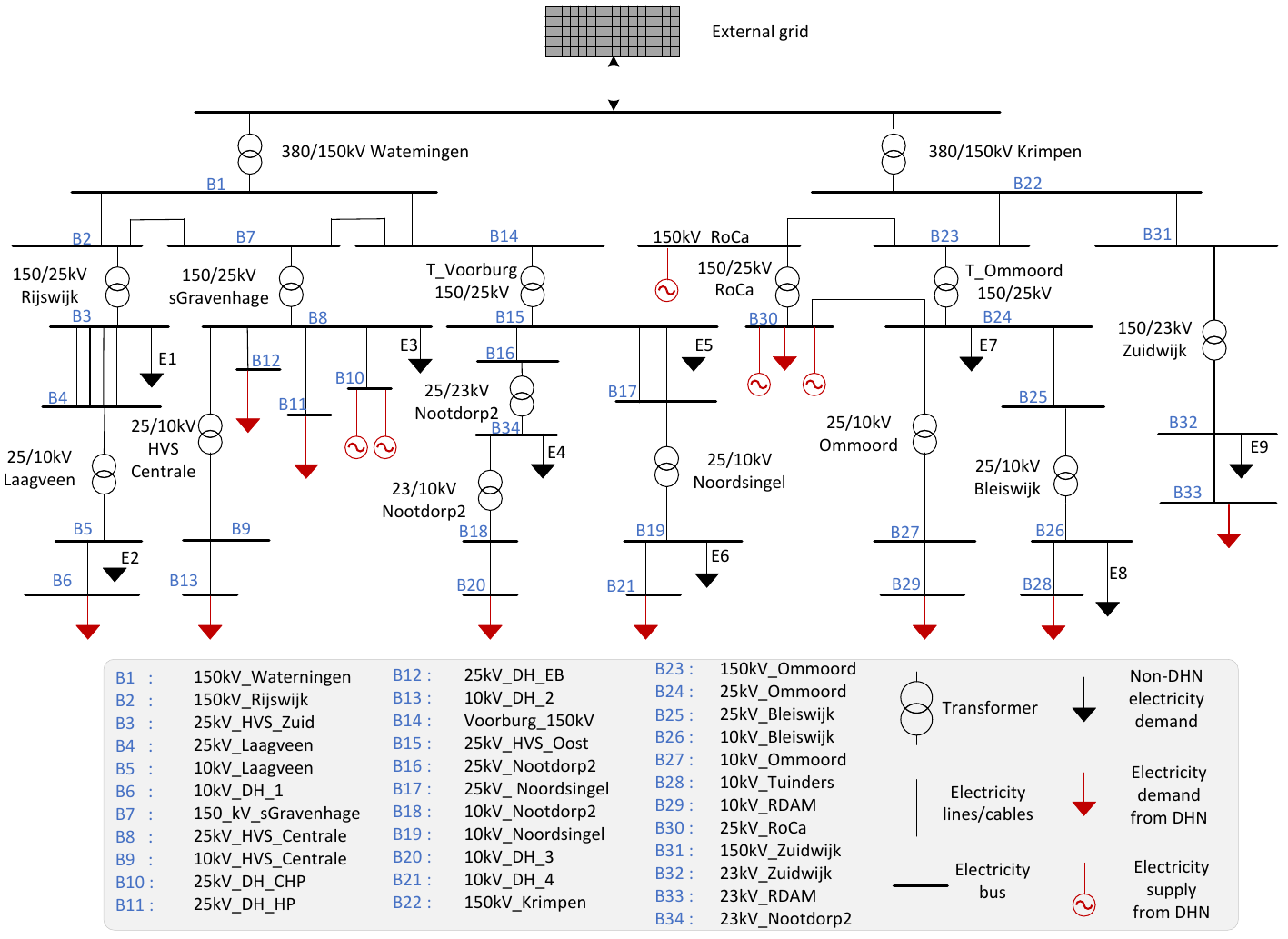}
    
    \subsubsection*{Figure S2. Electricity network}
    \phantomsection
    \label{electricity_network_fig}
\end{minipage}

\vspace{2em}

\noindent
\begin{minipage}{\linewidth}
    \centering
    \includegraphics[width=0.9\linewidth]{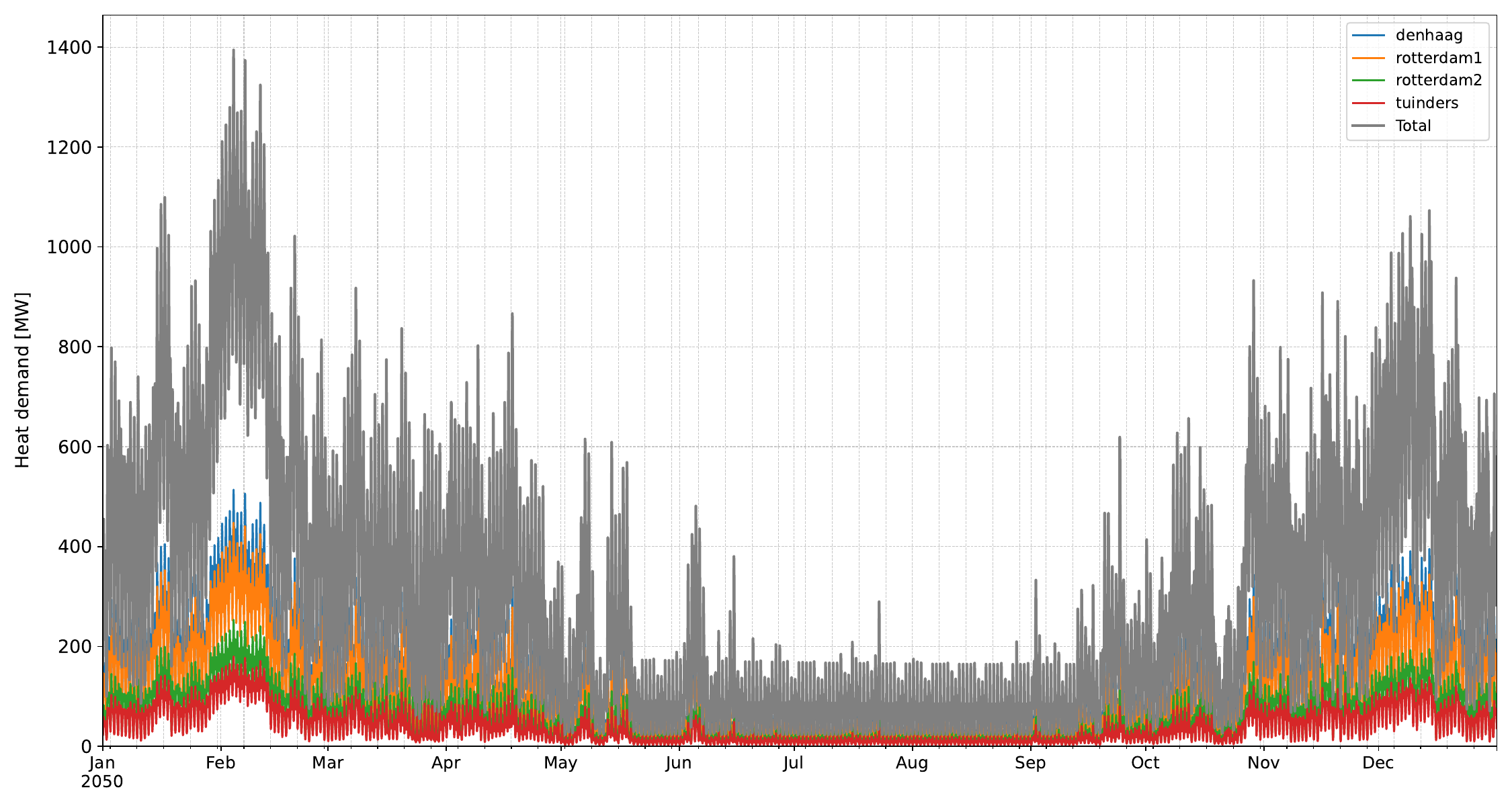}
    
    \subsubsection*{Figure S3. Projected heat demand by 2050}
    \phantomsection
    \label{heat_demand_fig}
\end{minipage}

\noindent
\begin{minipage}{\linewidth}
    \centering
    \includegraphics[width=0.9\linewidth]{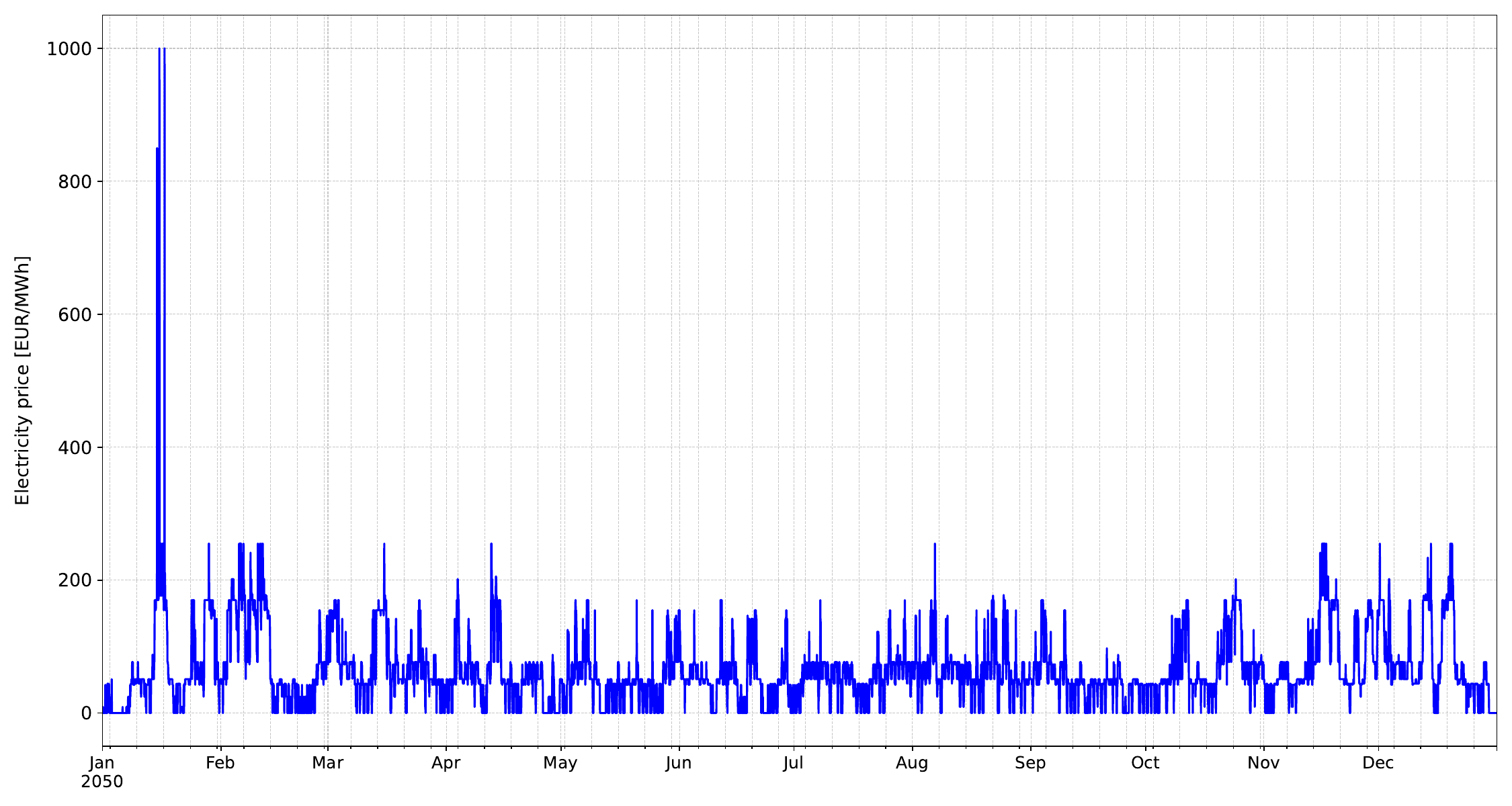}
    
    \subsubsection*{Figure S4. Electricity price}
    \phantomsection
    \label{electricity_price_fig}
\end{minipage}

\vspace{4em}

\noindent
\begin{minipage}{\linewidth}
    \centering
    \includegraphics[width=0.9\linewidth]{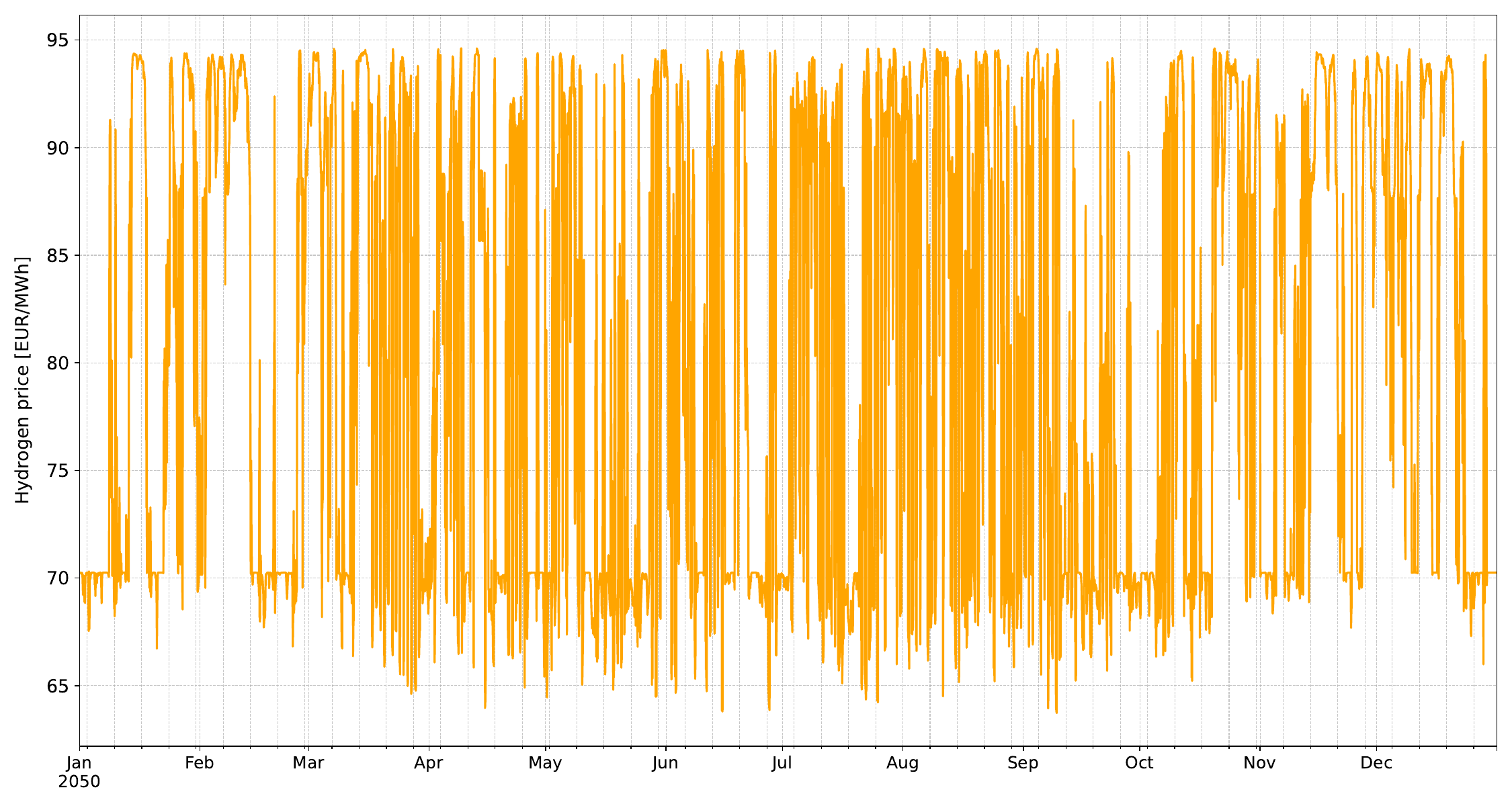}
    
    \subsubsection*{Figure S5. Hydrogen price}
    \phantomsection
    \label{hydrogen_price_fig}
\end{minipage}

\noindent
\begin{minipage}{\linewidth}
    \centering
    \includegraphics[width=0.9\linewidth]{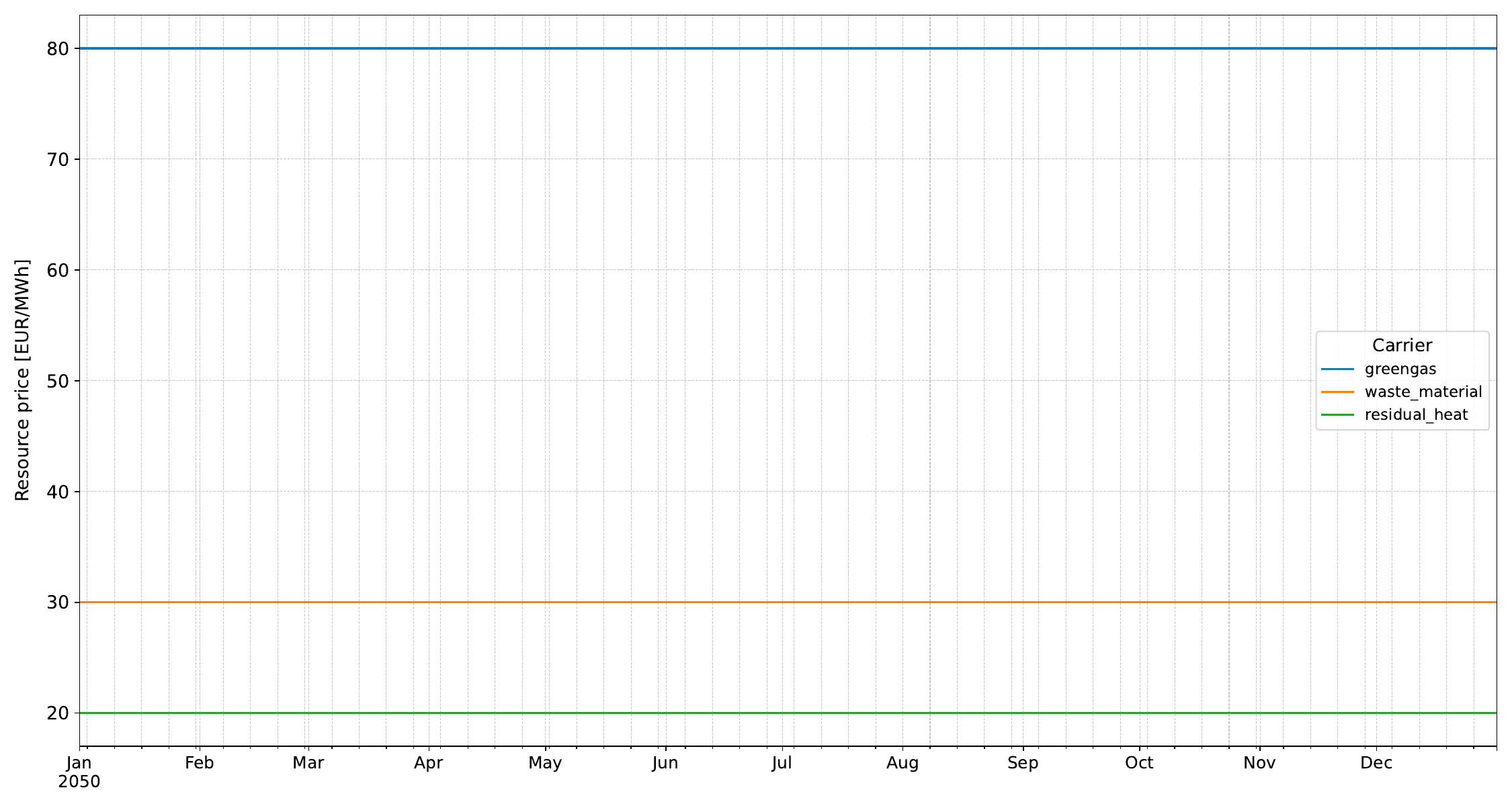}
    
    \subsubsection*{Figure S6. Green gas, waste material, and residual heat price}
    \phantomsection
    \label{static_prices_fig}
\end{minipage}

\vspace{4em}

\noindent
\begin{minipage}{\linewidth}
    \centering
    \includegraphics[width=0.9\linewidth]{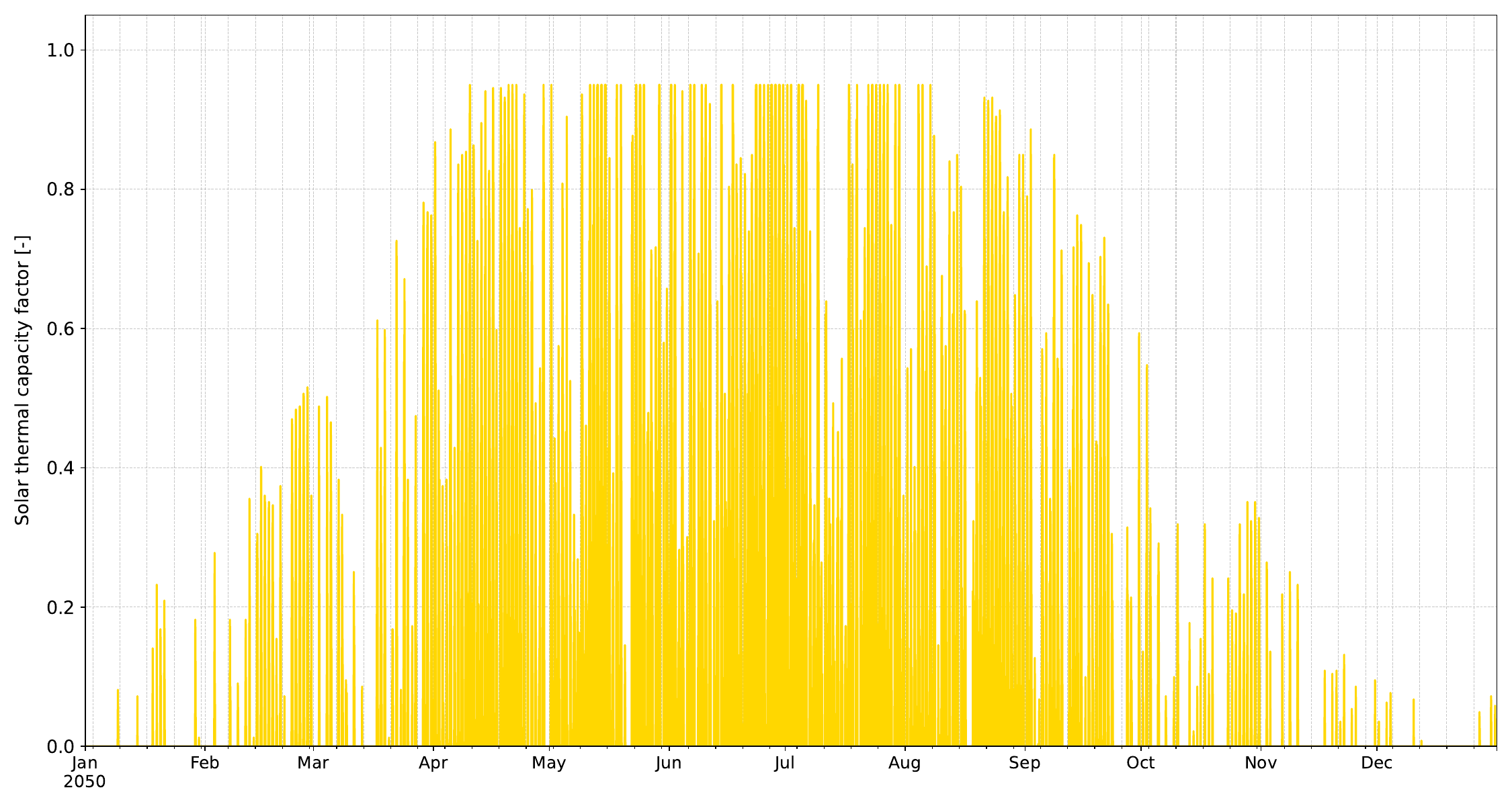}
    
    \subsubsection*{Figure S7. Solar thermal capacity factor}
    \phantomsection
    \label{solar_thermal_fig}
\end{minipage}

\vspace{2em}

\noindent
\begin{minipage}{\linewidth}
    \centering
    \includegraphics[width=0.9\linewidth]{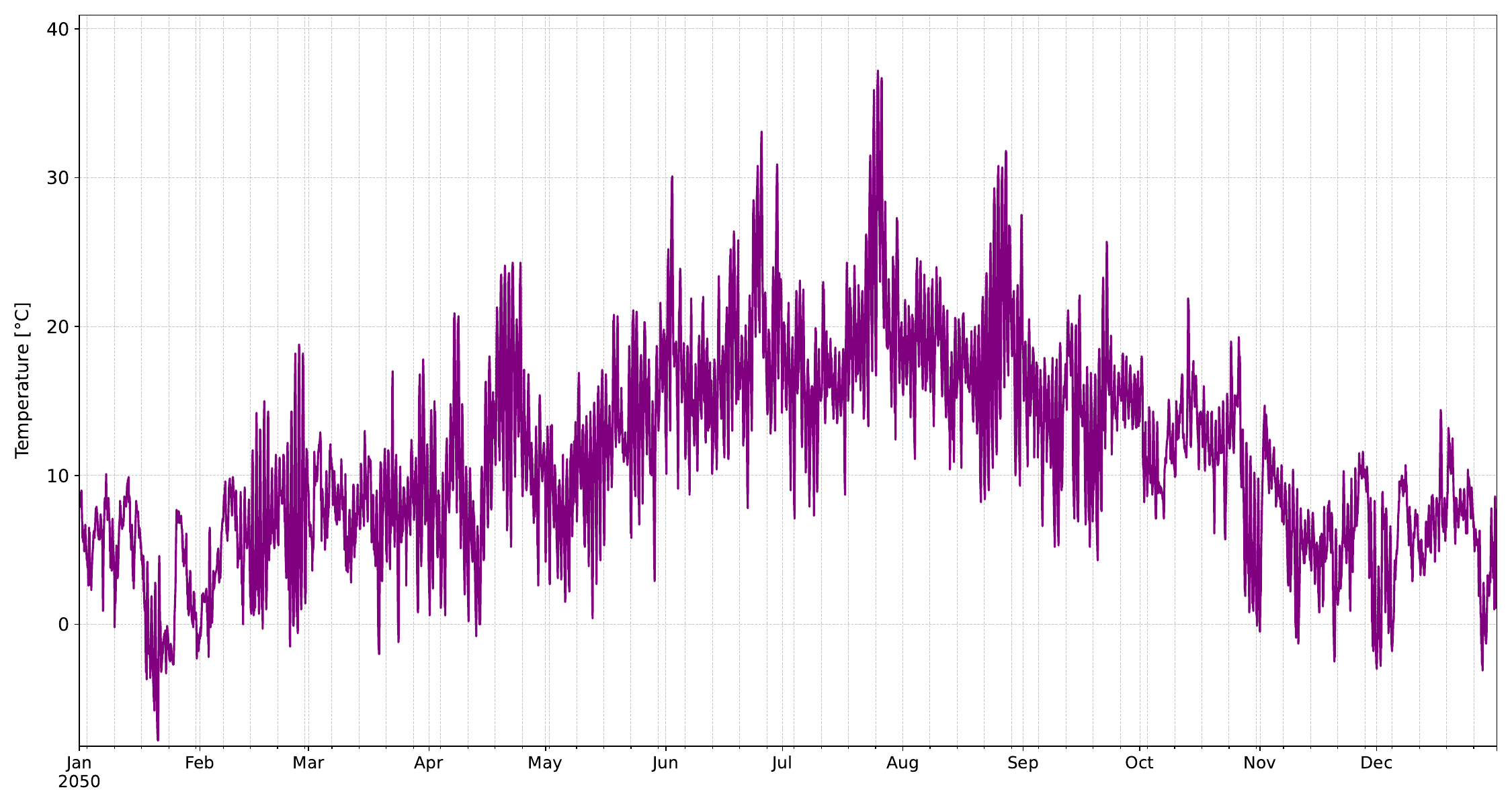}
    
    \subsubsection*{Figure S8. Ambient temperature}
    \phantomsection
    \label{temperature_fig}
\end{minipage}

\noindent
\begin{minipage}{\linewidth}
    \centering
    \includegraphics[width=0.9\linewidth]{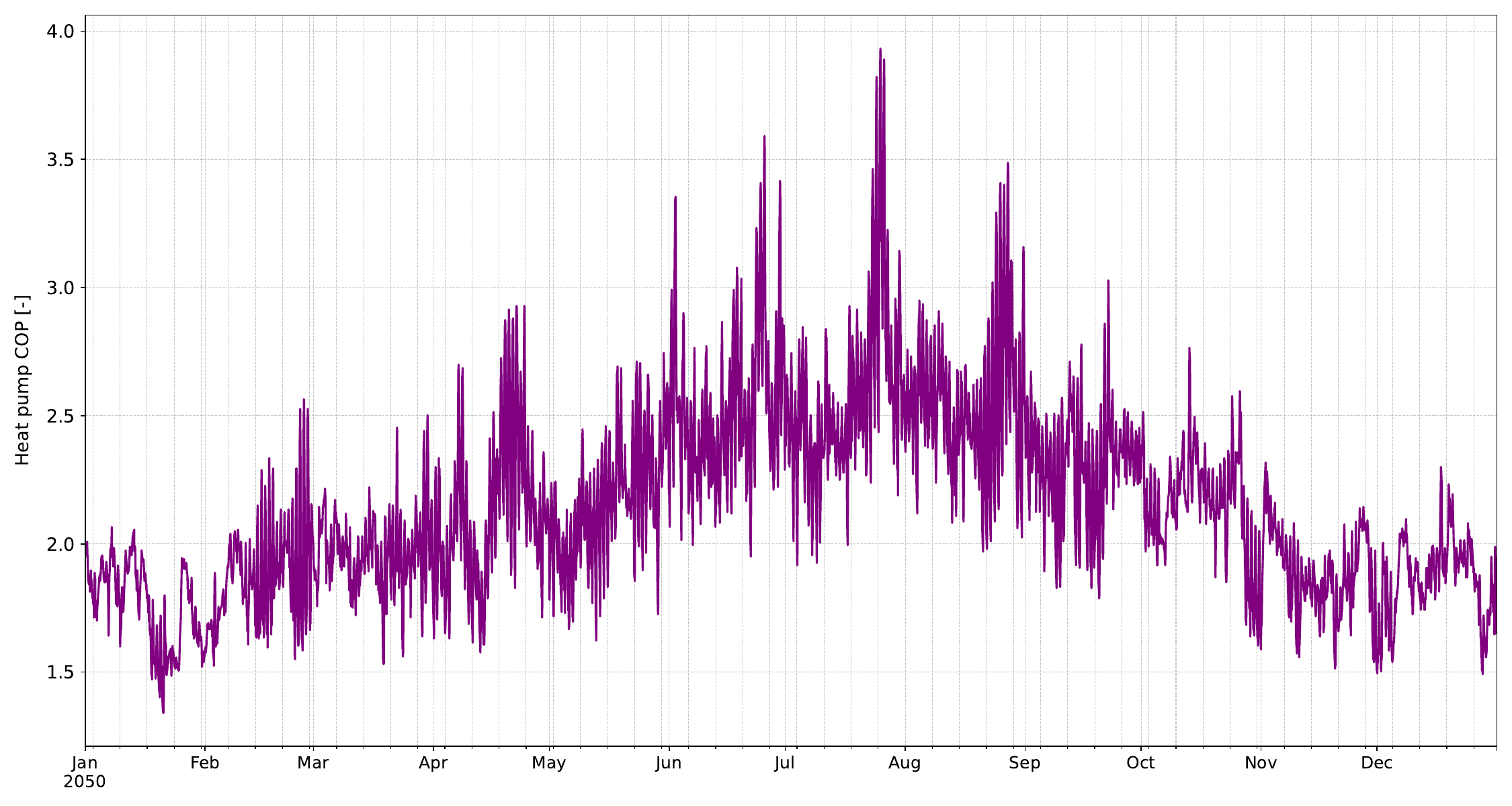}
    
    \subsubsection*{Figure S9. Air-source heat pump coefficient of performance}
    \phantomsection
    \label{heatpump_cop_fig}
\end{minipage}

\vspace{4em}

\noindent
\begin{minipage}{\linewidth}
    \centering
    \includegraphics[width=0.9\linewidth]{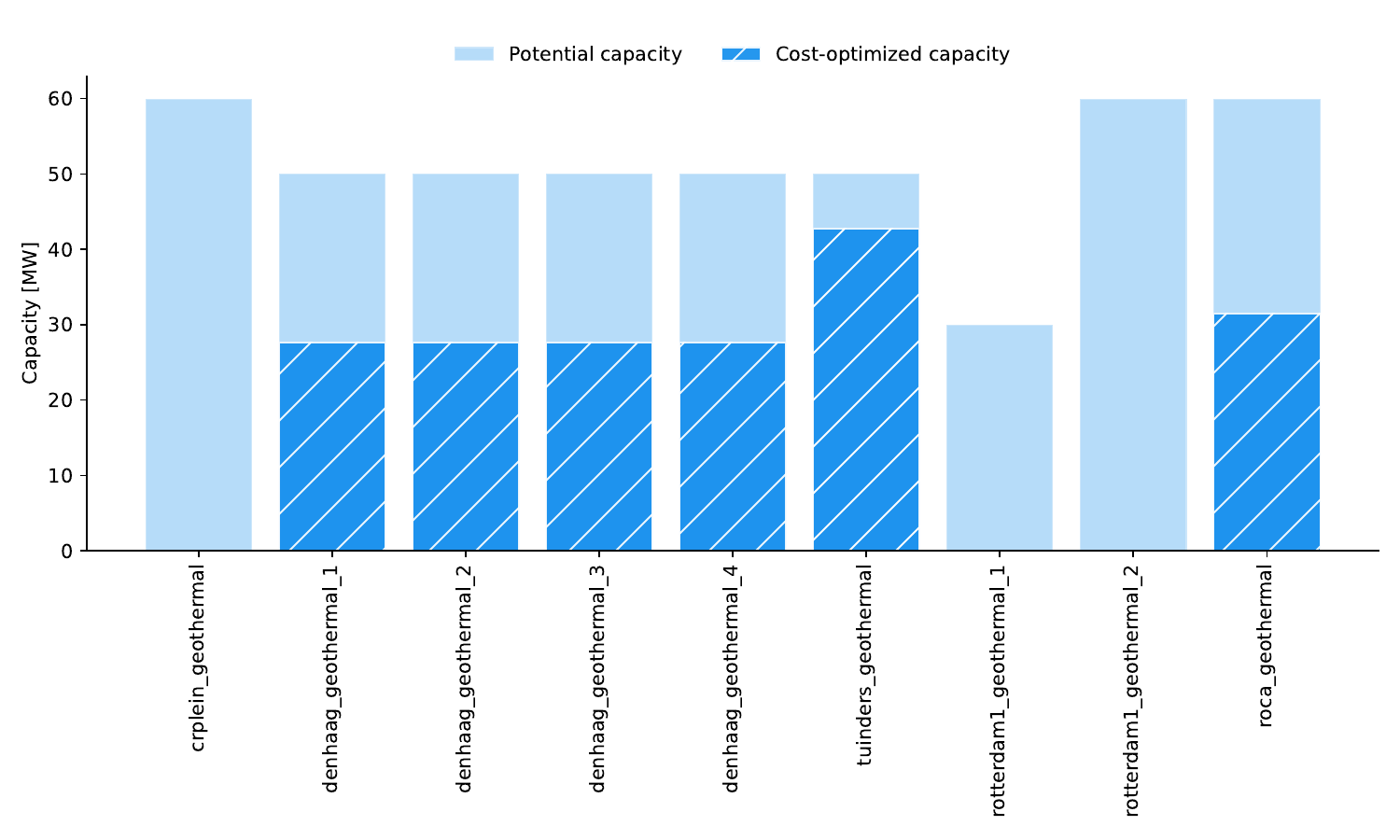}
    
    \subsubsection*{Figure S10. Spatial deployment of geothermal heat sources in the least-cost solution}
    \phantomsection
    \label{geothermal_deployment_lc}
\end{minipage}

\vspace{5em}

\noindent
\begin{minipage}{\linewidth}
    \centering
    \includegraphics[width=0.99\linewidth]{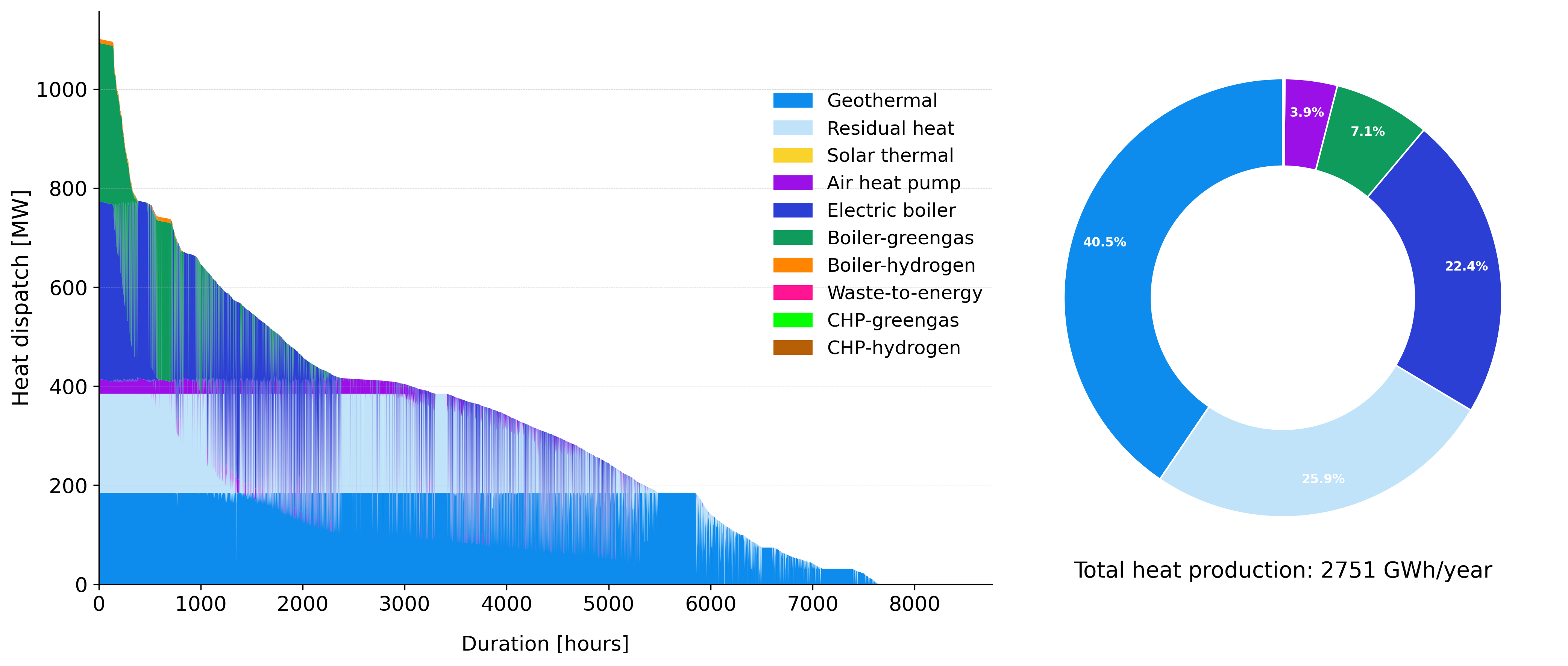}
    
    \subsubsection*{Figure S11. Heat dispatch duration curve and annual share of heat supply per technology category in the least-cost solution}
    \phantomsection
    \label{heat_dispatch_duration_curve}
\end{minipage}

\noindent
\begin{minipage}{\linewidth}
    \centering
    \includegraphics[width=0.9\linewidth]{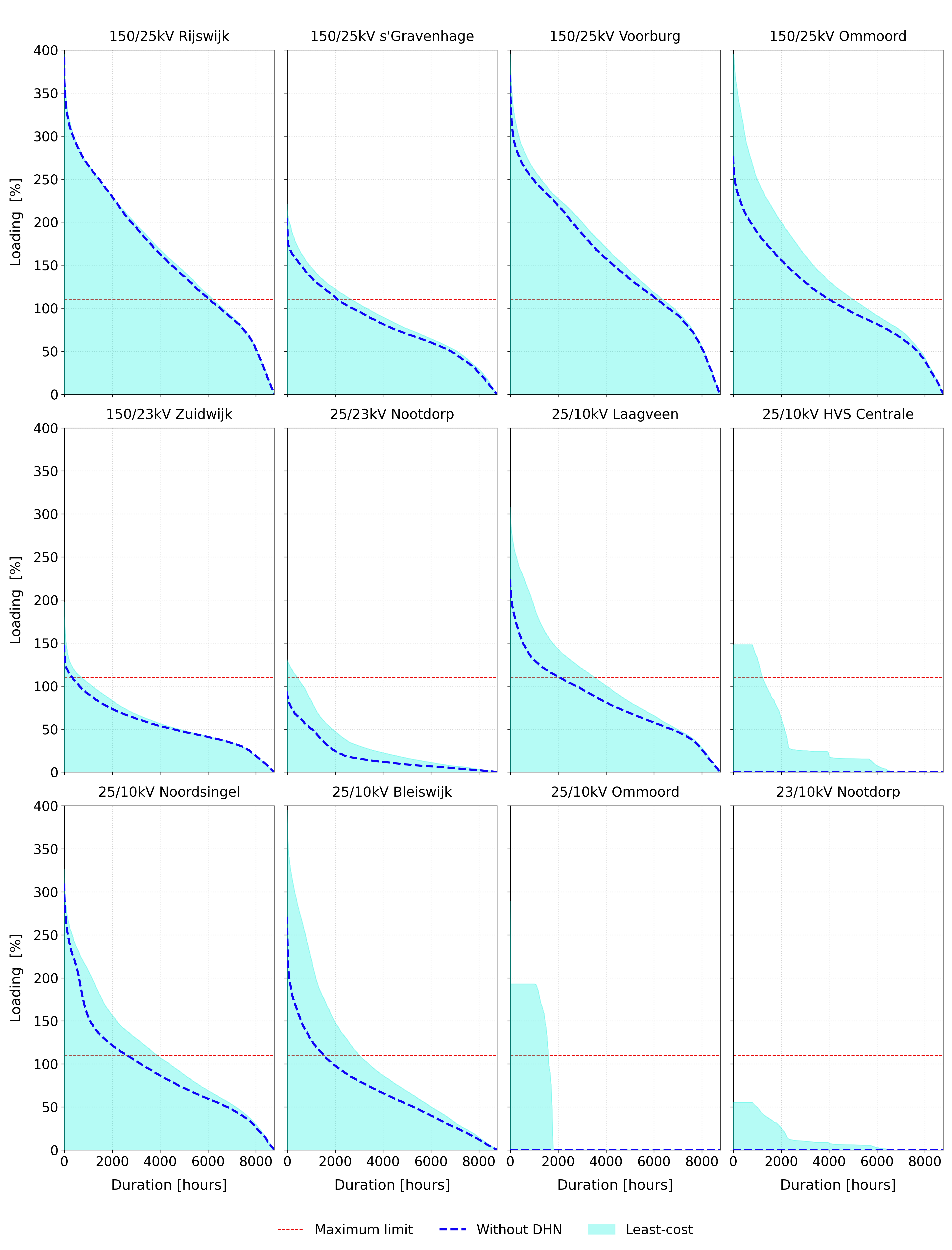}
    
    \subsubsection*{Figure S12. Loading duration curves of all transformers in the electricity network}
    \phantomsection
    \label{lc_loading_duration_all_trafos}
\end{minipage}

\noindent
\begin{minipage}{\linewidth}
    \centering
    \includegraphics[width=0.9\linewidth]{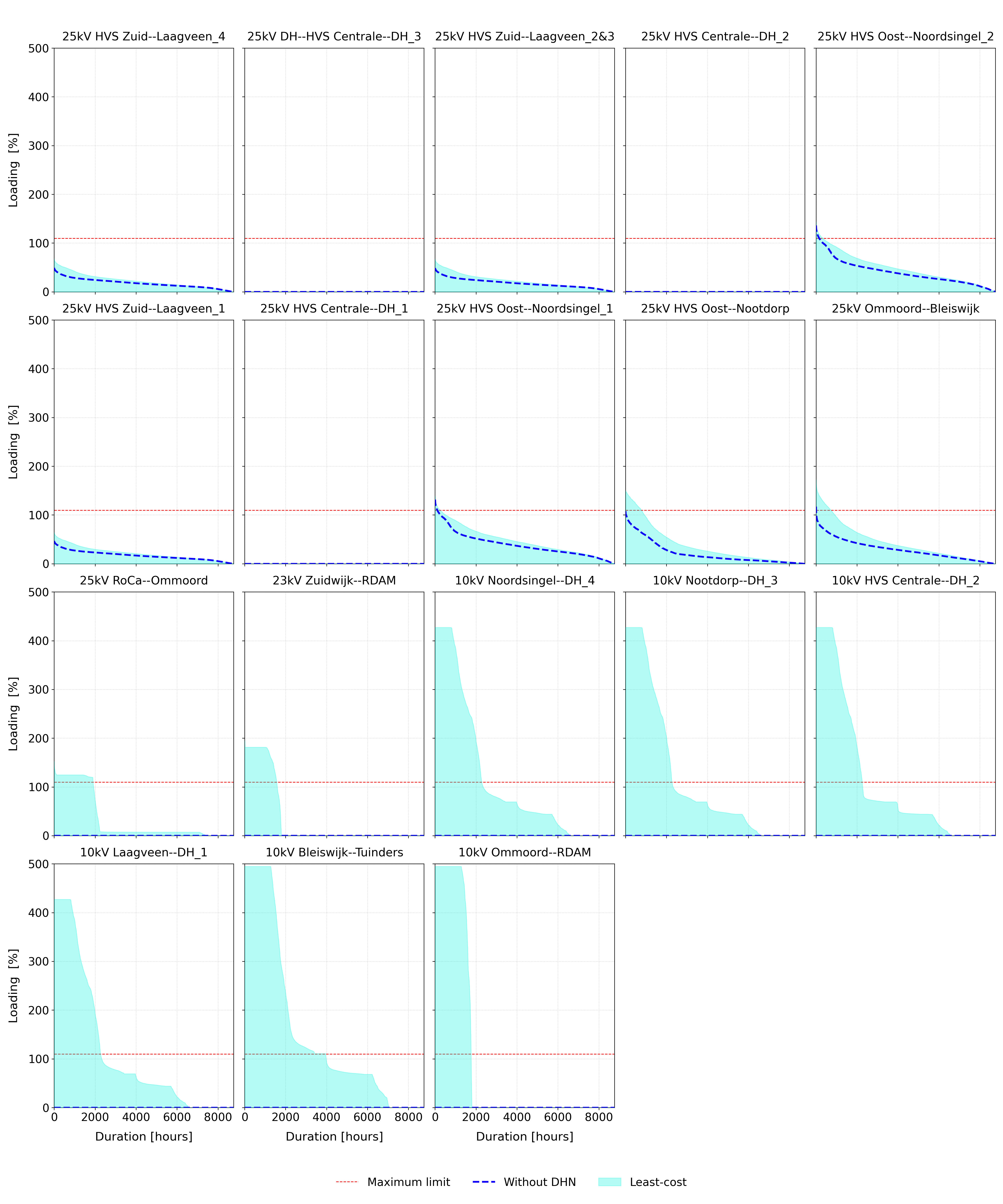}
    
    \subsubsection*{Figure S13. Loading duration curves of all lines in the electricity network}
    \phantomsection
    \label{lc_loading_duration_all_lines}
\end{minipage}

\noindent
\begin{minipage}{\linewidth}
    \centering
    \includegraphics[width=0.95\linewidth]{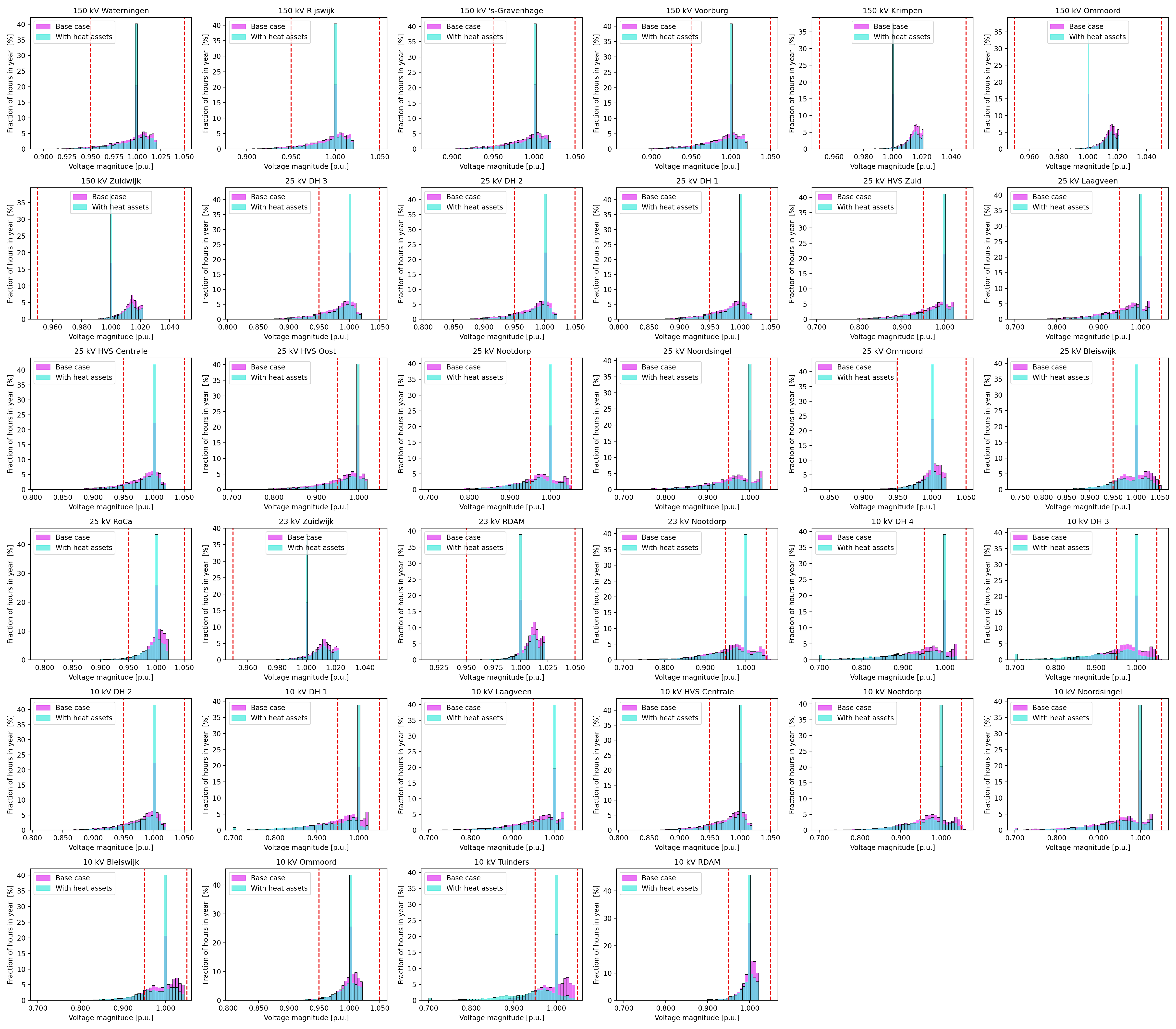}
    
    \subsubsection*{Figure S14. Voltage magnitude histograms of all buses in the electricity network}
    \phantomsection
    \label{lc_bus_voltage_histogram}
\end{minipage}

\noindent
\begin{minipage}{\linewidth}
    \centering
    \includegraphics[width=0.99\linewidth]{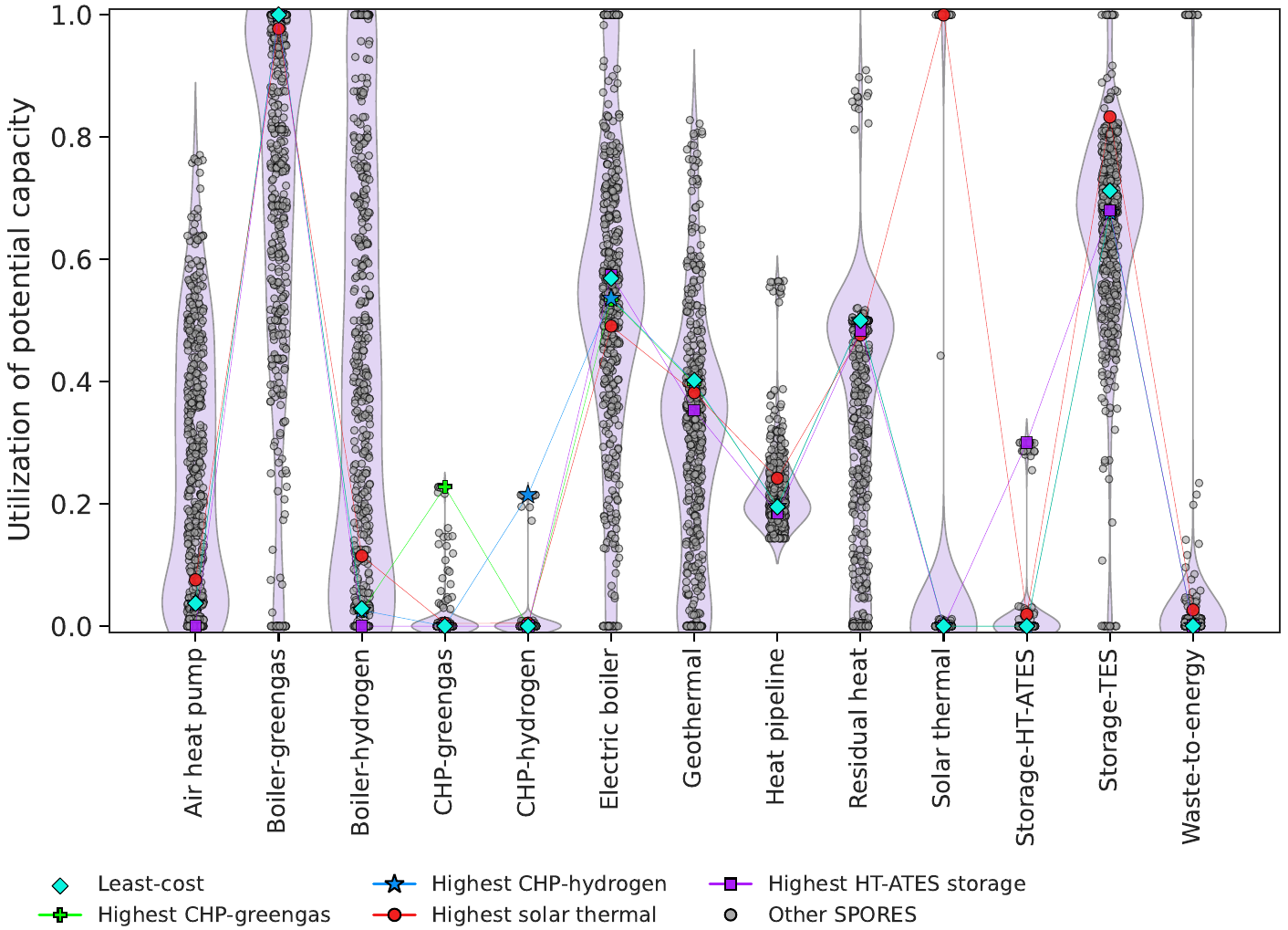}
    
    \subsubsection*{Figure S15. Trade-offs associated with maximizing technologies with low-utilization frequencies across SPORES}
    \phantomsection
    \label{low_deployment_frequency_techs}
\end{minipage}

\noindent
\begin{minipage}{\linewidth}
    \centering
    \includegraphics[width=0.99\linewidth]{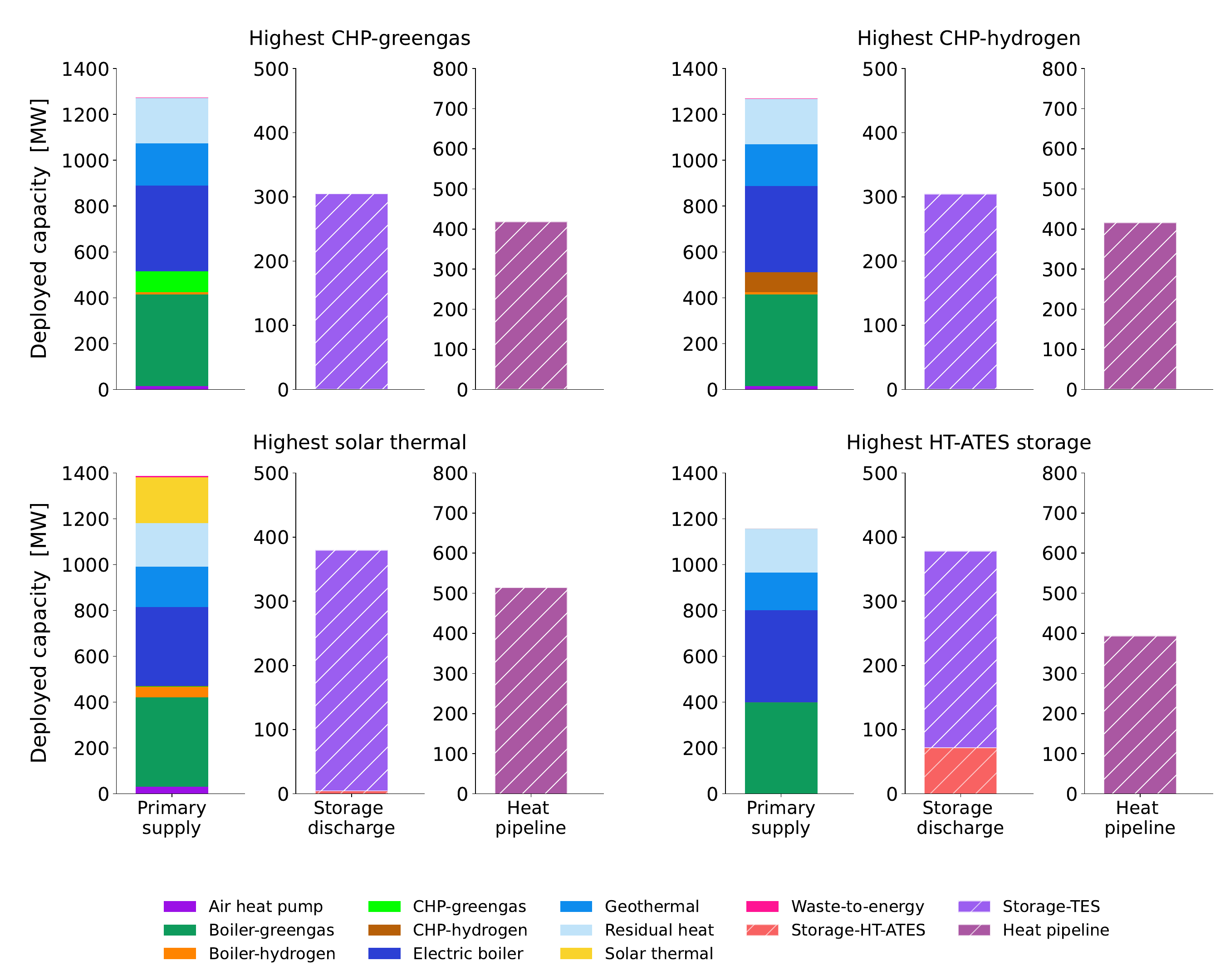}
    
    \subsubsection*{Figure S16. Technology configurations under maximum build-out of technologies with low-utilization frequencies across SPORES}
    \phantomsection
    \label{low_deploy_spores_technology_mix}
\end{minipage}

\noindent
\begin{minipage}{\linewidth}
    \centering
    \includegraphics[width=0.99\linewidth]{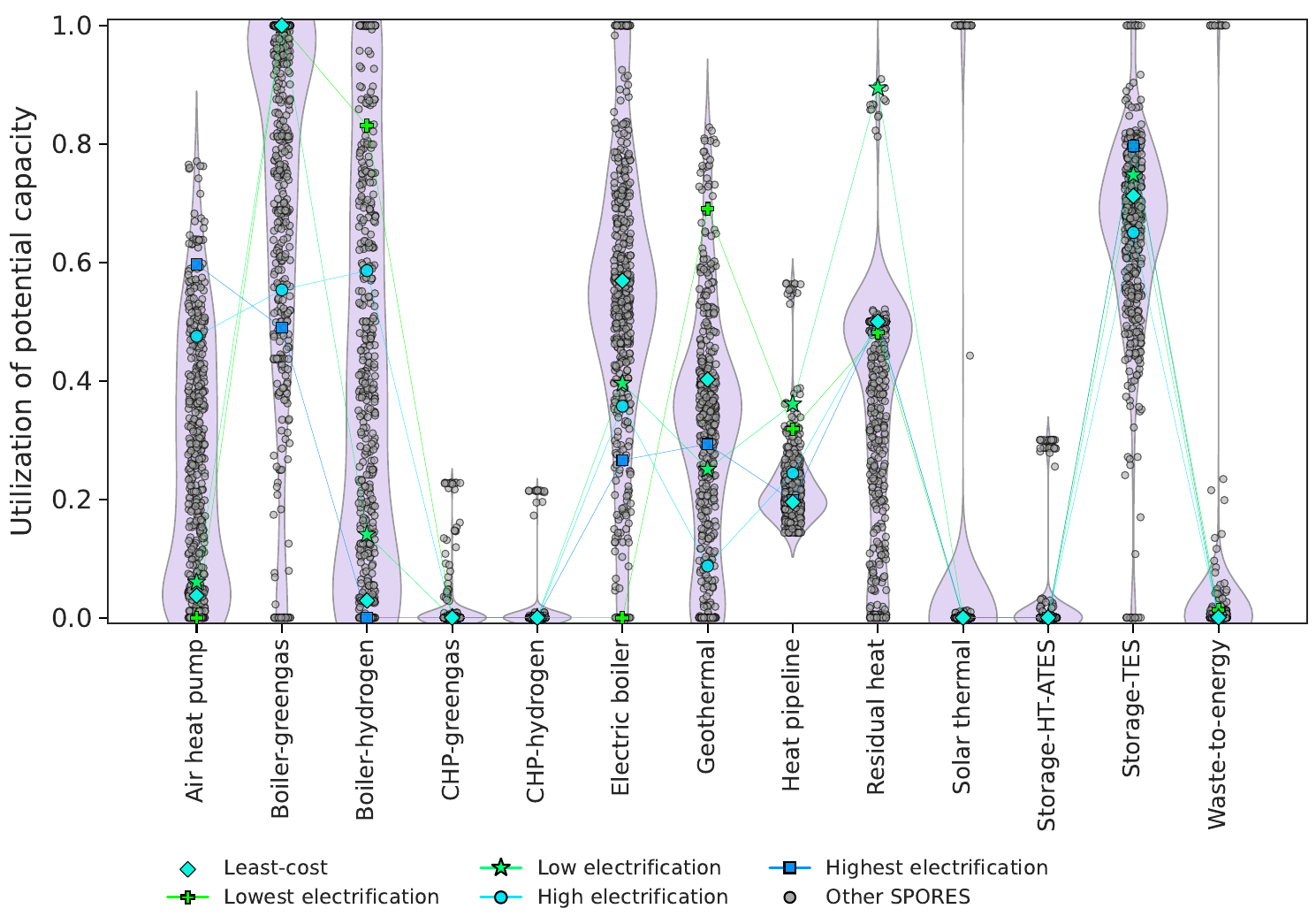}
    
    \subsubsection*{Figure S17. Trade-offs across highlighted SPORES with varying degrees of heat electrification}
    \phantomsection
    \label{electrification_techs_frequency}
\end{minipage}

\noindent
\begin{minipage}{\linewidth}
    \centering
    \includegraphics[width=0.99\linewidth]{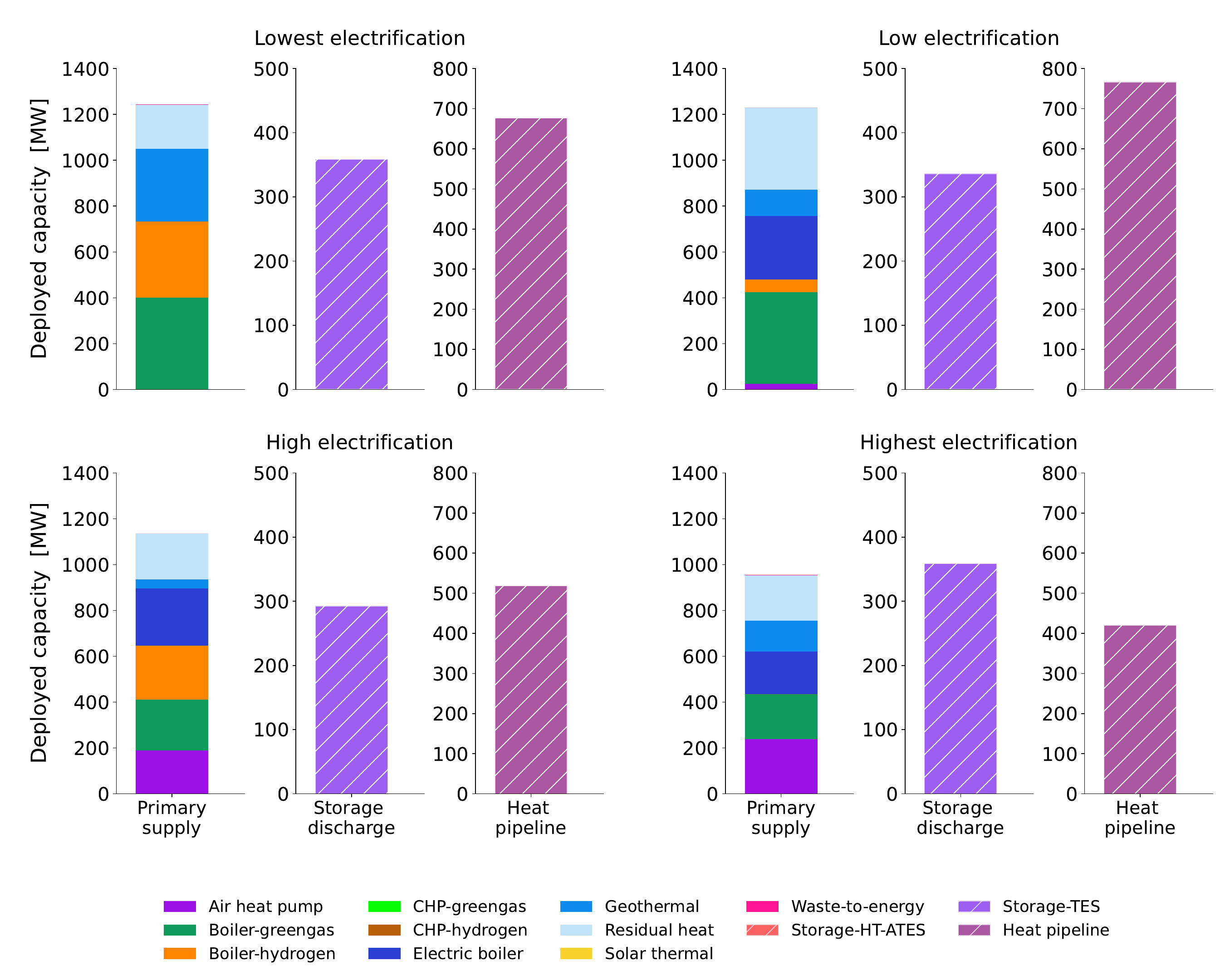}
    
    \subsubsection*{Figure S18. Technology configurations for highlighted SPORES with varying degrees of heat electrification}
    \phantomsection
    \label{electrification_spores_tech_mix}
\end{minipage}

\noindent
\begin{minipage}{\linewidth}
    \centering
    \includegraphics[width=0.9\linewidth]{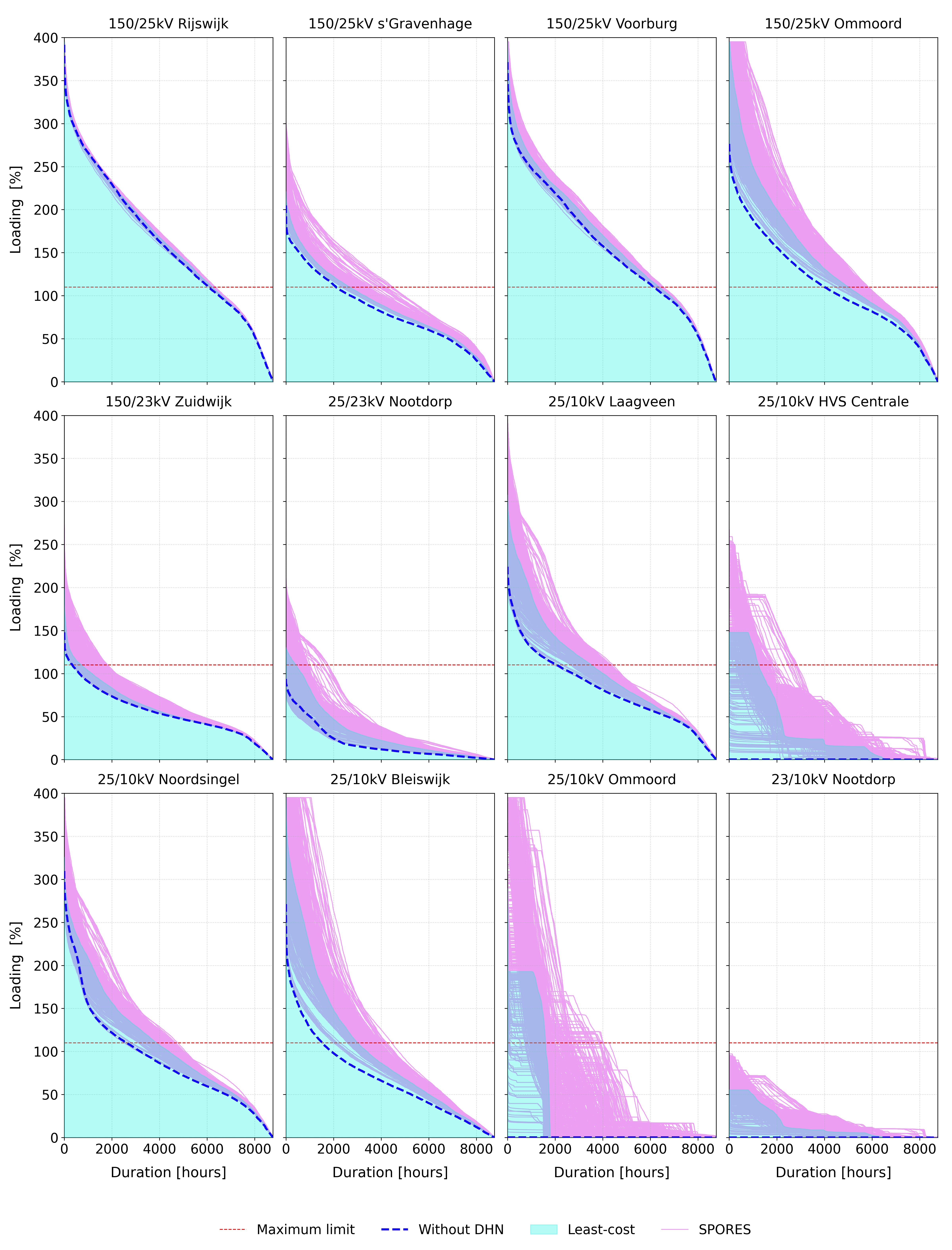}
    
    \subsubsection*{Figure S19. Range of transformer loading duration across all SPORES}
    \phantomsection
    \label{range_trafo_loading_all_spores}
\end{minipage}

\noindent
\begin{minipage}{\linewidth}
    \centering
    \includegraphics[width=0.9\linewidth]{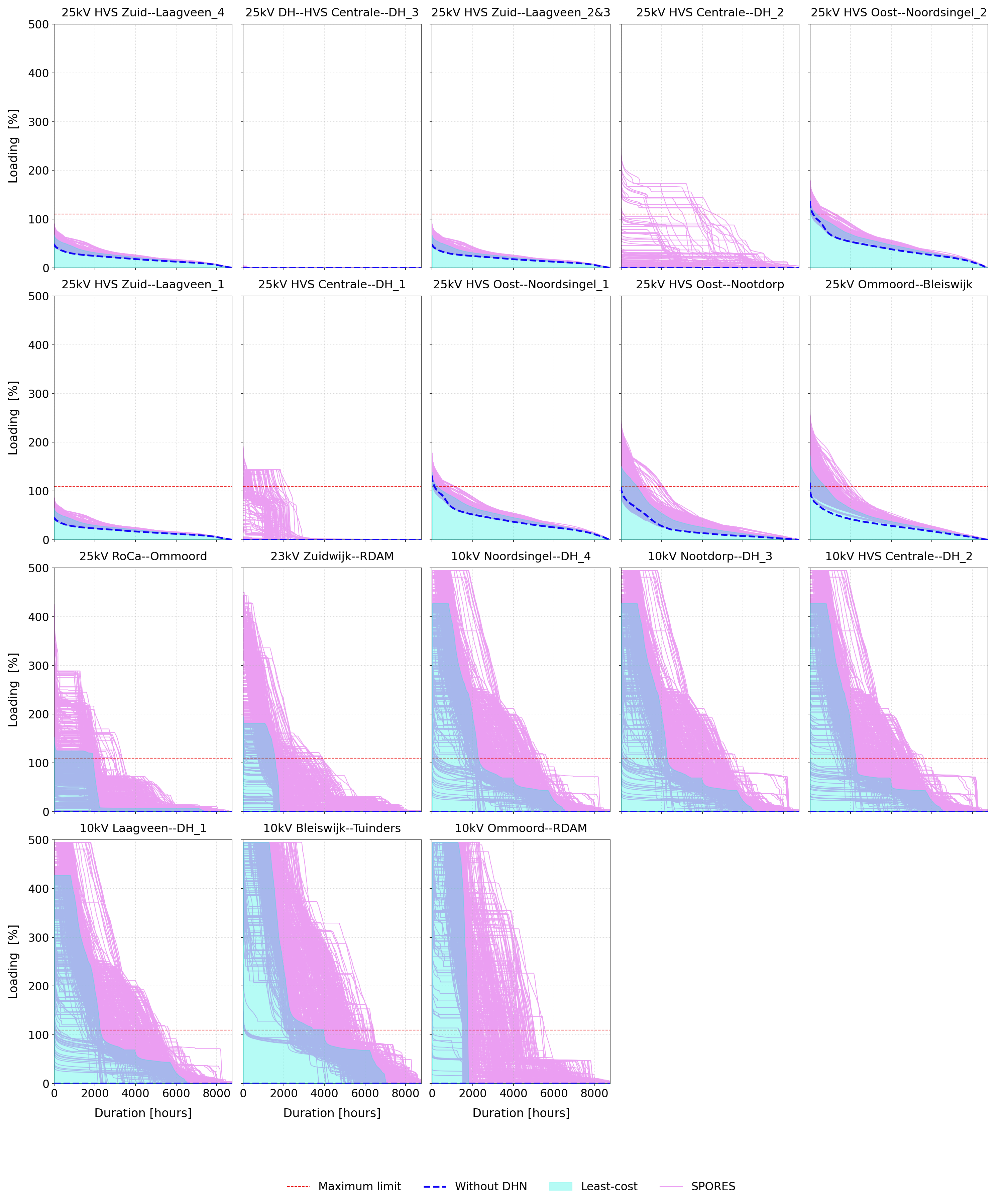}
    
    \subsubsection*{Figure S20. Range of line loading duration across all SPORES}
    \phantomsection
    \label{range_line_loading_all_spores}
\end{minipage}

\noindent
\begin{minipage}{\linewidth}
    \centering
    \includegraphics[width=0.9\linewidth]{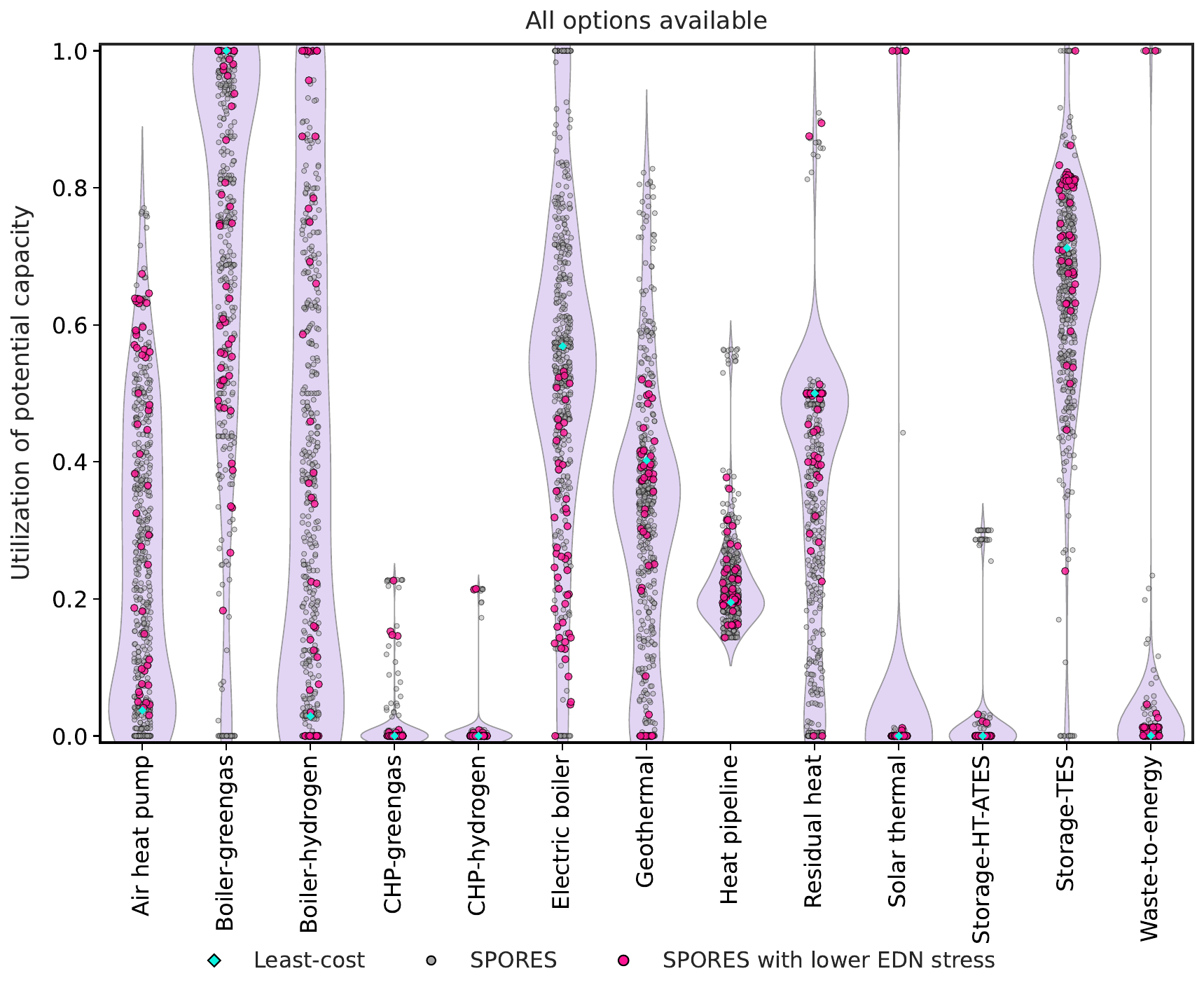}
    
    \subsubsection*{Figure S21. Frequency distribution of potential capacity utilization in the near-optimal decision space with lower grid loading SPORES highlighted: default scenario with 10\% cost slack}
    \phantomsection
    \label{unconstrained_maneuvering_space_with_grid_friendly_spores}
\end{minipage}

\noindent
\begin{minipage}{\linewidth}
    \centering
    \includegraphics[width=0.97\linewidth]{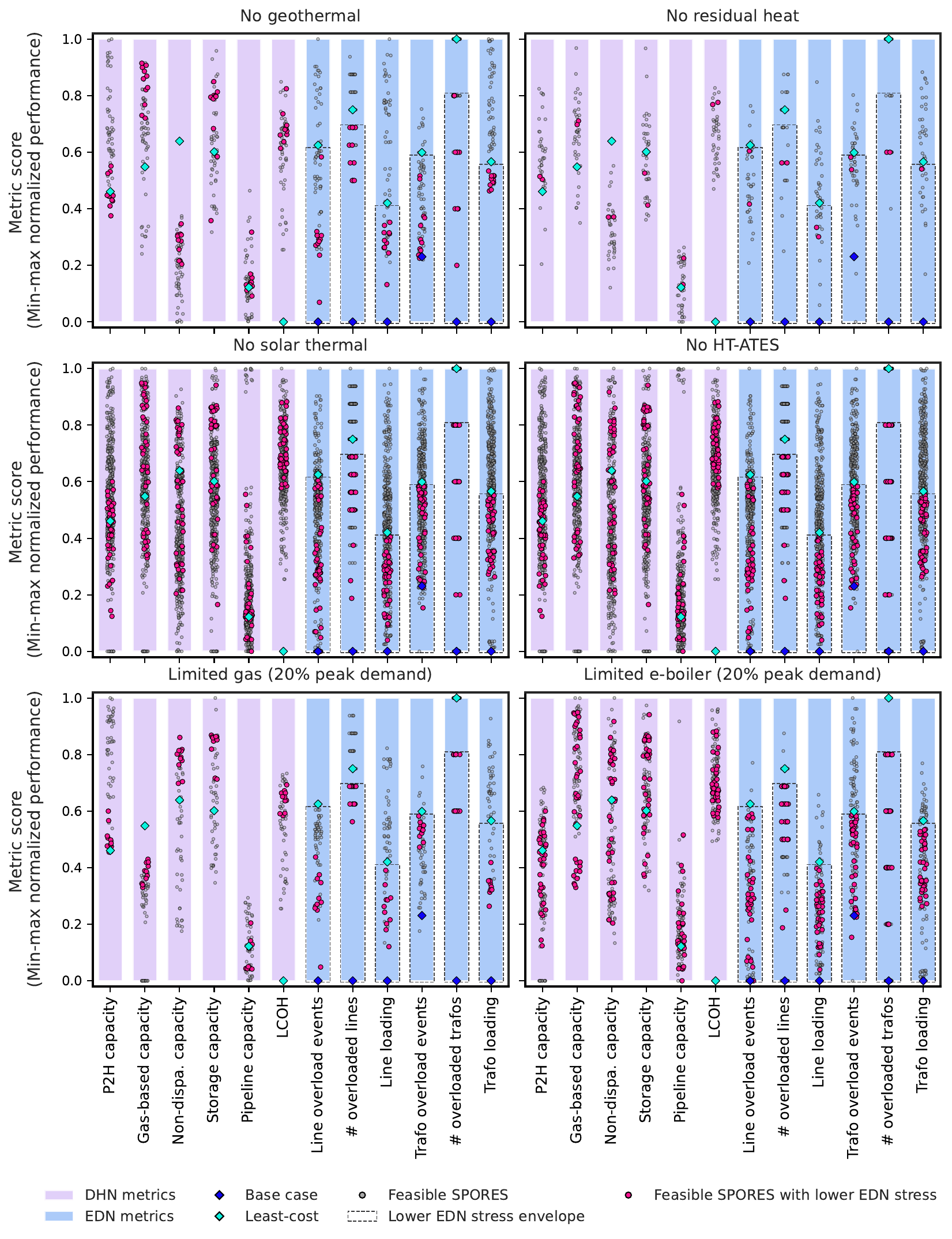}
    
    \subsubsection*{Figure S22. Integrated decision space under technology deployment constraints: default scenario with 10\% cost slack}
    \phantomsection
    \label{constrained_integrated_decision_space}
\end{minipage}

\section*{Sensitivity analysis results}\phantomsection\label{sensitivity_analysis_results}



\noindent
\begin{minipage}{\linewidth}
    \centering
    \includegraphics[width=0.65\linewidth]{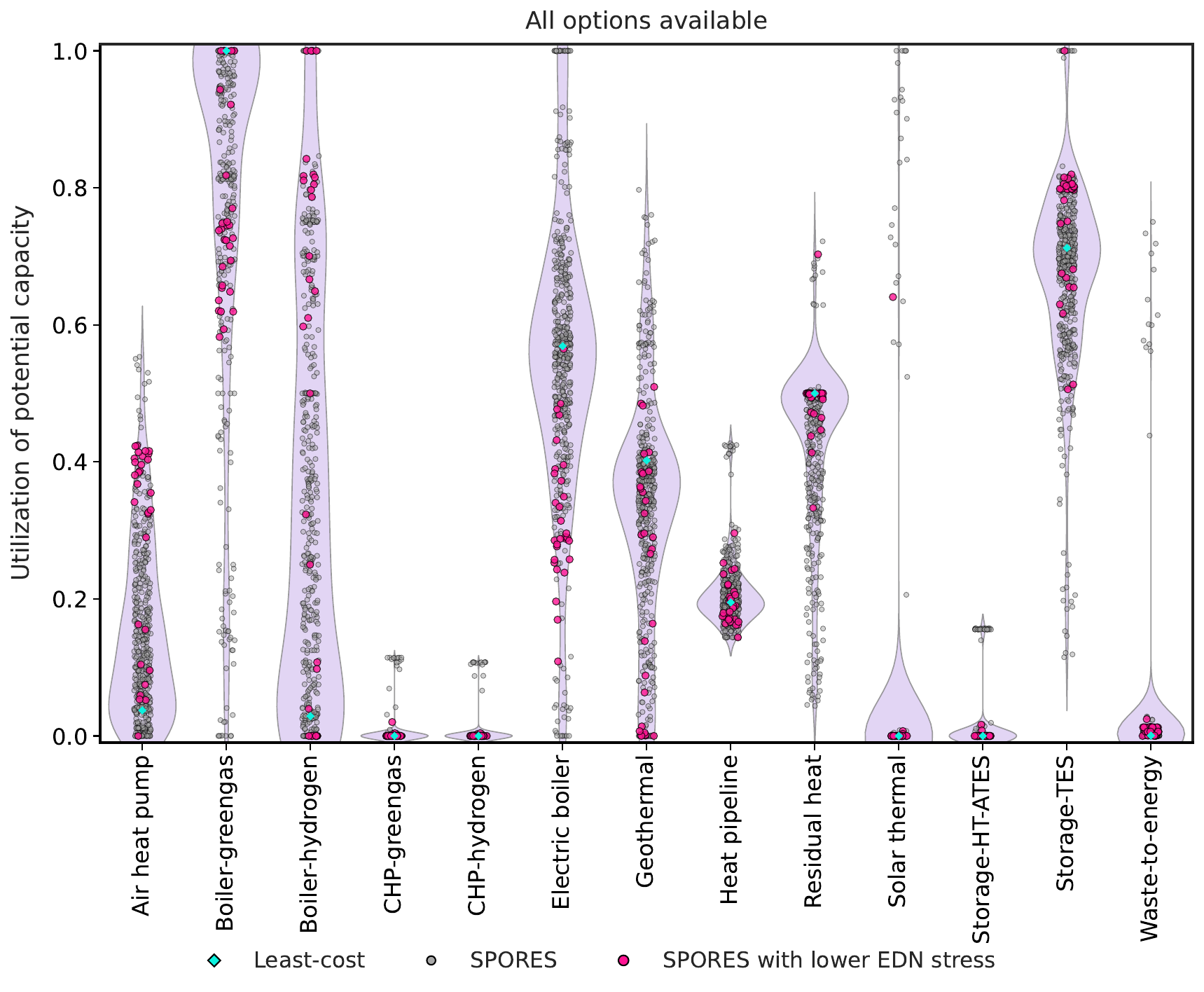}
    
    \subsubsection*{Figure S23. Frequency distribution of potential capacity utilization in the near-optimal decision space with lower grid loading SPORES highlighted: 5\% cost slack scenario}
    \phantomsection
    \label{05_unconstrained_maneuvering_space}
\end{minipage}

\vspace{1em}

\noindent
\begin{minipage}{\linewidth}
    \centering
    \includegraphics[width=0.65\linewidth]{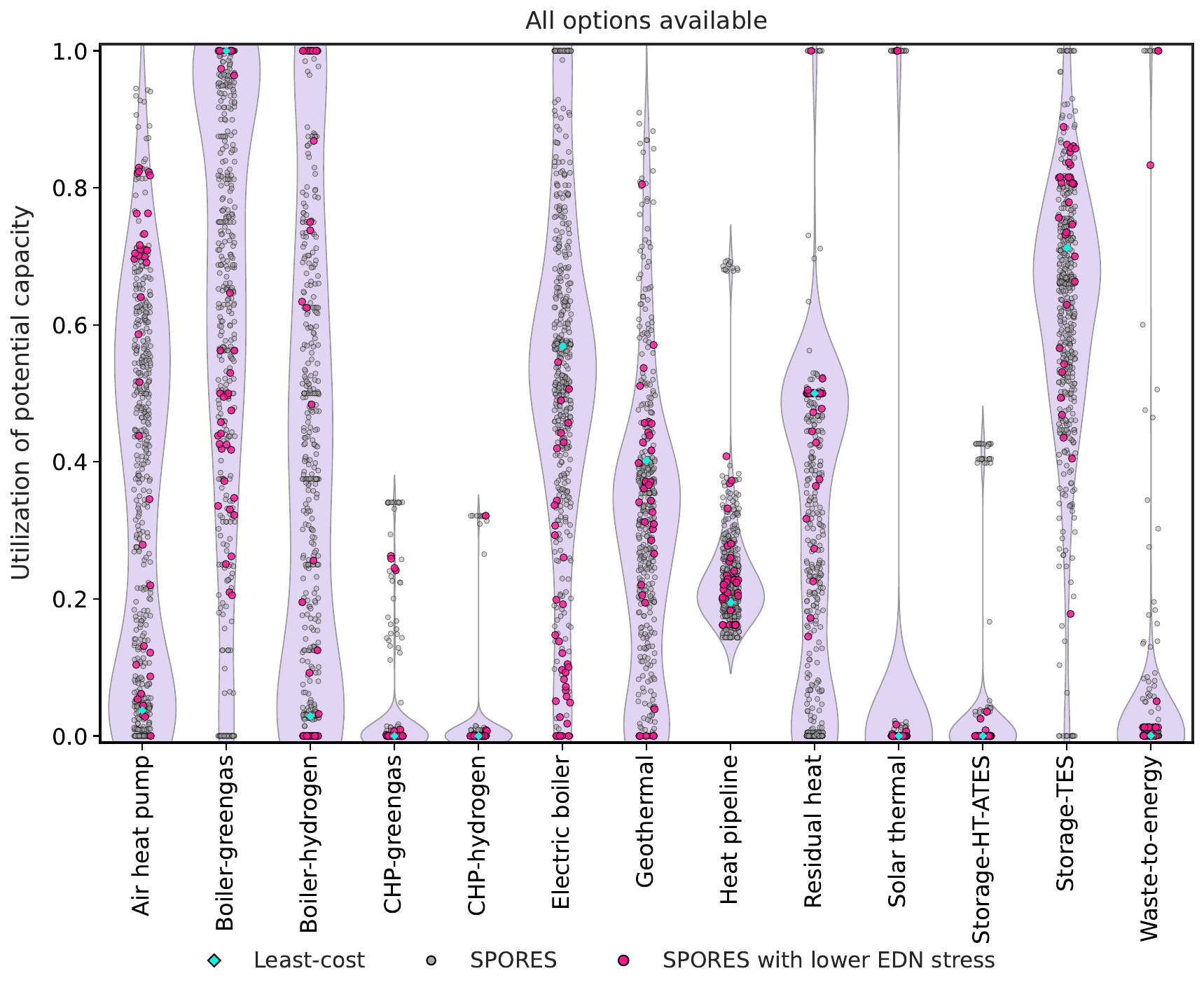}
    
    \subsubsection*{Figure S24. Frequency distribution of potential capacity utilization in the near-optimal decision space with lower grid loading SPORES highlighted: 15\% cost slack scenario}
    \phantomsection
    \label{15_unconstrained_maneuvering_space}
\end{minipage}

\noindent
\begin{minipage}{\linewidth}
    \centering
    \includegraphics[width=0.65\linewidth]{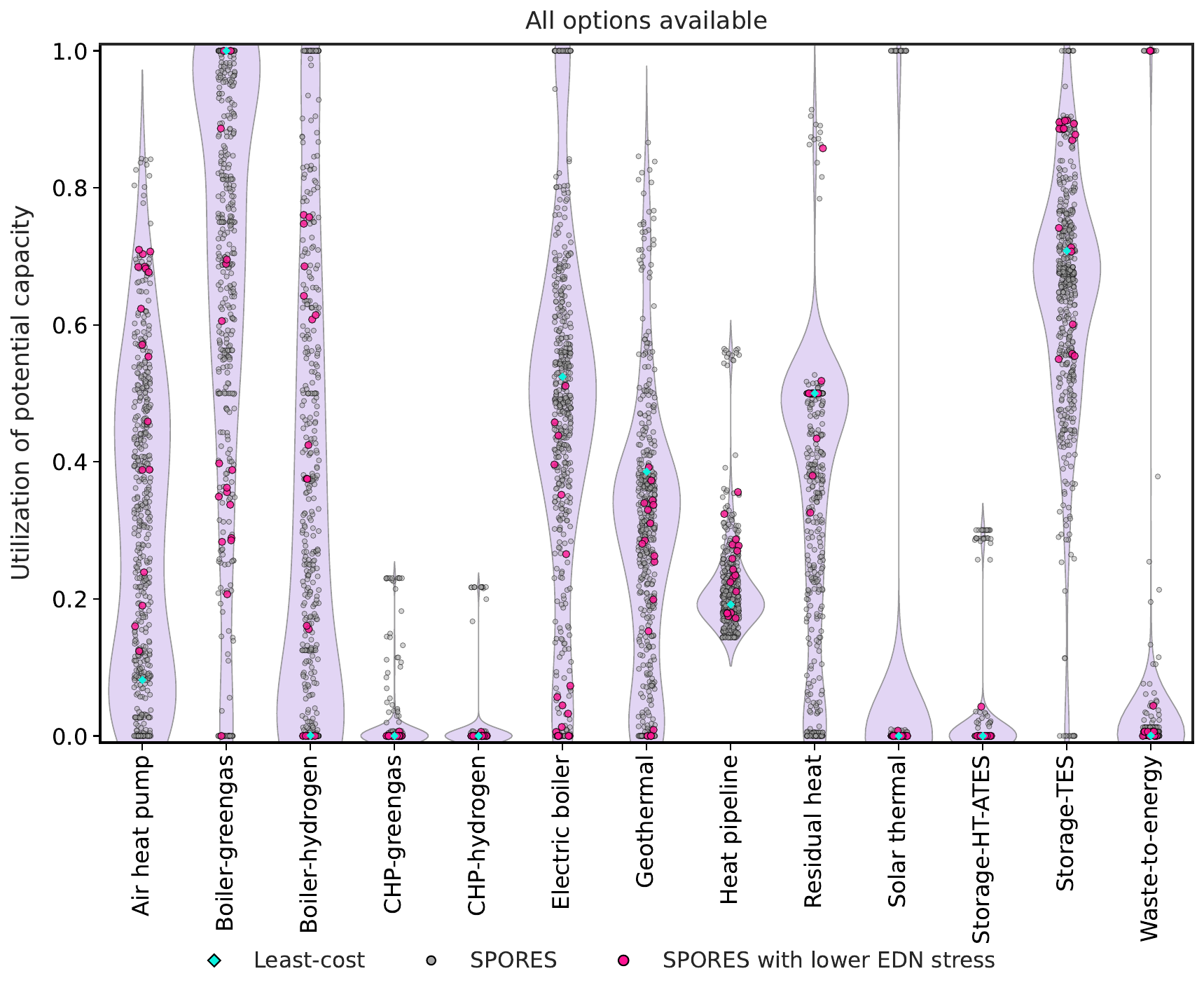}
    
    \subsubsection*{Figure S25. Frequency distribution of potential capacity utilization in the near-optimal decision space with lower grid loading SPORES highlighted: warm weather year scenario}
    \phantomsection
    \label{warm_weather_unconstrained_maneuvering_space}
\end{minipage}

\vspace{1em}

\noindent
\begin{minipage}{\linewidth}
    \centering
    \includegraphics[width=0.65\linewidth]{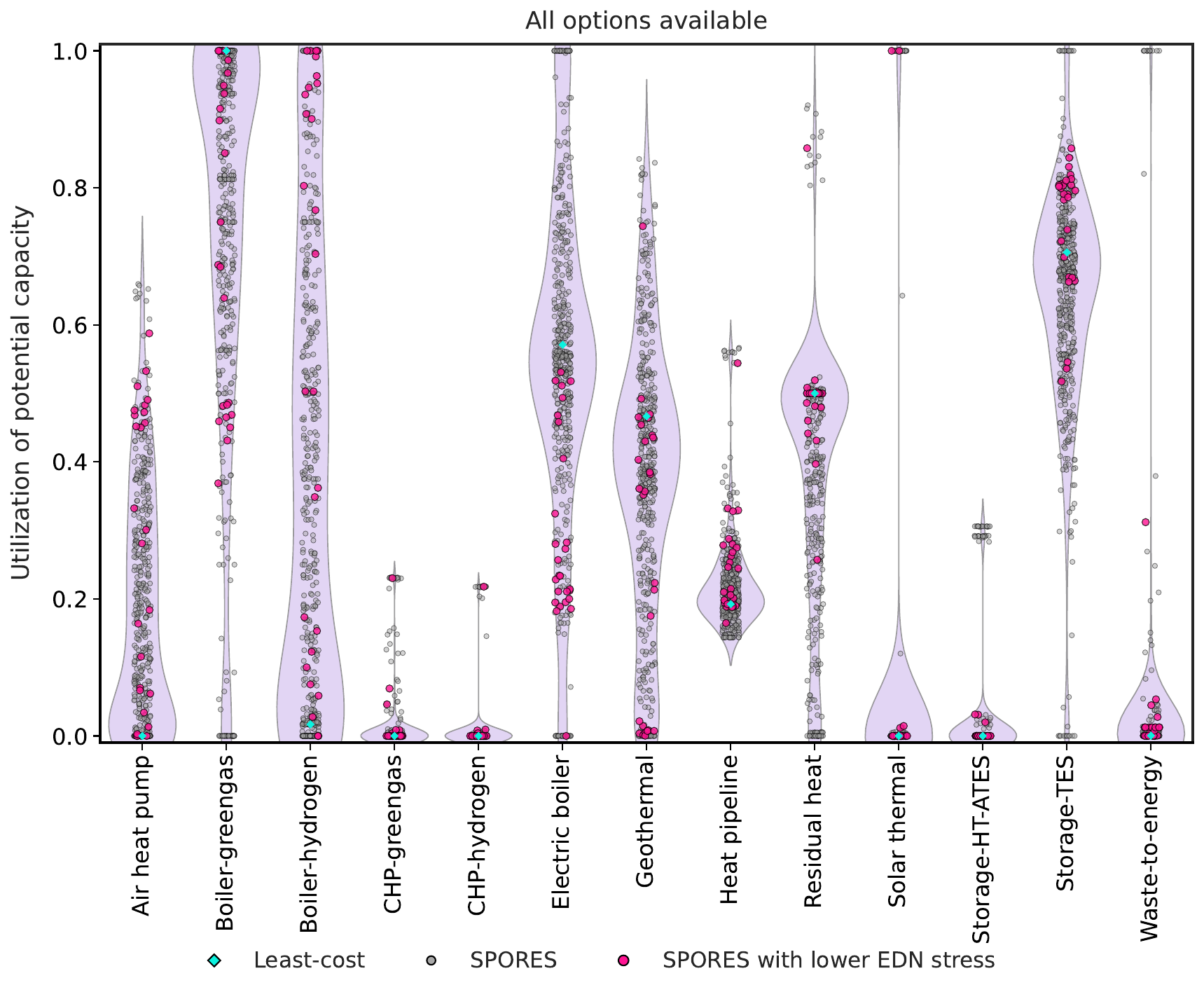}
    
    \subsubsection*{Figure S26. Frequency distribution of potential capacity utilization in the near-optimal decision space with lower grid loading SPORES highlighted: cold weather year scenario}
    \phantomsection
    \label{cold_weather_unconstrained_maneuvering_space}
\end{minipage}

\noindent
\begin{minipage}{\linewidth}
    \centering
    \includegraphics[width=0.65\linewidth]{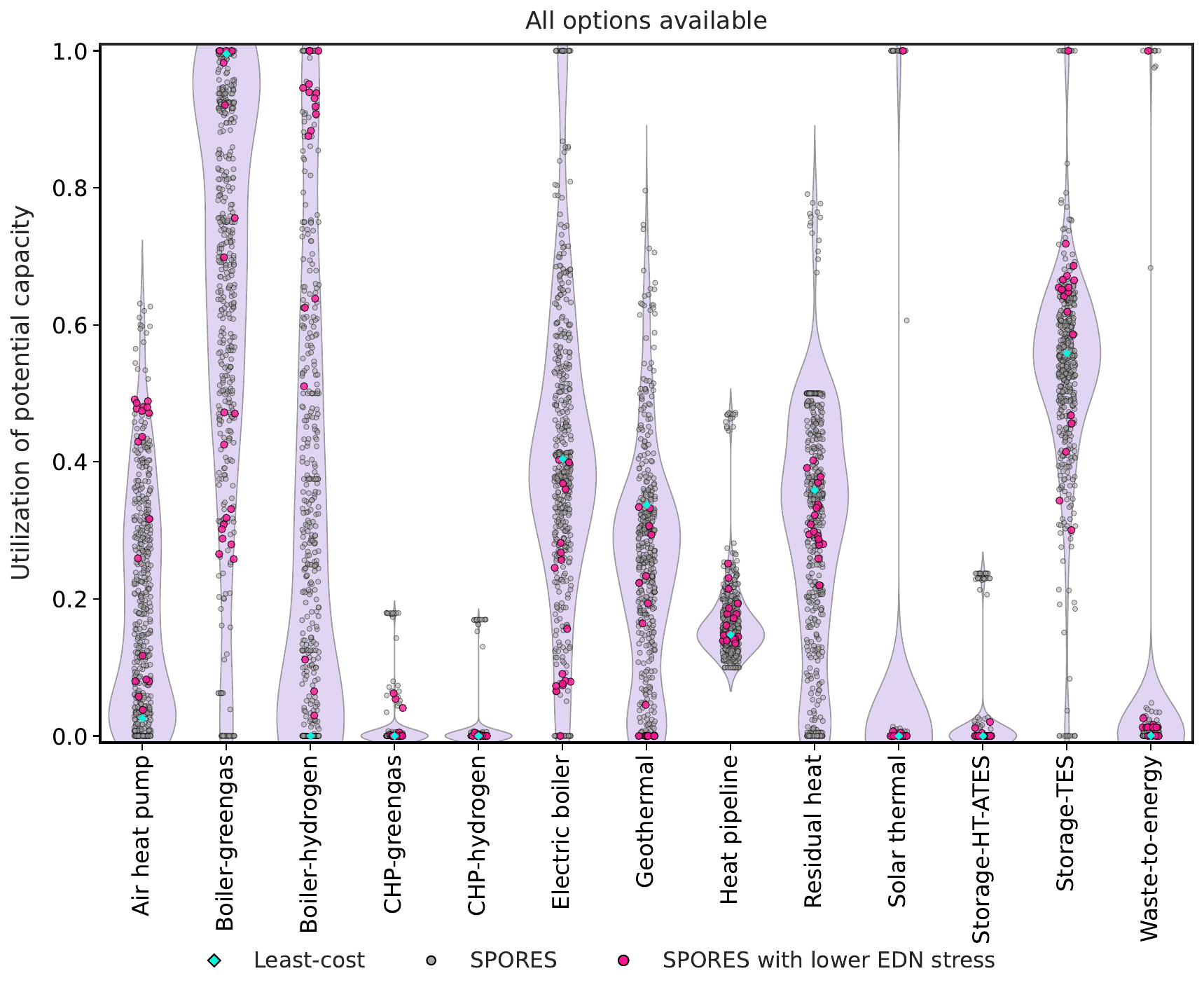}
    
    \subsubsection*{Figure S27. Frequency distribution of potential capacity utilization in the near-optimal decision space with lower grid loading SPORES highlighted: low heat demand scenario}
    \phantomsection
    \label{low_demand_unconstrained_maneuvering_space}
\end{minipage}

\vspace{1em}

\noindent
\begin{minipage}{\linewidth}
    \centering
    \includegraphics[width=0.65\linewidth]{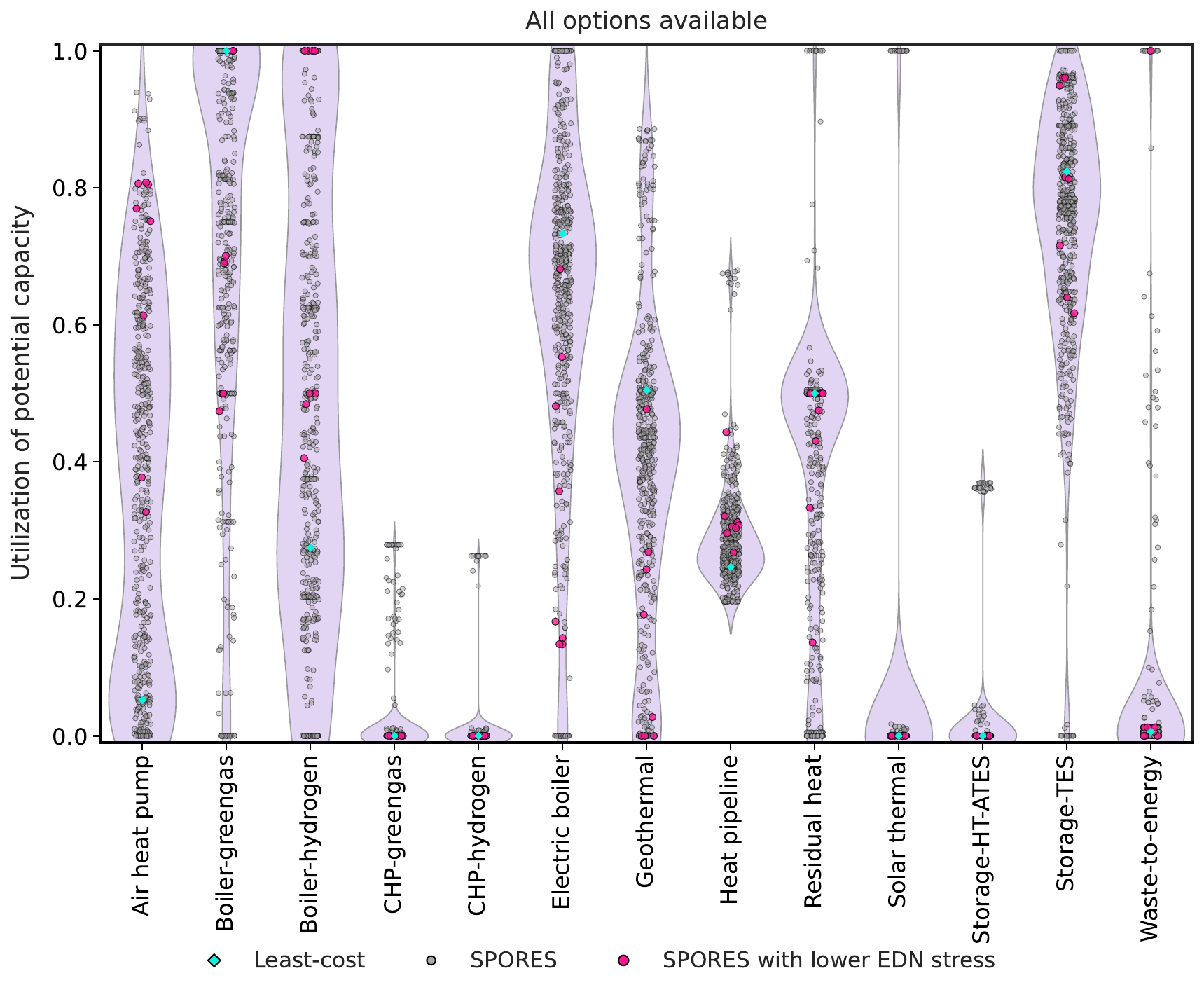}
    
    \subsubsection*{Figure S28. Frequency distribution of potential capacity utilization in the near-optimal decision space with lower grid loading SPORES highlighted: high heat demand scenario}
    \phantomsection
    \label{high_demand_unconstrained_maneuvering_space}
\end{minipage}

\noindent
\begin{minipage}{\linewidth}
    \centering
    \includegraphics[width=0.8\linewidth]{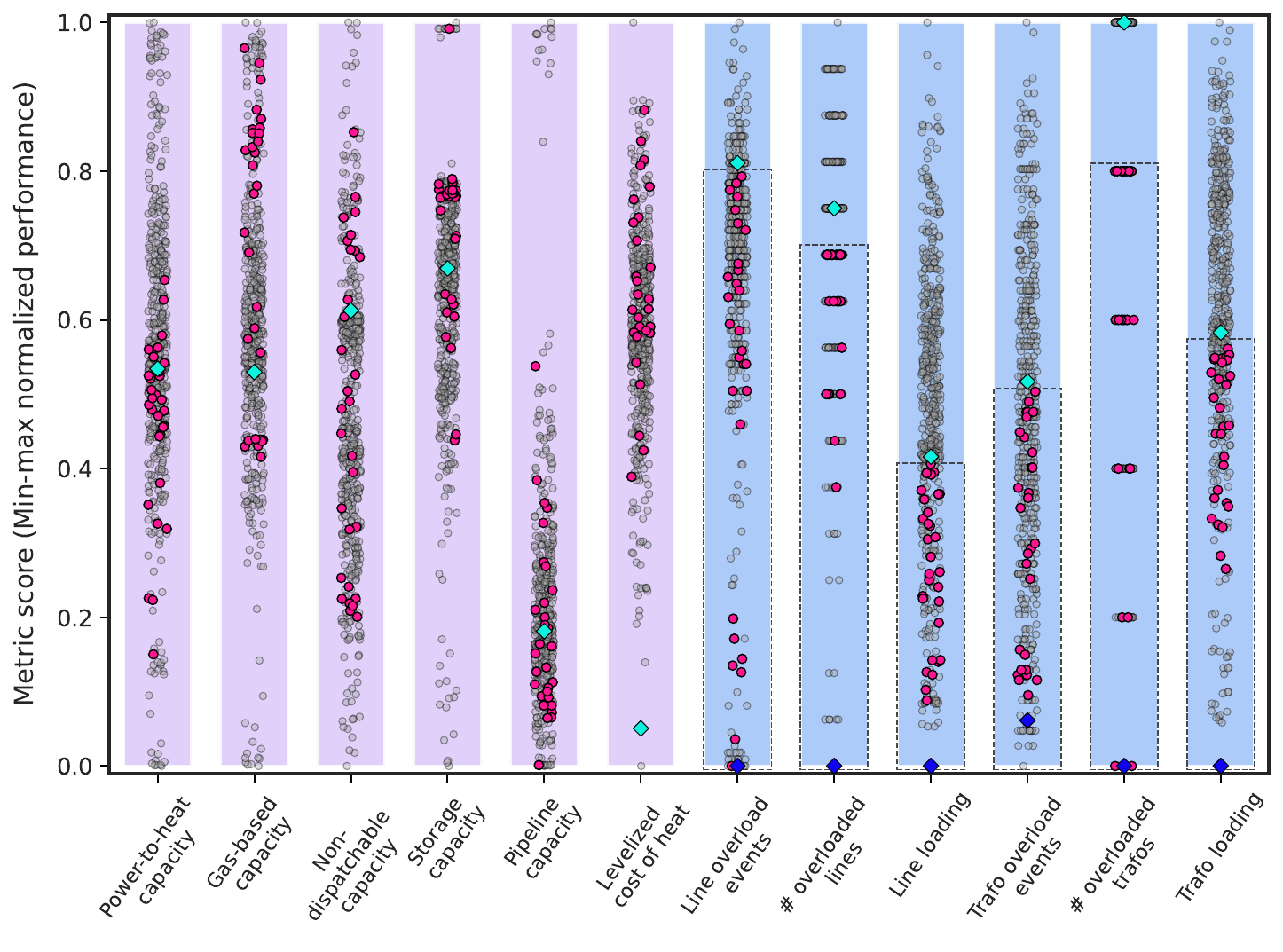}
    
    \subsubsection*{Figure S29. Integrated decision space: 5\% cost slack scenario}
    \phantomsection
    \label{05_integrated_decision_space}
\end{minipage}

\vspace{1em}

\noindent
\begin{minipage}{\linewidth}
    \centering
    \includegraphics[width=0.8\linewidth]{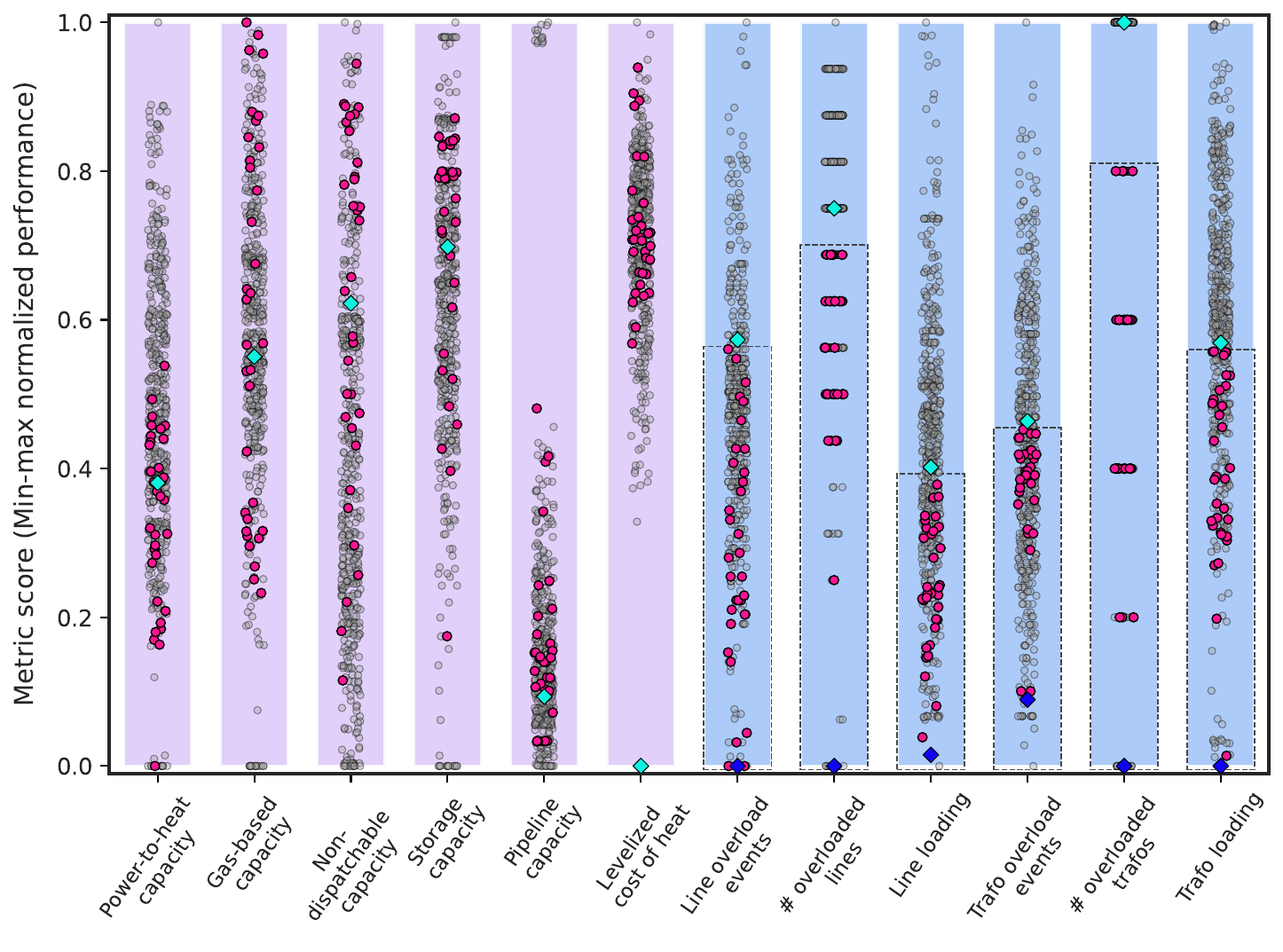}
    
    \subsubsection*{Figure S30. Integrated decision space: 15\% cost slack scenario}
    \phantomsection
    \label{15_integrated_decision_space}
\end{minipage}

\noindent
\begin{minipage}{\linewidth}
    \centering
    \includegraphics[width=0.8\linewidth]{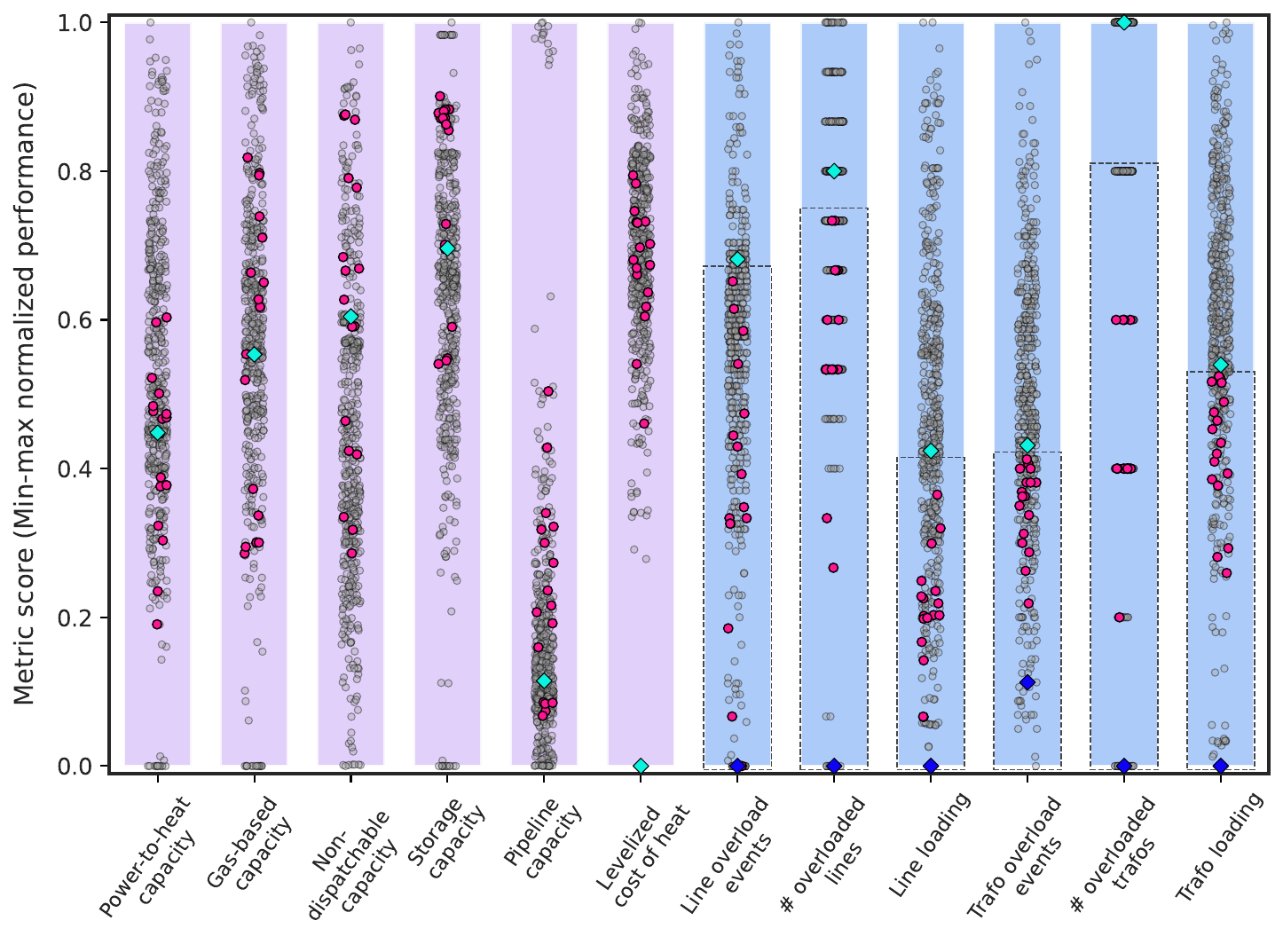}
    
    \subsubsection*{Figure S31. Integrated decision space: warm weather year scenario}
    \phantomsection
    \label{warm_weather_integrated_decision_space}
\end{minipage}

\vspace{1em}

\noindent
\begin{minipage}{\linewidth}
    \centering
    \includegraphics[width=0.8\linewidth]{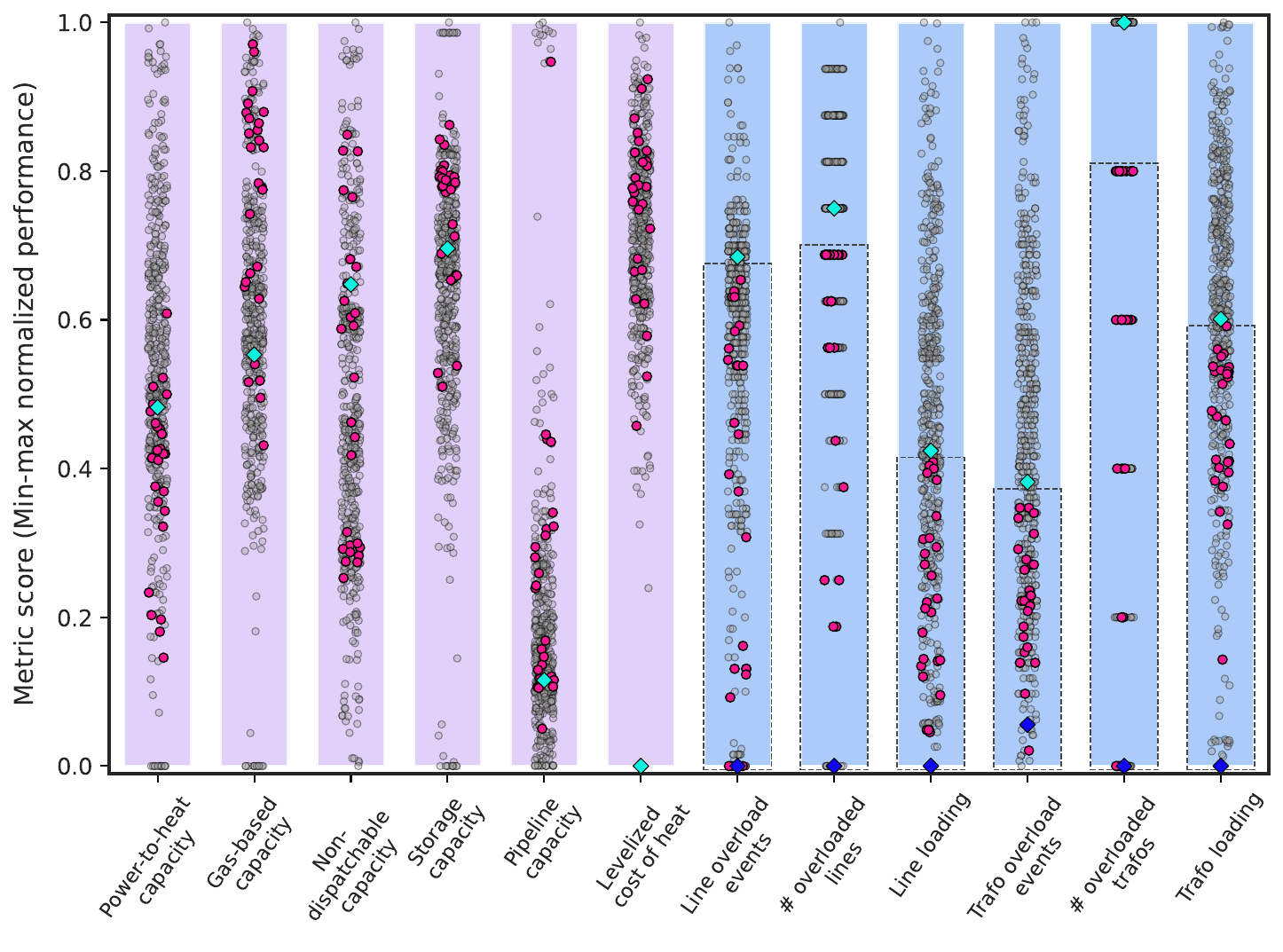}
    
    \subsubsection*{Figure S32. Integrated decision space: cold weather year scenario}
    \phantomsection
    \label{cold_weather_integrated_decision_space}
\end{minipage}

\noindent
\begin{minipage}{\linewidth}
    \centering
    \includegraphics[width=0.8\linewidth]{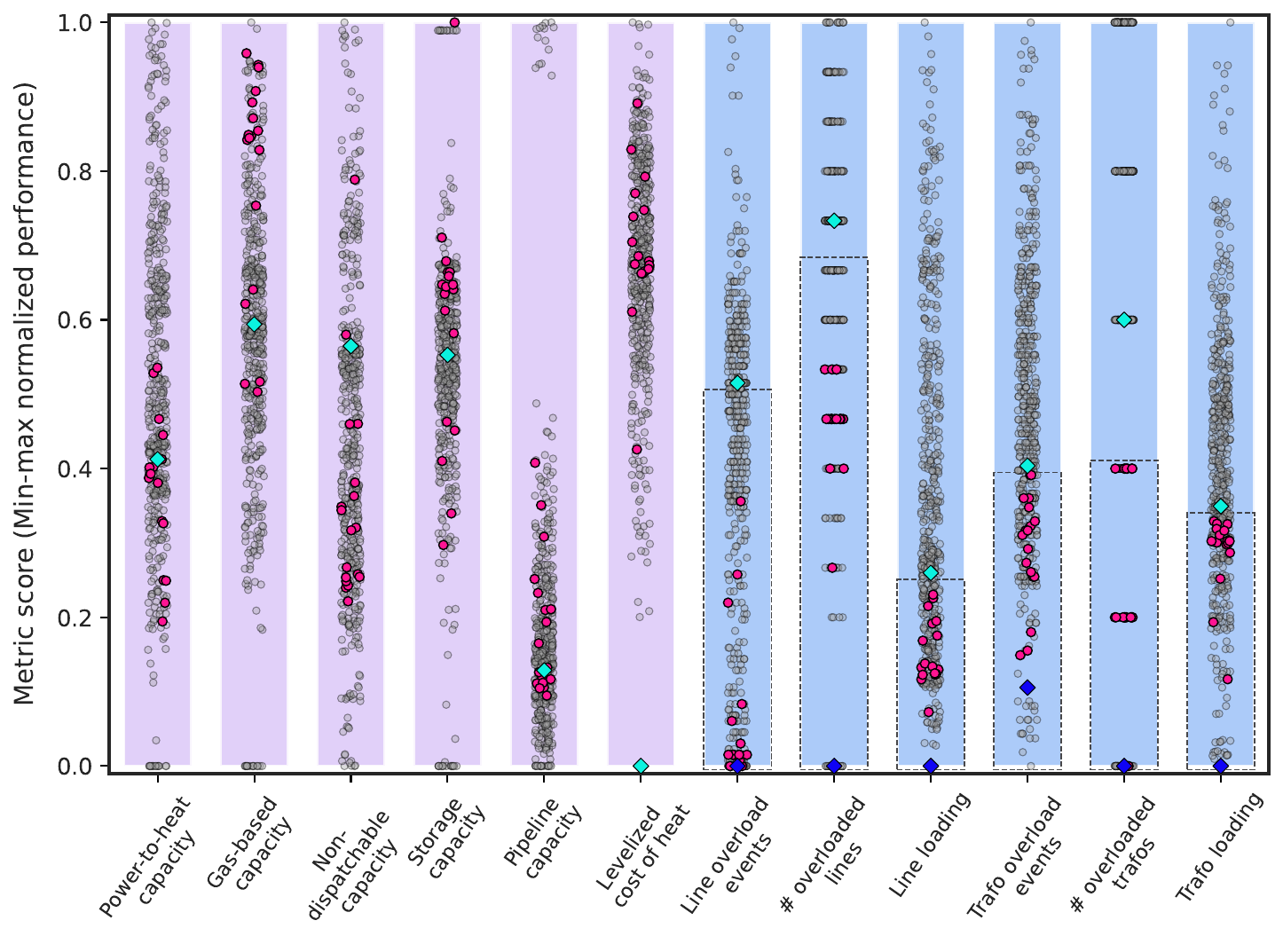}
    
    \subsubsection*{Figure S33. Integrated decision space: low heat demand scenario}
    \phantomsection
    \label{low_heat_demand_integrated_decision_space}
\end{minipage}

\vspace{1em}

\noindent
\begin{minipage}{\linewidth}
    \centering
    \includegraphics[width=0.8\linewidth]{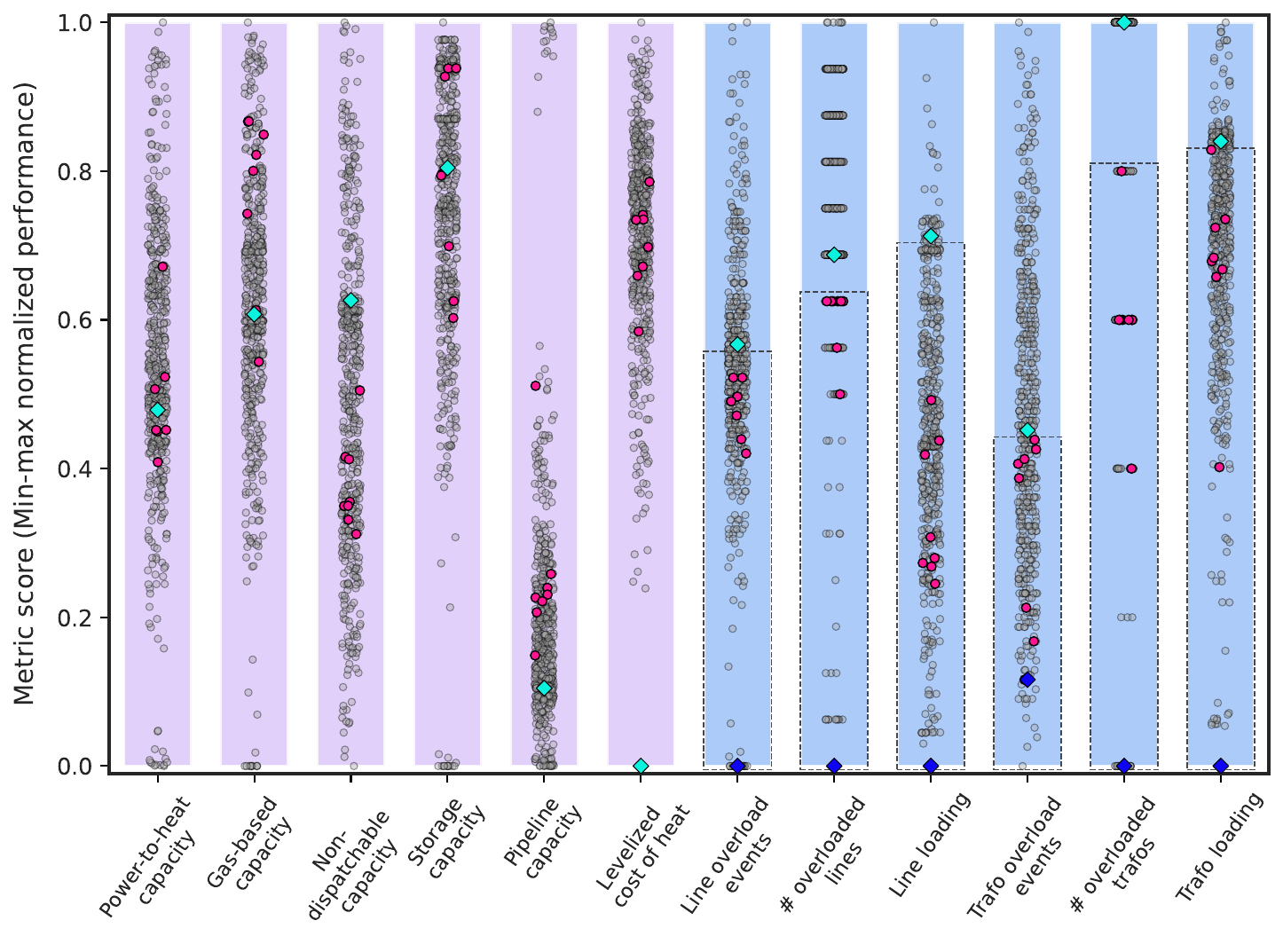}
    
    \subsubsection*{Figure S34. Integrated decision space: high heat demand scenario}
    \phantomsection
    \label{high_heat_demand_integrated_decision_space}
\end{minipage}

\noindent
\begin{minipage}{\linewidth}
    \centering
    \includegraphics[width=0.95\linewidth]{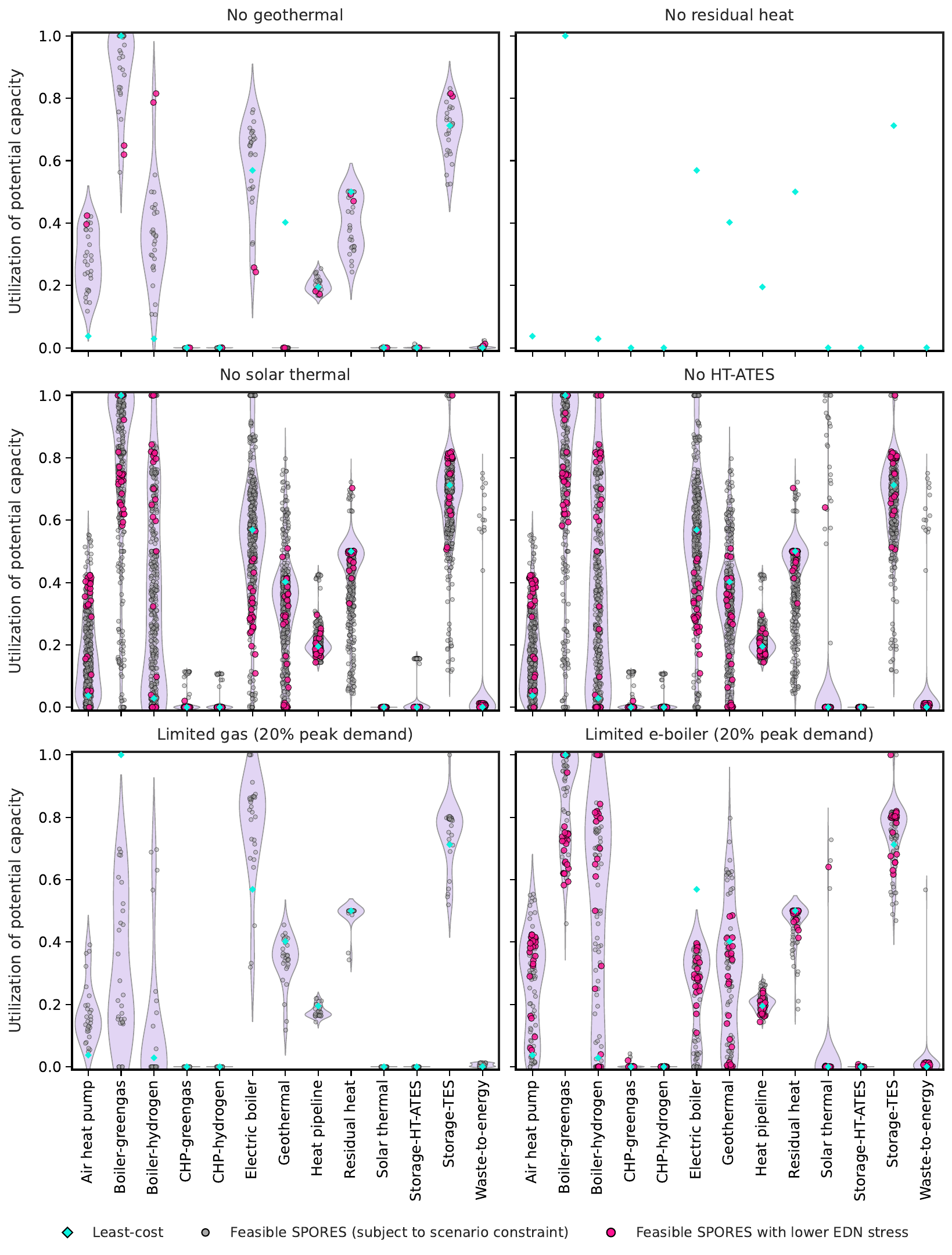}
    
    \subsubsection*{Figure S35. Trade-offs under local technology deployment constraints: 5\% cost slack scenario}
    \phantomsection
    \label{05_slack_constrained_maneuvering_space_fig}
\end{minipage}

\noindent
\begin{minipage}{\linewidth}
    \centering
    \includegraphics[width=0.95\linewidth]{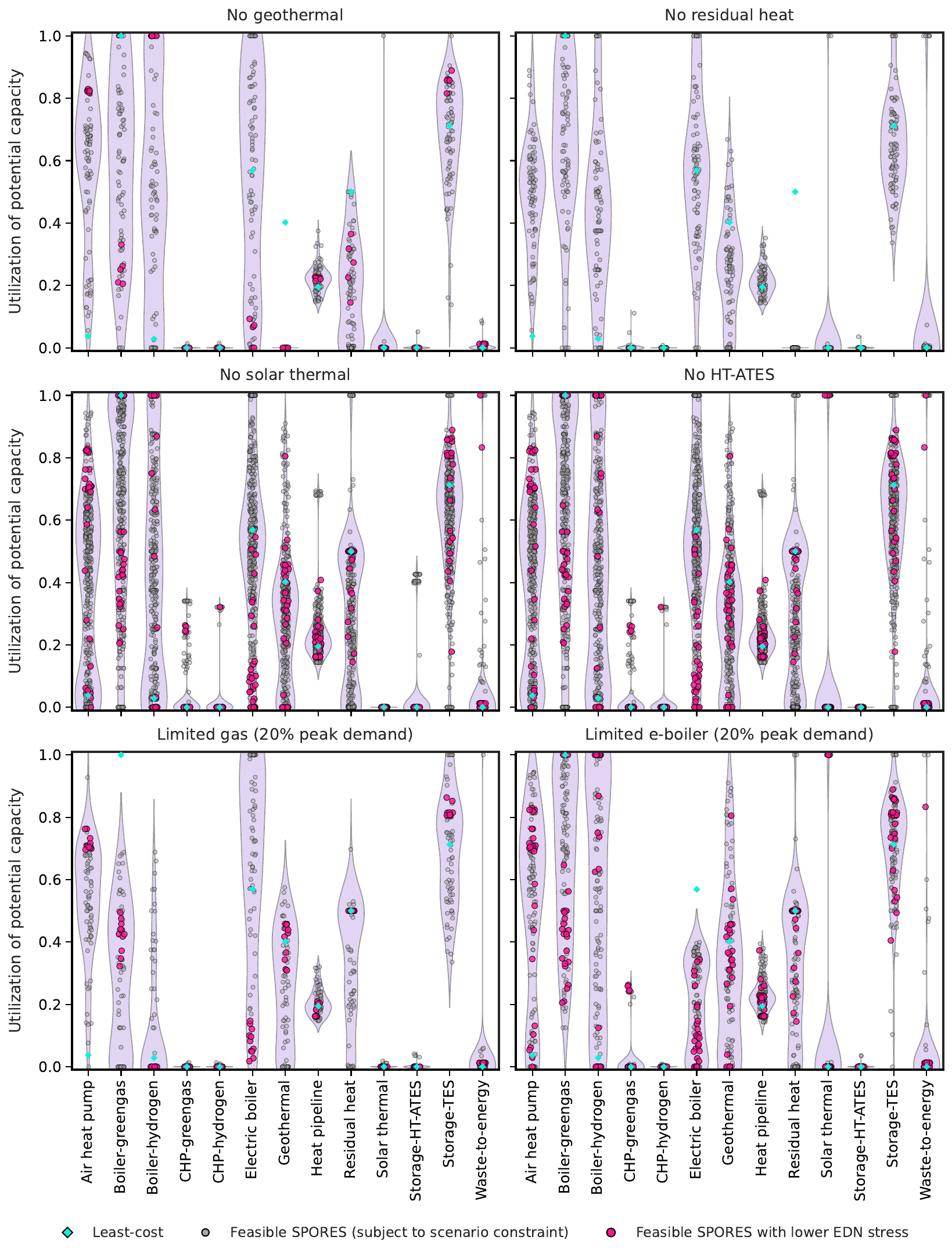}
    
    \subsubsection*{Figure S36. Trade-offs under local technology deployment constraints: 15\% cost slack scenario}
    \phantomsection
    \label{15_slack_constrained_maneuvering_space_fig}
\end{minipage}

\noindent
\begin{minipage}{\linewidth}
    \centering
    \includegraphics[width=0.95\linewidth]{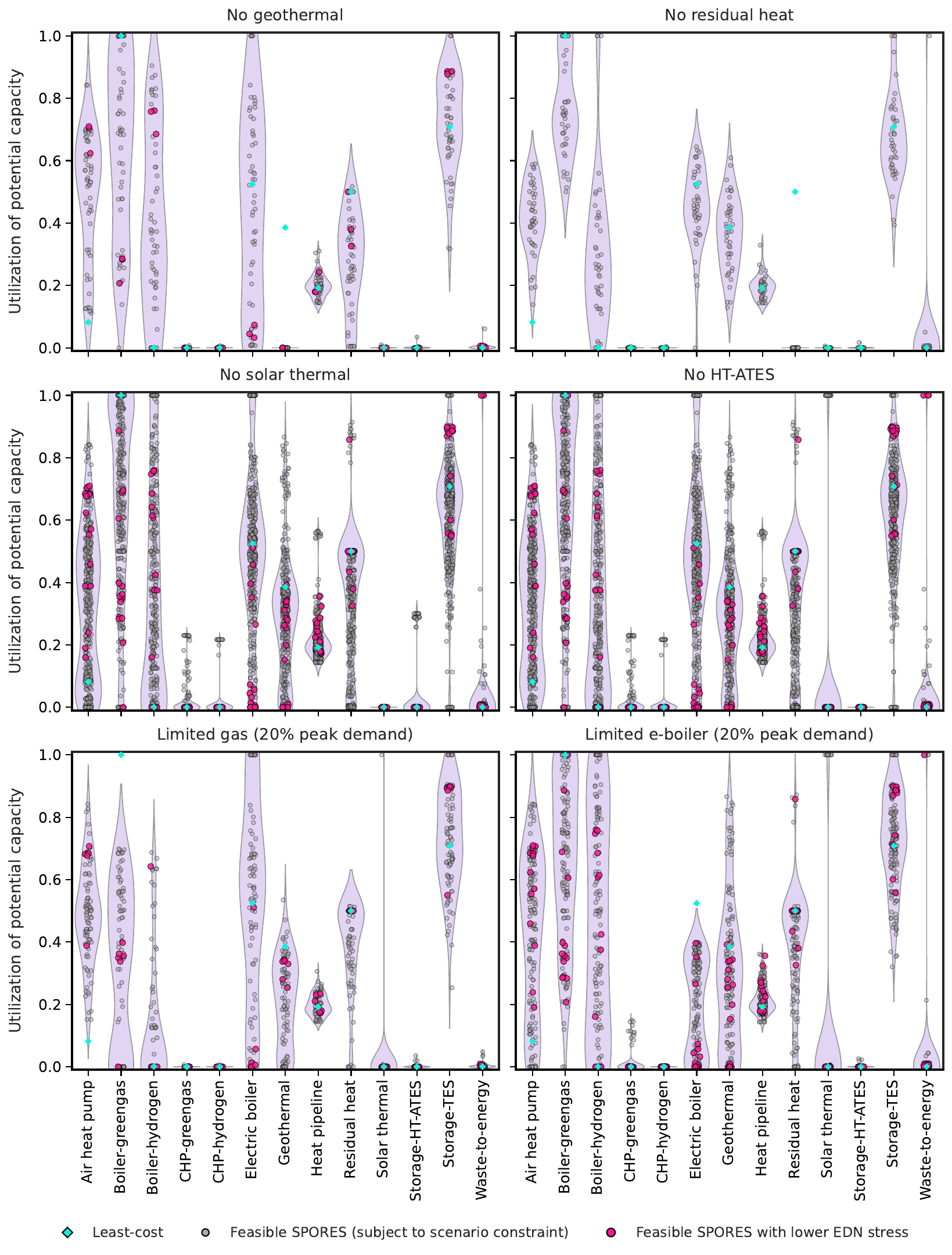}
    
    \subsubsection*{Figure S37. Trade-offs under local technology deployment constraints: warm weather year scenario}
    \phantomsection
    \label{weather_2004_constrained_maneuvering_space_fig}
\end{minipage}

\noindent
\begin{minipage}{\linewidth}
    \centering
    \includegraphics[width=0.95\linewidth]{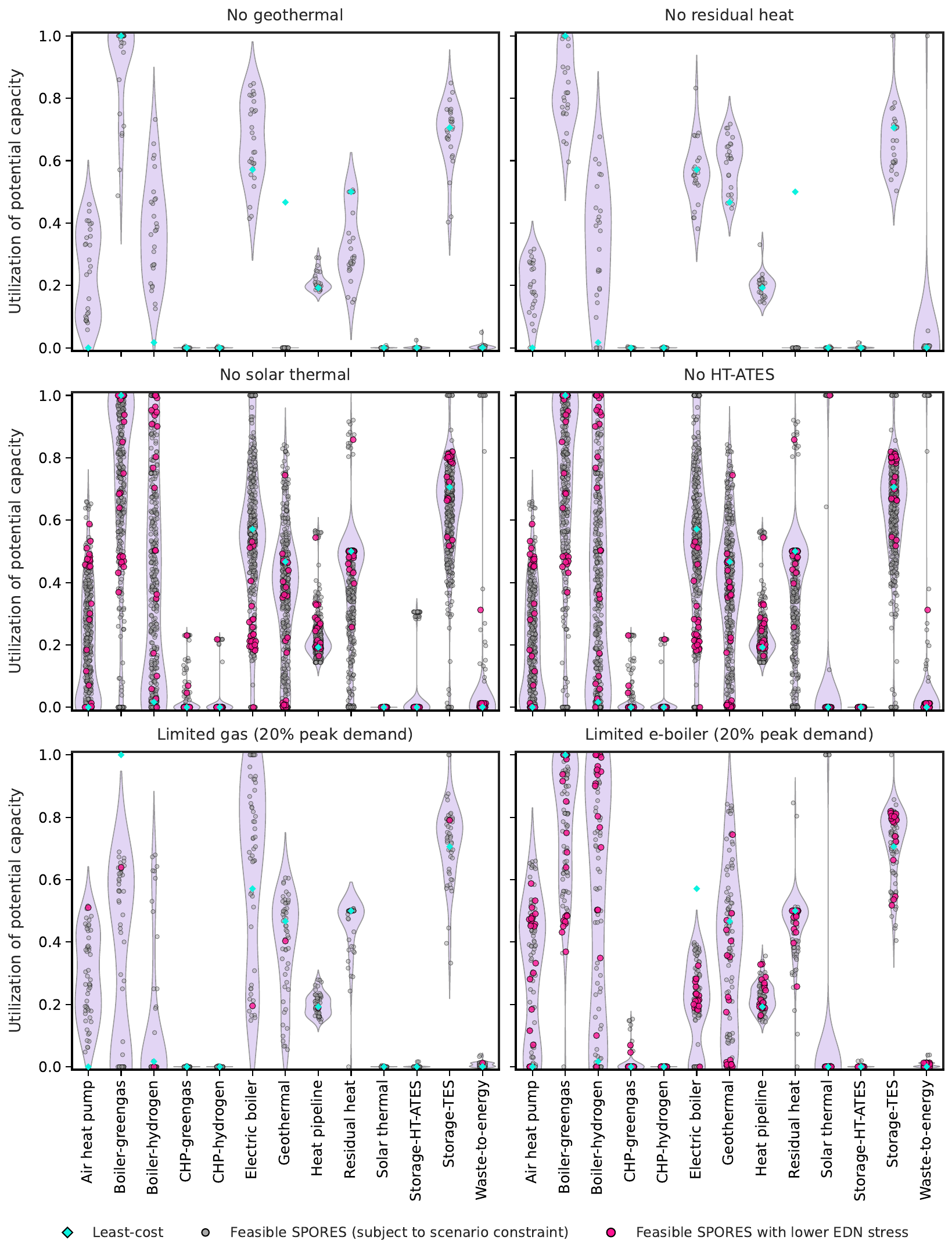}
    
    \subsubsection*{Figure S38. Trade-offs under local technology deployment constraints: cold weather year scenario}
    \phantomsection
    \label{weather_1987_constrained_maneuvering_space_fig}
\end{minipage}

\noindent
\begin{minipage}{\linewidth}
    \centering
    \includegraphics[width=0.95\linewidth]{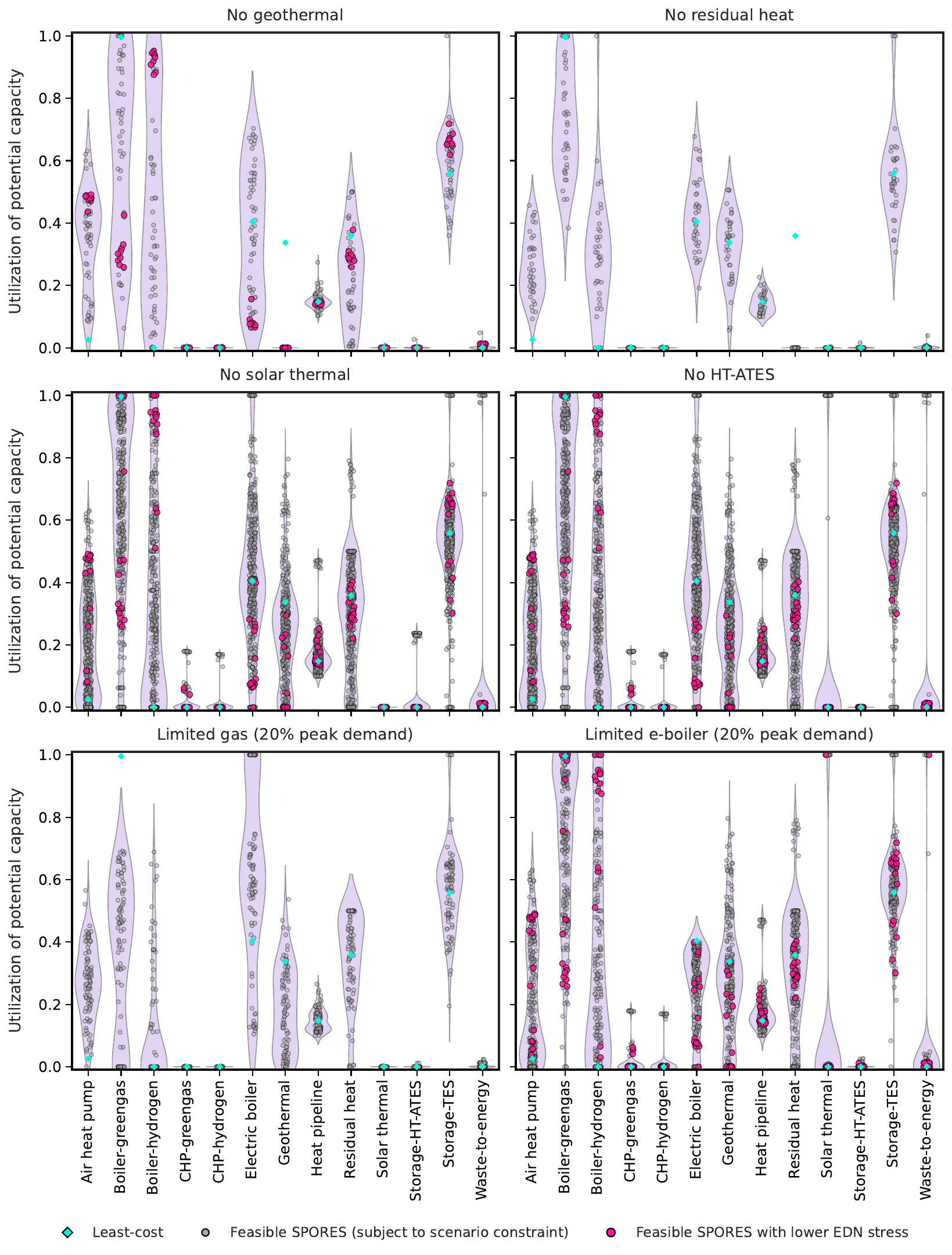}
    
    \subsubsection*{Figure S39. Trade-offs under local technology deployment constraints: low heat demand scenario}
    \phantomsection
    \label{low_demand_constrained_maneuvering_space_fig}
\end{minipage}

\noindent
\begin{minipage}{\linewidth}
    \centering
    \includegraphics[width=0.95\linewidth]{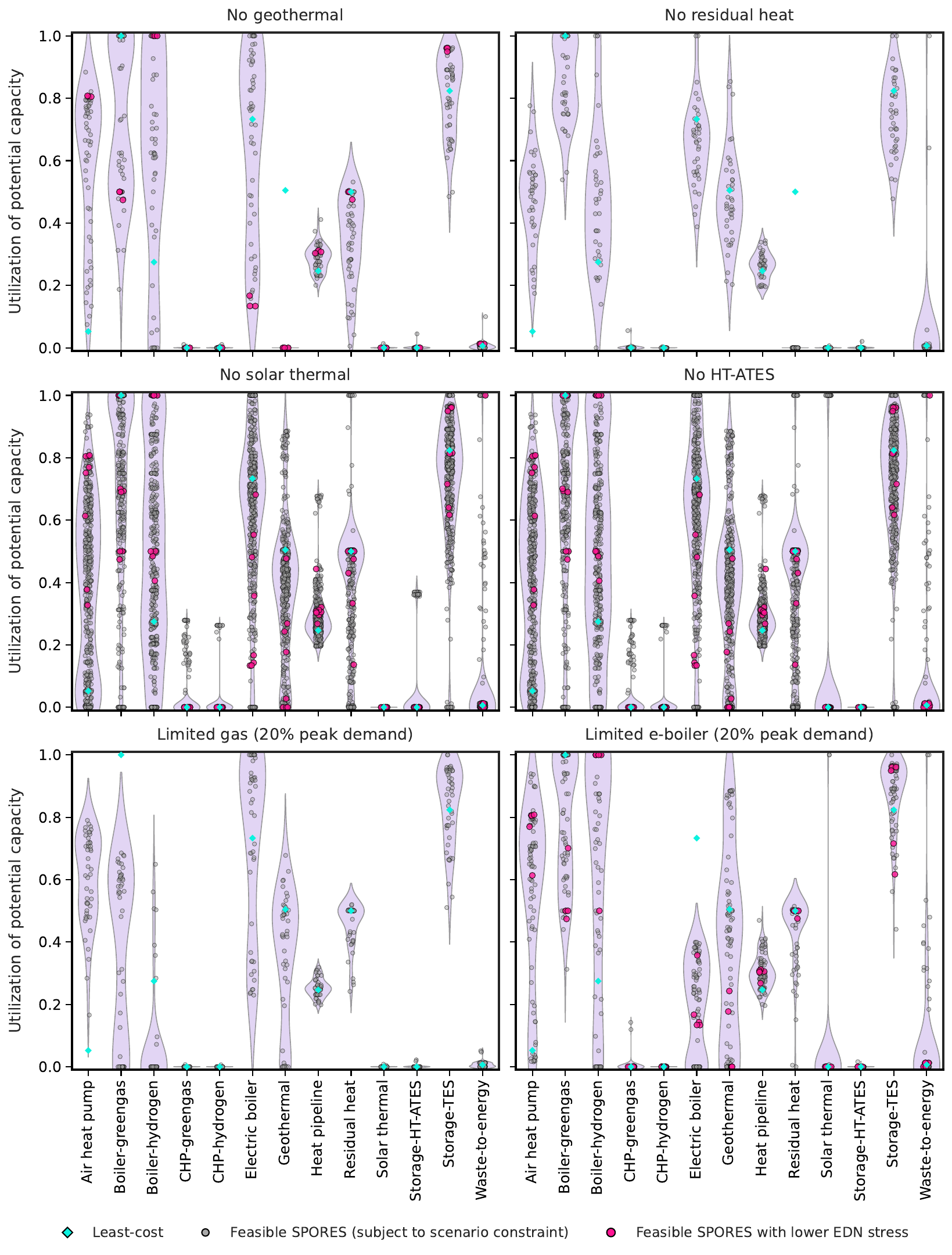}
    
    \subsubsection*{Figure S40. Trade-offs under local technology deployment constraints: high heat demand scenario}
    \phantomsection
    \label{high_demand_constrained_maneuvering_space_fig}
\end{minipage}

\noindent
\begin{minipage}{\linewidth}
    \centering
    \includegraphics[width=0.99\linewidth]{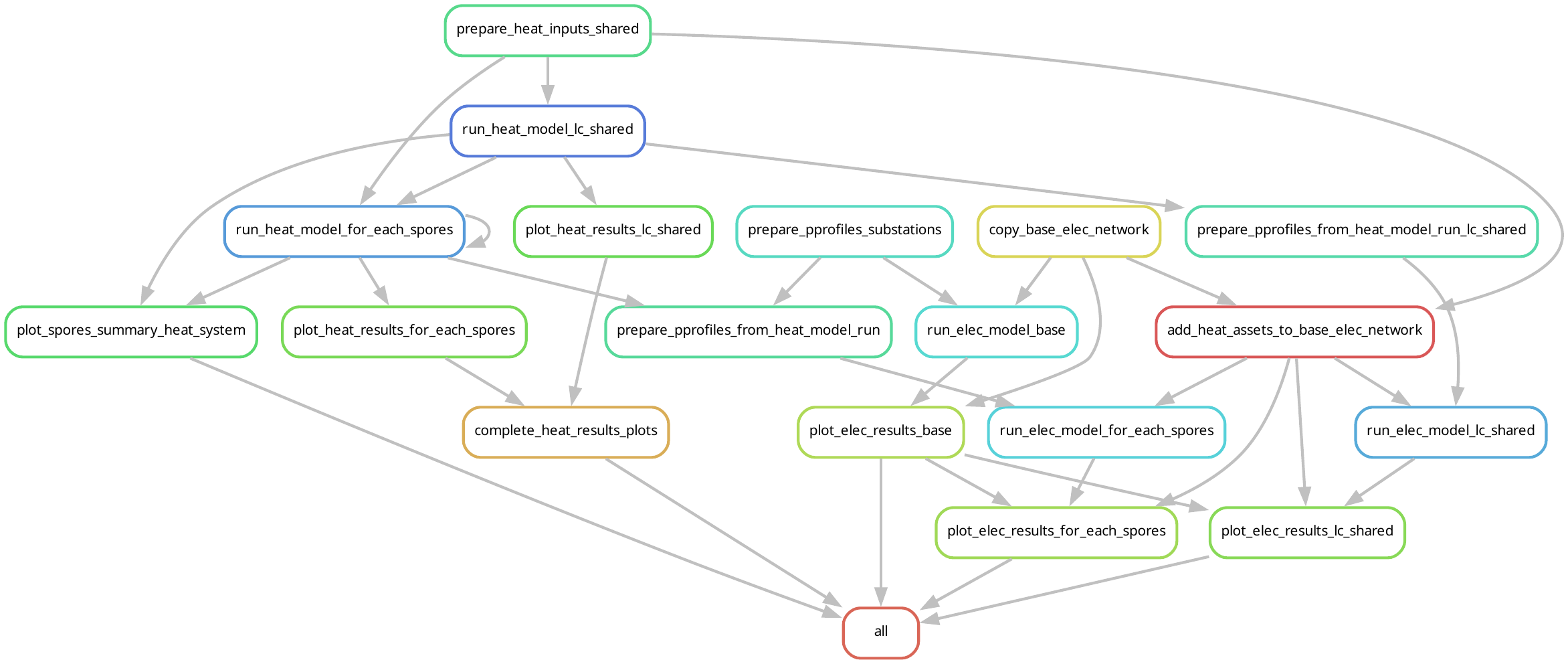}
    
    \subsubsection*{Figure S41. Directed acyclic graph of the model workflow for each scenario}
    \parbox{\linewidth}{%
     Directed acyclic graph of jobs executed for each scenario, illustrating the simplified high-level workflow from raw data processing to generation of SPORES, running the DHN model for each SPORE, executing the power flow simulations, and quantifying electricity network impacts. The loop rule \texttt{run\_heat\_model\_for\_each\_spores} corresponds to the generation of 515 SPORES per scenario, through intensification and diversification of DHN technology deployment.}
    \phantomsection
    \label{snakemake_dag_workflow_fig}
\end{minipage}

\vspace{3em}

\subsubsection*{Table S1. District heating technology parameters}\phantomsection\label{dhn_technology_params_table}

\vspace{-2em}

\small
\begin{longtable}{p{4.2cm} p{5cm} p{4.2cm} p{3cm}}

\label{tab:tech_parameters} \\

\toprule
Technology & Parameter & Value & Unit \\
\midrule
\endfirsthead

\toprule
Technology & Parameter & Value & Unit \\
\midrule
\endhead

\midrule
\multicolumn{4}{r}{Continued on next page} \\
\midrule
\endfoot

\bottomrule
\endlastfoot

Central air-source heat pump 
& investment & 2,039,000 & €/MW\_th \\
& COP & temperature-dependent & -- \\
& lifetime & 15 & years \\
\addlinespace

Central CHP (green gas) 
& investment cost & 2,000,000 & €/MW\_th \\
& heat efficiency & 0.45 & -- \\
& electric efficiency & 0.45 & -- \\
& lifetime & 15 & years \\
\addlinespace

Central CHP (hydrogen) 
& investment cost & 2,100,000 & €/MW\_th \\
& heat efficiency & 0.45 & -- \\
& electric efficiency & 0.45 & -- \\
& lifetime & 15 & years \\
\addlinespace

Central electric boiler 
& investment cost & 305,000 & €/MW\_th \\
& efficiency & 0.9 & -- \\
& lifetime & 15 & years \\
\addlinespace

Central gas boiler (green gas) 
& investment cost & 200,000 & €/MW\_th \\
& efficiency & 0.8 & -- \\
& lifetime & 15 & years \\
\addlinespace

Central gas boiler (hydrogen)
& investment cost & 270,000 & €/MW\_th \\
& efficiency & 0.7 & -- \\
& lifetime & 15 & years \\
\addlinespace

Central (deep) geothermal heat 
& investment cost & 2,500,000 & €/MW\_th \\
& lifetime & 15 & years \\
& Seasonal performance factor & 6 & -- \\
\addlinespace

Central solar thermal heat
& investment cost & 435,000 & €/MW\_th \\
& lifetime & 15 & years \\
\addlinespace

High-temperature residual heat
& investment cost & 1,200,000 & €/MW\_th \\
& lifetime & 15 & years \\
\addlinespace

Central short-term storage (TES)
& investment cost & 250,000 & €/MW \\
& energy-to-power ratio & 6 & hours \\
& charge efficiency & 0.99 & -- \\
& discharge efficiency & 0.99 & -- \\
& lifetime & 30 & years \\
\addlinespace

Seasonal storage (HT-ATES) 
& investment cost & 199,550 & €/MW \\
& charge efficiency & 1 & -- \\
& discharge efficiency & 1 & -- \\
& round-trip efficiency & 0.7 & -- \\
& full load hours & 3,000 & hours \\
& Seasonal performance factor & 50 & -- \\
& lifetime & 30 & years \\
\addlinespace

Heat distribution pipeline (high-temperature network) 
& investment cost & 100 & €/MW/m \\
& maximum capacity & 250 & MW \\
& losses & 0.00001 & \%/m \\
& lifetime & 30 & years \\
\addlinespace

Waste incinerator (waste-to-energy) 
& investment cost & 600,000 & €/MW \\
& efficiency & 0.80 & -- \\
& lifetime & 15 & years \\

\end{longtable}

\end{document}